%% file: main.tex
 \documentclass[final,3p,times,preprint]{elsarticle}

\UseRawInputEncoding

\usepackage{comment}
\usepackage{rotating}
\usepackage{caption}
\usepackage{subcaption}
\usepackage{tablefootnote}
\usepackage{amssymb}
\usepackage{amsmath}
\usepackage{longtable}

\usepackage[ruled]{algorithm2e}
\SetKwInput{KwData}{Input}
\SetKwInput{KwResult}{Updated}

\SetCommentSty{mycommfont}
\SetArgSty{textup}

\usepackage{hyperref}
\hypersetup
{
    colorlinks=true,
    linkcolor=blue,
    filecolor=magenta,      
    urlcolor=cyan,
}
\usepackage{multirow}
\usepackage{hhline}
\usepackage{tikz}
\usepackage{soul}

\usetikzlibrary{fadings}
\usetikzlibrary{patterns}
\usetikzlibrary{shadows.blur}
\usetikzlibrary{shapes}

\usepackage{framed} 
\usepackage{multicol} 
\usepackage{nomencl} 
\makenomenclature
\renewcommand*\nompreamble{\begin{multicols}{2}}
\renewcommand*\nompostamble{\end{multicols}}

\makeatletter
\def\smallunderbrace#1{\mathop{\vtop{\m@th\ialign{##\crcr
   $\hfil\displaystyle{#1}\hfil$\crcr
   \noalign{\kern3\p@\nointerlineskip}%
   \tiny\upbracefill\crcr\noalign{\kern3\p@}}}}\limits}
\makeatother

\journal{Computer Physics Communications}

\begin{document}

\begin{frontmatter}



\title{GPU-Native Adaptive Mesh Refinement with Application to Lattice Boltzmann Simulations}


\author[inst1]{Khodr Jaber\corref{email1}}
\cortext[email1]{Corresponding author}
\ead{khodr.jaber@mail.utoronto.ca}

\affiliation[inst1]{organization={Department of Mechanical and Industrial Engineering, University of Toronto},
addressline={5 King's College Rd}, 
            city={Toronto},
            postcode={M5S 3G8}, 
            state={ON},
       country={Canada}
            }
,
\affiliation[inst2]{organization={Department of Mechanical, Industrial and Aerospace Engineering,  Concordia University  },
addressline={1515 Ste-Catherine St. W.}, 
            city={Montreal},
            postcode={H3G 1M8}, 
            state={QC},
            country={Canada}
            }            

\author[inst2]{Ebenezer E. Essel}
\author[inst1]{Pierre E. Sullivan}



\begin{abstract}
%
%
%
%
Adaptive Mesh Refinement (AMR) enables efficient computation of flows by providing high resolution in critical regions while allowing for coarsening in areas where fine detail is unnecessary. While early AMR software packages relied solely on CPU parallelization, the widespread adoption of heterogeneous computing systems has led to GPU-accelerated implementations. In these hybrid approaches, simulation data typically resides on the GPU, and mesh management and adaptation occur exclusively on the CPU, necessitating frequent data transfers between them. A more efficient strategy is to adapt and maintain the entire mesh structure exclusively on the GPU, eliminating these transfers. Because of its inherent parallelism, the Lattice Boltzmann Method (LBM) has been widely implemented in hybrid AMR frameworks. This work presents a GPU-native algorithm for AMR using a block-based forest of octrees approach, implemented in both two and three dimensions as open-source C++/CUDA code. The implementation includes a Lattice Boltzmann solver for weakly compressible flow, though the underlying grid refinement procedure is compatible with any solver operating on cell-centered block-based grids. The lid-driven cavity and flow past a square cylinder benchmarks validate the algorithm's effectiveness across multiple velocity sets in both single- and double-precision. Tests conducted on consumer and datacenter-grade GPUs demonstrate its versatility across different hardware platforms.

\noindent Link to repository: \url{https://github.com/KhodrJ/AGAL}
\end{abstract}




\end{frontmatter}



\input{p0_literature}
\input{p1_amr}
\input{p2_solver.tex}
\input{p3_tests.tex}
\input{p4_conclusion.tex}




\section{Acknowledgements}

The authors gratefully acknowledge support from the Natural Sciences and Engineering Research Council of Canada, Canadian Microelectronics Corporation and the Digital Research Alliance of Canada.


\appendix

\input{pA_appendix.tex}

\input{nomenclature}

\printnomenclature
\bibliographystyle{elsarticle-num} 
\bibliography{refs}





\end{document}

%% file: p0_literature.tex
\section{Introduction}
\label{sec:intro}
In computational fluid dynamics (CFD), the Lattice Boltzmann Method (LBM) has evolved from a simple algorithm for weakly-compressible hydrodynamics at moderate Reynolds number to a more advanced family of schemes capable of simulating turbulence \cite{Hou1996, Yu2002, Hossain2015, Gaedtke2018}, free-surface flows \cite{Watanabe2021, Watanabe2023, Plewinski2024}, and phase-field simulations in material science \cite{Sakane2022, Sakane2022i, Sakane2023}. The LBM models the Navier-Stokes equations from a mesoscopic description of the flow physics, and possesses advantages in contrast with conventional Navier-Stokes solvers (e.g., the finite difference and volume method) such as a natural treatment of advection, fully explicit temporal integration without the need to invert a linear system of equations per time step \cite{LBM_Book}, and natural parallelism in the form of an embarrassingly parallel collision step and a streaming step which can exploit techniques such as pointer shifting \cite{Mohrhard2019}. The natural parallelism of the LBM has been exploited to implement a variety of central processing units (CPUs), graphics processing units (GPUs), and hybrid parallelization techniques to reduce the computational cost of simulations, resulting in the development of several publicly available codes \cite{waLBerla, ESPResSo, Palabos, FluidX3D, TCLB, HemeLB, Neon}. The uniform grid requirement imposed by the discretization strategy of the standard LBM is well-suited for GPU implementations. However, it is insufficient for applications where greater resolution is needed only in a few regions of the domain (e.g., turbulent flows where the turbulent boundary layer needs to be resolved). 
Increasing the resolution locally requires refining the whole domain, dramatically increasing computational overhead and potentially rendering the flow simulation infeasible due to memory constraints. This has motivated the development of local grid refinement schemes for the LBM to enable simulation on non-uniform grids \cite{Filippova1998,Lin2000,Dupuis2003,Rohde2006,Chen2006,Tolke2006,Geier2009,Lagrava2012,Fakhari2014,Schukmann2023}.
\begin{figure}[t]
    \centering
   \input{tikz/velocitysets.tikz}
    \caption{Visualization of discretization of particle velocity space into discrete velocity sets. Three sets are shown: D2Q9 on the left, D3Q19, and D3Q27 on the right (the latter shares the same velocity vectors as the former but requires, in addition to them, the velocities pointing through the cell corners. Velocity sets were adapted from \cite{LBM_Book}.}
    \label{fig:p0_velocitysets}
\end{figure}

Adaptive mesh refinement (AMR) \cite{Berger1984, Berger1988, Berger1989} enables targeted refinement in the domain at runtime according to user-defined criteria (e.g., vorticity magnitude and the Q-criterion in hydrodynamics \cite{Fakhari2014}). Numerous AMR software packages have been developed~\cite{PARAMESH, Chombo, SAMRAI, FLASH, Dendro, Octor, EnzoP, p4est, Daino, GAMER2, AMReX}. The technique has been applied to disciplines in physics and engineering such as shallow-water equations modeling of tsunamis, dam breaks and tidal waves \cite{Saetra2015, Asuncion2017}, gas dynamics \cite{Beckingsale2014, Giuliani2019, Menshov2020}, phase-field modeling for simulation of dendrites \cite{Sakane2022, Sakane2022i}, and free-surface flows \cite{Watanabe2021, Watanabe2023}. It is common to categorize the approaches to AMR as either octree or block-structured adaptive mesh refinement (SAMR) \cite{Dubey2014, Dubey2021}. In octree AMR, cells are organized hierarchically in an octree whose nodes correspond to the blocks so that explicit parent-child relationships exist. In contrast, cells can be organized in any logically rectangular shape in a SAMR as long as they reside on the same level in the hierarchy, and explicit parent-child relationships do not exist. AMR approaches have also been categorized as cell-, block- and patch-based AMR approaches \cite{Asuncion2017, Giuliani2019, Qin2019, Dunning2020} in which the patches are defined as arbitrary collections of clustered cells, and blocks are defined as a particular case of square/cubic patches. Cell-based AMR leads to a tree structure where nodes correspond to individual cells. The distinction between these terms can vary between communities. For example, Schornbaum and R{\"u}de \cite{Schornbaum2018} describe a block-structured AMR where the domain is partitioned as a forest of octrees similar to \cite{EnzoP, Burstedde2013}. 
This work adopts the cell-and block-based AMR terminology defined in the second categorization, which encompasses block-based octree AMR while consistently using patch-based and SAMR concepts.

Recent advancements in General-Purpose GPU (GPGPU) programming and the increased availability of hybrid CPU-GPU computing systems have motivated GPU-accelerated AMR, where computations over the solution fields are offloaded to the GPU. Packages that offer GPU support include Daino \cite{Daino}, GAMER-2 \cite{GAMER2}, AMReX \cite{AMReX}, and waLBerla \cite{waLBerla}. Early GPU-based AMR packages would handle most of the AMR algorithm on the CPU while offloading suitable computations, requiring costly copies of data between the CPU and GPU, \cite{Wang2010, GAMER}. It is now common for AMR packages to permanently retain the data on the GPU to be processed entirely therein due to the increased availability of virtual RAM on modern GPUs, reducing these copies significantly. However, management and dynamic adaptation of the mesh are considered challenging to parallelize on the GPU \cite{Luo2016, Qin2019, Chowdhury2023}, and specific subroutines may require CPU execution depending on the chosen AMR approach. Compute kernels in a SAMR approach are highly suited for GPU execution due to the regular structure of the patches on which the data resides; however, the clustering algorithm involved during refinement is not straightforward to port to a GPU and involves the CPU at some point \cite{Saetra2015, Beckingsale2015, Qin2019, Dunning2020}. Recursive data structures can define and dynamically modify quad/octree meshes (e.g., node insertion/removal, neighbor-search). However, these are ill-suited for deployment on GPUs, where contiguity in memory is crucial for performance. As a result, the mesh structure often continues to be hosted and managed, in whole or part, on the CPU; this strategy has been reported explicitly in some AMR packages such as Daino \cite{Daino} and GAMER-2 \cite{GAMER2}, and in recent applications \cite{Qin2019, Watanabe2021, Sakane2022, Sakane2022i, Zaghi2023}.

Some have gone further, moving the mesh structure and adaptation routines to the GPU in addition to the solution data to achieve a GPU-native implementation of AMR that eliminates data transfers \cite{Luo2016, Alhadeff2016, Giuliani2019, Pavlukhin2019, Menshov2020, Wang2024}. Two common concerns are generally addressed in doing so: 1) linear representation of the mesh that can enable efficient refinement/coarsening, and 2) a strategy to deal with 'gaps' that inevitably form in these linear data structures after coarsening (since the data no longer corresponds to a region in the grid). Cell-based AMR schemes for the 2D Euler equations were developed in which the mesh adaptation and solver routines were implemented entirely on the GPU by Luo et al. \cite{Luo2016}, for unstructured quadrilateral meshes, and Giuliani and Krivodonova \cite{Giuliani2019}, for unstructured triangular meshes. The mesh was described with a cell-edge decomposition and represented with integer lists that facilitated refinement, coarsening, and 'recycling' of the coarsened cells/edges indices for reuse in later refinements. Menshov and Pavlukhin \cite{Pavlukhin2019, Menshov2020} developed a cell-based GPU-native AMR algorithm for the same system of equations using a linear octree represented with a Z-order Morton space-filling curve (SFC) \cite{Morton1966}. They eliminated gaps with a defragmentation strategy based on a prefix-scan operation. Wang et al. \cite{Wang2024} developed a cell-based GPU-native AMR scheme for the flux reconstruction method based on a linear octree \cite{Sundar2008} represented with a Z-order SFC. Algorithms for tree manipulation were redesigned, including routines for building mesh connectivity and 2:1 balancing \cite{Sundar2008} (which ensures that a coarse octant neighboring a fine one is no more than double its size) since previous CPU-based algorithms \cite{Sundar2008,p4est,Weinzierl2011} utilized data structures which were not well-suited for GPU execution.

Several open-source codes have successfully implemented the GPU-parallel LBM with static and dynamic grid refinement. Palabos \cite{Palabos} recently introduced a multi-GPU backend to the original repository that implements the former. waLBerla \cite{waLBerla} employs a forest of octrees approach similar to that of Burstedde et al. \cite{Burstedde2008, p4est} which is 
organized so that nodes corresponding to blocks in the grid can be distributed via MPI to different processes \cite{Godenschwager2013, Schornbaum2016, Schornbaum2018}. waLBerla provides GPU support by mirroring CPU data on the device to enable complete kernel execution, with copies back to the CPU only for post-processing and I/O. ESPResSo \cite{ESPResSo}, a package for soft matter systems that had initially employed regular Cartesian grids, was extended \cite{Lahnert2016, Mehl2019} to enable AMR with the p4est framework \cite{p4est}. ESPResSo has also integrated waLBerla for its hydrodynamics. The underlying forest-of-octrees AMR frameworks of ESPResSo and waLBerla were designed for extreme-scale AMR on CPU-based supercomputers, involving parallelization and distribution of the grid via MPI across many processes, with scalability demonstrated over $\mathcal{O}(10^5)$ processes \cite{p4est, Schornbaum2018, Schornbaum2018Thesis}. MPI-based distribution is necessary when executing on compute clusters with distributed memory, even if the cluster offers GPU acceleration. Continued CPU-parallel mesh adaptation is naturally attractive, especially when such scalability has been demonstrated. However, sub-optimal performance is possible when porting without optimization specifically for GPU execution. This was observed, for example, by Mahmoud et al. \cite{Mahmoud2024}, who implemented the LBM with a static grid refinement scheme optimized specifically for single-GPU execution (in contrast with the previous packages where a previous CPU code was ported to the GPU) in the Neon package \cite{Neon}. Their code was compared with Palabos and waLBerla in test cases employing a single GPU, and they observed a reduction in the total simulation time. Other recent AMR-LBM works, such as the phase-field simulations of Sakane et al. \cite{Sakane2022, Sakane2022i, Sakane2023} mention the use of the framework of Schive et al. implemented in the GAMER-2 package \cite{GAMER2}.

We present a novel block-based forest of octrees approach to AMR hosted and managed entirely on a single GPU. Integer index arrays explicitly identify the nodes of the octrees to facilitate mesh adaptation. The refinement and coarsening algorithm consists of parallel subroutines (such as sorts and copies) implemented in Thrust \cite{Thrust} (part of Nvidia's CUDA Core Compute Libraries \cite{CCCL}) and additional compute kernels designed straightforwardly to: 1) utilize explicit neighbor-link storage, shared memory and arithmetic to update connectivity, and 2) to revert violating blocks flagged for refinement/coarsening to enforce 2:1 balancing. 
This paper also describes a solver based on the Lattice Boltzmann Method for weakly compressible flow, compatible with the mesh arrangement in 2D and 3D. Linear and cubic interpolation communicate data from the coarse grid to the fine grid along the refinement interface, while basic averaging transfers data in the reverse direction. The streaming step uses shared memory for data transfers between blocks on the same grid level. Separate C++/CUDA scripts implement the refinement and coarsening algorithm (\texttt{mesh\_amr.cu}) and solver (\texttt{solver\_lbm*.cu}) to enable utilization with other LBM codes or numerical solvers in future work. Two benchmark problems validate the solver: the 2D/3D lid-driven cavity and the 2D flow past a square cylinder. We report the performance of the code with respect to the distribution of subroutine execution times for the AMR scheme, node updates per second for the solver, and speedup provided relative to uniform grids with equivalent effective resolution.

The paper is structured as follows. Section \ref{sec:lbm} introduces the Lattice Boltzmann Method. Section \ref{sec:amr} provides an overview of the data structures and memory access patterns, and details the refinement and coarsening algorithm and the coarse-fine grid communication routines.
Section \ref{sec:solver} outlines the recursive time-stepping algorithm, the kernels for streaming and collision, and computation of the refinement criterion. Section \ref{sec:tests} reports the results of the lid-driven cavity and flow past a square cylinder test cases. Section \ref{sec:conc} presents our conclusion.

\section{Lattice Boltzmann Method}\label{sec:lbm}
A time step for the LBM can be expressed as
\begin{align}
    f_p(t+\Delta t,\textbf{x}+\textbf{c}_p\Delta x) - f_p(t,\textbf{x}) = \Delta t \Omega(t,\textbf{x}),
\end{align}

\noindent where $\Omega$ is the collision operator, $f_p$ are the density distribution functions (DDFs) corresponding to the particle velocity vectors $\textbf{c}_p = (\Delta x/\Delta t)\textbf{e}_p$, $t$ and $\textbf{x}$ are the current time and spatial location, and $\Delta t$ is the time step. The $f_p$ are arranged on a lattice of equal spacing $\Delta x$ in all directions with edges that align with the Cartesian axes. The spatial and temporal steps are set equal so that the increments $\textbf{x} + \textbf{c}_p \Delta t$ lie exactly on the neighboring lattice nodes. The analytical expression for the operator is complex and is usually replaced with a simpler model, such as the Bhatnagar-Gross-Crook (BGK) model \cite{Bhatnagar1954} involving linear relaxation of the DDFs at a rate $\tau$ towards equilibrium distributions obtained from a discretization of the Maxwell-Boltzmann distribution. The multiple-relaxation-time (MRT) operator \cite{Lallemand2000, dHumieres2002} allows the kinematic and bulk viscosity to be specified independently. The BGK model (also referred to as the single-relaxation-time model) is used in this study and is given by:
\begin{align}
    f_p(t+\Delta t,\textbf{x}+\textbf{c}_p\Delta x) &= f_p(t,\textbf{x}) -\frac{\Delta t}{\tau} \bigg(f_p(t,\textbf{x}) - f_p^{\text{eq.}}(t,\textbf{x})\bigg), \\
    f_p^{\text{eq.}}(t,\textbf{x}) &= w_p \rho\bigg( 1 + \frac{\textbf{u}\cdot\textbf{c}_p}{c_s^2} - \frac{(\textbf{u}\cdot\textbf{c}_p)^2}{2c_s^4} + \frac{\textbf{u}\cdot\textbf{u}}{2c_s^2} \bigg),
\end{align}
where $c_s$, denoted the lattice speed of sound, is given by $c_s = (1/\sqrt{3})(\Delta x/\Delta t$, $f_p^{\text{eq.}}$ are the discrete equilibrium distributions, and $w_p$ are quadrature weights corresponding to the abscissae $c_p$. The $w_p$ and $\textbf{c}_p$ form a so-called velocity set with the notation D$n$Q$m$ where $n$ and $m$ are the problem dimension and velocity set size (these are $N_D$ and $N_Q$, respectively). This work considers three velocity sets: D2Q9, D3Q19 and D3Q27. These are illustrated in Figure \ref{fig:p0_velocitysets}. The moments are computed via Gauss-Hermite quadrature in discrete $f_p$ by:
\begin{align}
    \rho(t,\textbf{x}) = &\sum_{p=1}^{N_Q} f_p(t,\textbf{x}), \quad\quad \rho\textbf{u}(t,\textbf{x}) = \sum_{p=1}^{N_Q} f_p (t,\textbf{x}) \textbf{c}_p.
\end{align}

Boundary conditions specified in terms of the macroscopic properties $\rho$, $\textbf{u}$ must be translated to equivalent conditions on the DDFs. The form and implementation of these conditions depend on the chosen grid type. These are classified as wet-node schemes when nodes coincide with cell vertices along the boundary and link-wise schemes otherwise \cite{LBM_Book}. The latter are used for the current cell-centered grids where the boundary lies in-between nodes, while wet-node schemes are used for vertex-centered grids. Two boundary conditions are implemented for the test cases: bounce-back conditions where a known velocity $\textbf{u}_{\text{w}}$ is specified at the boundary to implement no-slip and inlet conditions, and anti-bounce-back conditions were a specified pressure field $p_{\text{w}}$ implements an outflow condition. These boundary conditions are, respectively, given by \cite{LBM_Book}:
\begin{align}
    \text{Known \ $\textbf{u}_w$}: \quad f_{\overline{i}} &= f_i^* - 2w_i \rho_w \frac{\textbf{c}_i \cdot \textbf{u}_w}{c_s^2}, \\
    \text{Known \ $\textbf{p}_w$}: \quad f_{\overline{i}} &= -f_i^* + 2w_i \rho_w \bigg(1 + \frac{\textbf{c}_i \cdot \textbf{u}_w}{2c_s^4} - \frac{\textbf{u}_w \cdot \textbf{u}_w}{2c_s^2}\bigg).
\end{align}

\noindent Density is related to pressure $p$ by $(p-\overline{p}) = c_s^2 (\rho-\overline{\rho})$ in the incompressible model, where $\overline{p}$ and $\overline{\rho}$ are baseline values for pressure and density, respectively. 
We set density $\rho_w$ equal to $1$ in both boundary conditions and use reference values $\overline{p}=0$ and $\overline{\rho}=1$.

%% file: tikz/velocitysets.tikz
\tikzset{every picture/.style={line width=0.75pt}} 

\begin{tikzpicture}[x=0.75pt,y=0.75pt,yscale=-1,xscale=1]

\draw [color={rgb, 255:red, 159; green, 0; blue, 18 }  ,draw opacity=1 ]   (527.05,64.28) -- (410.95,199.72) ;
\draw [shift={(409,202)}, rotate = 310.6] [fill={rgb, 255:red, 159; green, 0; blue, 18 }  ,fill opacity=1 ][line width=0.08]  [draw opacity=0] (3.57,-1.72) -- (0,0) -- (3.57,1.72) -- cycle    ;
\draw [shift={(529,62)}, rotate = 130.6] [fill={rgb, 255:red, 159; green, 0; blue, 18 }  ,fill opacity=1 ][line width=0.08]  [draw opacity=0] (3.57,-1.72) -- (0,0) -- (3.57,1.72) -- cycle    ;
\draw [color={rgb, 255:red, 159; green, 0; blue, 18 }  ,draw opacity=1 ]   (430.49,64.6) -- (507.51,199.4) ;
\draw [shift={(509,202)}, rotate = 240.26] [fill={rgb, 255:red, 159; green, 0; blue, 18 }  ,fill opacity=1 ][line width=0.08]  [draw opacity=0] (3.57,-1.72) -- (0,0) -- (3.57,1.72) -- cycle    ;
\draw [shift={(429,62)}, rotate = 60.26] [fill={rgb, 255:red, 159; green, 0; blue, 18 }  ,fill opacity=1 ][line width=0.08]  [draw opacity=0] (3.57,-1.72) -- (0,0) -- (3.57,1.72) -- cycle    ;
\draw [color={rgb, 255:red, 159; green, 0; blue, 18 }  ,draw opacity=1 ]   (526.32,160.66) -- (411.68,103.34) ;
\draw [shift={(409,102)}, rotate = 26.57] [fill={rgb, 255:red, 159; green, 0; blue, 18 }  ,fill opacity=1 ][line width=0.08]  [draw opacity=0] (3.57,-1.72) -- (0,0) -- (3.57,1.72) -- cycle    ;
\draw [shift={(529,162)}, rotate = 206.57] [fill={rgb, 255:red, 159; green, 0; blue, 18 }  ,fill opacity=1 ][line width=0.08]  [draw opacity=0] (3.57,-1.72) -- (0,0) -- (3.57,1.72) -- cycle    ;
\draw [color={rgb, 255:red, 159; green, 0; blue, 18 }  ,draw opacity=1 ]   (506.6,103.8) -- (431.4,160.2) ;
\draw [shift={(429,162)}, rotate = 323.13] [fill={rgb, 255:red, 159; green, 0; blue, 18 }  ,fill opacity=1 ][line width=0.08]  [draw opacity=0] (3.57,-1.72) -- (0,0) -- (3.57,1.72) -- cycle    ;
\draw [shift={(509,102)}, rotate = 143.13] [fill={rgb, 255:red, 159; green, 0; blue, 18 }  ,fill opacity=1 ][line width=0.08]  [draw opacity=0] (3.57,-1.72) -- (0,0) -- (3.57,1.72) -- cycle    ;
\draw [line width=0.75]  [dash pattern={on 0.84pt off 2.51pt}]  (479,162) -- (479,62) ;
\draw [line width=0.75]  [dash pattern={on 0.84pt off 2.51pt}]  (429,112) -- (529,112) ;
\draw   (409,102) -- (509,102) -- (509,202) -- (409,202) -- cycle ;
\draw    (409,102) -- (429,62) ;
\draw   (429,62) -- (529,62) -- (529,162) -- (429,162) -- cycle ;
\draw [color={rgb, 255:red, 128; green, 128; blue, 128 }  ,draw opacity=1 ]   (477.66,114.68) -- (460.34,149.32) ;
\draw [shift={(459,152)}, rotate = 296.57] [fill={rgb, 255:red, 128; green, 128; blue, 128 }  ,fill opacity=1 ][line width=0.08]  [draw opacity=0] (3.57,-1.72) -- (0,0) -- (3.57,1.72) -- cycle    ;
\draw [shift={(479,112)}, rotate = 116.57] [fill={rgb, 255:red, 128; green, 128; blue, 128 }  ,fill opacity=1 ][line width=0.08]  [draw opacity=0] (3.57,-1.72) -- (0,0) -- (3.57,1.72) -- cycle    ;
\draw    (409,202) -- (429,162) ;
\draw    (509,202) -- (529,162) ;
\draw    (509,102) -- (529,62) ;
\draw [color={rgb, 255:red, 128; green, 128; blue, 128 }  ,draw opacity=1 ]   (469.42,132.42) -- (516,132.03) ;
\draw [shift={(519,132)}, rotate = 179.51] [fill={rgb, 255:red, 128; green, 128; blue, 128 }  ,fill opacity=1 ][line width=0.08]  [draw opacity=0] (3.57,-1.72) -- (0,0) -- (3.57,1.72) -- cycle    ;
\draw [color={rgb, 255:red, 128; green, 128; blue, 128 }  ,draw opacity=1 ]   (469.42,132.42) -- (422,132.03) ;
\draw [shift={(419,132)}, rotate = 0.48] [fill={rgb, 255:red, 128; green, 128; blue, 128 }  ,fill opacity=1 ][line width=0.08]  [draw opacity=0] (3.57,-1.72) -- (0,0) -- (3.57,1.72) -- cycle    ;
\draw [color={rgb, 255:red, 128; green, 128; blue, 128 }  ,draw opacity=1 ]   (469,85) -- (469,179) ;
\draw [shift={(469,182)}, rotate = 270] [fill={rgb, 255:red, 128; green, 128; blue, 128 }  ,fill opacity=1 ][line width=0.08]  [draw opacity=0] (3.57,-1.72) -- (0,0) -- (3.57,1.72) -- cycle    ;
\draw [shift={(469,82)}, rotate = 90] [fill={rgb, 255:red, 128; green, 128; blue, 128 }  ,fill opacity=1 ][line width=0.08]  [draw opacity=0] (3.57,-1.72) -- (0,0) -- (3.57,1.72) -- cycle    ;
\draw [color={rgb, 255:red, 128; green, 128; blue, 128 }  ,draw opacity=1 ]   (516.88,84.12) -- (421.12,179.88) ;
\draw [shift={(419,182)}, rotate = 315] [fill={rgb, 255:red, 128; green, 128; blue, 128 }  ,fill opacity=1 ][line width=0.08]  [draw opacity=0] (3.57,-1.72) -- (0,0) -- (3.57,1.72) -- cycle    ;
\draw [shift={(519,82)}, rotate = 135] [fill={rgb, 255:red, 128; green, 128; blue, 128 }  ,fill opacity=1 ][line width=0.08]  [draw opacity=0] (3.57,-1.72) -- (0,0) -- (3.57,1.72) -- cycle    ;
\draw [color={rgb, 255:red, 128; green, 128; blue, 128 }  ,draw opacity=1 ]   (421.12,84.12) -- (516.88,179.88) ;
\draw [shift={(519,182)}, rotate = 225] [fill={rgb, 255:red, 128; green, 128; blue, 128 }  ,fill opacity=1 ][line width=0.08]  [draw opacity=0] (3.57,-1.72) -- (0,0) -- (3.57,1.72) -- cycle    ;
\draw [shift={(419,82)}, rotate = 45] [fill={rgb, 255:red, 128; green, 128; blue, 128 }  ,fill opacity=1 ][line width=0.08]  [draw opacity=0] (3.57,-1.72) -- (0,0) -- (3.57,1.72) -- cycle    ;
\draw [color={rgb, 255:red, 128; green, 128; blue, 128 }  ,draw opacity=1 ]   (478.58,64.97) -- (459.42,199.03) ;
\draw [shift={(459,202)}, rotate = 278.13] [fill={rgb, 255:red, 128; green, 128; blue, 128 }  ,fill opacity=1 ][line width=0.08]  [draw opacity=0] (3.57,-1.72) -- (0,0) -- (3.57,1.72) -- cycle    ;
\draw [shift={(479,62)}, rotate = 98.13] [fill={rgb, 255:red, 128; green, 128; blue, 128 }  ,fill opacity=1 ][line width=0.08]  [draw opacity=0] (3.57,-1.72) -- (0,0) -- (3.57,1.72) -- cycle    ;
\draw [color={rgb, 255:red, 128; green, 128; blue, 128 }  ,draw opacity=1 ]   (459.95,104.85) -- (478.05,159.15) ;
\draw [shift={(479,162)}, rotate = 251.57] [fill={rgb, 255:red, 128; green, 128; blue, 128 }  ,fill opacity=1 ][line width=0.08]  [draw opacity=0] (3.57,-1.72) -- (0,0) -- (3.57,1.72) -- cycle    ;
\draw [shift={(459,102)}, rotate = 71.57] [fill={rgb, 255:red, 128; green, 128; blue, 128 }  ,fill opacity=1 ][line width=0.08]  [draw opacity=0] (3.57,-1.72) -- (0,0) -- (3.57,1.72) -- cycle    ;
\draw [color={rgb, 255:red, 128; green, 128; blue, 128 }  ,draw opacity=1 ]   (526.15,112.95) -- (411.85,151.05) ;
\draw [shift={(409,152)}, rotate = 341.57] [fill={rgb, 255:red, 128; green, 128; blue, 128 }  ,fill opacity=1 ][line width=0.08]  [draw opacity=0] (3.57,-1.72) -- (0,0) -- (3.57,1.72) -- cycle    ;
\draw [shift={(529,112)}, rotate = 161.57] [fill={rgb, 255:red, 128; green, 128; blue, 128 }  ,fill opacity=1 ][line width=0.08]  [draw opacity=0] (3.57,-1.72) -- (0,0) -- (3.57,1.72) -- cycle    ;
\draw [color={rgb, 255:red, 128; green, 128; blue, 128 }  ,draw opacity=1 ]   (506.32,150.66) -- (431.68,113.34) ;
\draw [shift={(429,112)}, rotate = 26.57] [fill={rgb, 255:red, 128; green, 128; blue, 128 }  ,fill opacity=1 ][line width=0.08]  [draw opacity=0] (3.57,-1.72) -- (0,0) -- (3.57,1.72) -- cycle    ;
\draw [shift={(509,152)}, rotate = 206.57] [fill={rgb, 255:red, 128; green, 128; blue, 128 }  ,fill opacity=1 ][line width=0.08]  [draw opacity=0] (3.57,-1.72) -- (0,0) -- (3.57,1.72) -- cycle    ;
\draw [line width=0.75]    (409,212) -- (509,212) ;
\draw    (409,208) -- (409,216) ;
\draw    (509,208) -- (509,216) ;
\draw    (523,202) -- (543,162) ;
\draw    (519,202) -- (527,202) ;
\draw    (539,162) -- (547,162) ;
\draw    (399,102) -- (399,202) ;
\draw    (395,102) -- (403,102) ;
\draw    (395,202) -- (403,202) ;
\draw    (579,162) -- (579,134) ;
\draw [shift={(579,132)}, rotate = 90] [color={rgb, 255:red, 0; green, 0; blue, 0 }  ][line width=0.75]    (4.37,-1.32) .. controls (2.78,-0.56) and (1.32,-0.12) .. (0,0) .. controls (1.32,0.12) and (2.78,0.56) .. (4.37,1.32)   ;
\draw    (579,162) -- (605,162) ;
\draw [shift={(607,162)}, rotate = 180] [color={rgb, 255:red, 0; green, 0; blue, 0 }  ][line width=0.75]    (4.37,-1.32) .. controls (2.78,-0.56) and (1.32,-0.12) .. (0,0) .. controls (1.32,0.12) and (2.78,0.56) .. (4.37,1.32)   ;
\draw    (579,162) -- (569.89,180.21) ;
\draw [shift={(569,182)}, rotate = 296.57] [color={rgb, 255:red, 0; green, 0; blue, 0 }  ][line width=0.75]    (4.37,-1.32) .. controls (2.78,-0.56) and (1.32,-0.12) .. (0,0) .. controls (1.32,0.12) and (2.78,0.56) .. (4.37,1.32)   ;
\draw [line width=0.75]  [dash pattern={on 0.84pt off 2.51pt}]  (419,82) -- (519,82) ;
\draw [line width=0.75]  [dash pattern={on 0.84pt off 2.51pt}]  (419,182) -- (519,182) ;
\draw [line width=0.75]  [dash pattern={on 0.84pt off 2.51pt}]  (459,202) -- (479,162) ;
\draw [line width=0.75]  [dash pattern={on 0.84pt off 2.51pt}]  (459,102) -- (479,62) ;
\draw [line width=0.75]  [dash pattern={on 0.84pt off 2.51pt}]  (519,182) -- (519,82) ;
\draw [line width=0.75]  [dash pattern={on 0.84pt off 2.51pt}]  (419,182) -- (419,82) ;
\draw [line width=0.75]  [dash pattern={on 0.84pt off 2.51pt}]  (509,152) -- (529,112) ;
\draw [line width=0.75]  [dash pattern={on 0.84pt off 2.51pt}]  (409,152) -- (429,112) ;
\draw [line width=0.75]  [dash pattern={on 0.84pt off 2.51pt}]  (409,152) -- (509,152) ;
\draw [line width=0.75]  [dash pattern={on 0.84pt off 2.51pt}]  (459,202) -- (459,102) ;
\draw   (149,87) -- (249,87) -- (249,187) -- (149,187) -- cycle ;
\draw [line width=0.75]    (149,197) -- (249,197) ;
\draw    (149,193) -- (149,201) ;
\draw    (249,193) -- (249,201) ;
\draw    (139,87) -- (139,187) ;
\draw    (135,87) -- (143,87) ;
\draw    (135,187) -- (143,187) ;
\draw [color={rgb, 255:red, 128; green, 128; blue, 128 }  ,draw opacity=1 ]   (199,90) -- (199,184) ;
\draw [shift={(199,187)}, rotate = 270] [fill={rgb, 255:red, 128; green, 128; blue, 128 }  ,fill opacity=1 ][line width=0.08]  [draw opacity=0] (3.57,-1.72) -- (0,0) -- (3.57,1.72) -- cycle    ;
\draw [shift={(199,87)}, rotate = 90] [fill={rgb, 255:red, 128; green, 128; blue, 128 }  ,fill opacity=1 ][line width=0.08]  [draw opacity=0] (3.57,-1.72) -- (0,0) -- (3.57,1.72) -- cycle    ;
\draw [color={rgb, 255:red, 128; green, 128; blue, 128 }  ,draw opacity=1 ]   (246.88,89.12) -- (151.12,184.88) ;
\draw [shift={(149,187)}, rotate = 315] [fill={rgb, 255:red, 128; green, 128; blue, 128 }  ,fill opacity=1 ][line width=0.08]  [draw opacity=0] (3.57,-1.72) -- (0,0) -- (3.57,1.72) -- cycle    ;
\draw [shift={(249,87)}, rotate = 135] [fill={rgb, 255:red, 128; green, 128; blue, 128 }  ,fill opacity=1 ][line width=0.08]  [draw opacity=0] (3.57,-1.72) -- (0,0) -- (3.57,1.72) -- cycle    ;
\draw [color={rgb, 255:red, 128; green, 128; blue, 128 }  ,draw opacity=1 ]   (151.12,89.12) -- (246.88,184.88) ;
\draw [shift={(249,187)}, rotate = 225] [fill={rgb, 255:red, 128; green, 128; blue, 128 }  ,fill opacity=1 ][line width=0.08]  [draw opacity=0] (3.57,-1.72) -- (0,0) -- (3.57,1.72) -- cycle    ;
\draw [shift={(149,87)}, rotate = 45] [fill={rgb, 255:red, 128; green, 128; blue, 128 }  ,fill opacity=1 ][line width=0.08]  [draw opacity=0] (3.57,-1.72) -- (0,0) -- (3.57,1.72) -- cycle    ;
\draw [color={rgb, 255:red, 128; green, 128; blue, 128 }  ,draw opacity=1 ]   (152,137) -- (246,137) ;
\draw [shift={(249,137)}, rotate = 180] [fill={rgb, 255:red, 128; green, 128; blue, 128 }  ,fill opacity=1 ][line width=0.08]  [draw opacity=0] (3.57,-1.72) -- (0,0) -- (3.57,1.72) -- cycle    ;
\draw [shift={(149,137)}, rotate = 0] [fill={rgb, 255:red, 128; green, 128; blue, 128 }  ,fill opacity=1 ][line width=0.08]  [draw opacity=0] (3.57,-1.72) -- (0,0) -- (3.57,1.72) -- cycle    ;
\draw    (89,186) -- (89,158) ;
\draw [shift={(89,156)}, rotate = 90] [color={rgb, 255:red, 0; green, 0; blue, 0 }  ][line width=0.75]    (4.37,-1.32) .. controls (2.78,-0.56) and (1.32,-0.12) .. (0,0) .. controls (1.32,0.12) and (2.78,0.56) .. (4.37,1.32)   ;
\draw    (89,186) -- (115,186) ;
\draw [shift={(117,186)}, rotate = 180] [color={rgb, 255:red, 0; green, 0; blue, 0 }  ][line width=0.75]    (4.37,-1.32) .. controls (2.78,-0.56) and (1.32,-0.12) .. (0,0) .. controls (1.32,0.12) and (2.78,0.56) .. (4.37,1.32)   ;

\draw (452,216.4) node [anchor=north west][inner sep=0.75pt]  [font=\tiny]  {$2\Delta x$};
\draw (380,148.4) node [anchor=north west][inner sep=0.75pt]  [font=\tiny]  {$2\Delta x$};
\draw (536,178.4) node [anchor=north west][inner sep=0.75pt]  [font=\tiny]  {$2\Delta x$};
\draw (602,152.4) node [anchor=north west][inner sep=0.75pt]  [font=\tiny]  {$x$};
\draw (566,182.4) node [anchor=north west][inner sep=0.75pt]  [font=\tiny]  {$z$};
\draw (576,122.4) node [anchor=north west][inner sep=0.75pt]  [font=\tiny]  {$y$};
\draw (520,128.4) node [anchor=north west][inner sep=0.75pt]  [font=\tiny]  {$1$};
\draw (410,128.4) node [anchor=north west][inner sep=0.75pt]  [font=\tiny]  {$2$};
\draw (462,74.4) node [anchor=north west][inner sep=0.75pt]  [font=\tiny]  {$3$};
\draw (470,184.4) node [anchor=north west][inner sep=0.75pt]  [font=\tiny]  {$4$};
\draw (452,154.4) node [anchor=north west][inner sep=0.75pt]  [font=\tiny]  {$5$};
\draw (480,104.4) node [anchor=north west][inner sep=0.75pt]  [font=\tiny]  {$6$};
\draw (520,80.4) node [anchor=north west][inner sep=0.75pt]  [font=\tiny]  {$7$};
\draw (418,184.4) node [anchor=north west][inner sep=0.75pt]  [font=\tiny]  {$8$};
\draw (510,152.4) node [anchor=north west][inner sep=0.75pt]  [font=\tiny]  {$9$};
\draw (430,118.4) node [anchor=north west][inner sep=0.75pt]  [font=\tiny]  {$10$};
\draw (448,92.4) node [anchor=north west][inner sep=0.75pt]  [font=\tiny]  {$11$};
\draw (480,162.4) node [anchor=north west][inner sep=0.75pt]  [font=\tiny]  {$12$};
\draw (472,136.4) node [anchor=north west][inner sep=0.75pt]  [font=\tiny]  {$0$};
\draw (518,184.4) node [anchor=north west][inner sep=0.75pt]  [font=\tiny]  {$13$};
\draw (408,76.4) node [anchor=north west][inner sep=0.75pt]  [font=\tiny]  {$14$};
\draw (530,108.4) node [anchor=north west][inner sep=0.75pt]  [font=\tiny]  {$15$};
\draw (410,154.4) node [anchor=north west][inner sep=0.75pt]  [font=\tiny]  {$16$};
\draw (446,194.4) node [anchor=north west][inner sep=0.75pt]  [font=\tiny]  {$18$};
\draw (474,52.4) node [anchor=north west][inner sep=0.75pt]  [font=\tiny]  {$17$};
\draw (192,201.4) node [anchor=north west][inner sep=0.75pt]  [font=\tiny]  {$2\Delta x$};
\draw (120,133.4) node [anchor=north west][inner sep=0.75pt]  [font=\tiny]  {$2\Delta x$};
\draw (112,176.4) node [anchor=north west][inner sep=0.75pt]  [font=\tiny]  {$x$};
\draw (86,146.4) node [anchor=north west][inner sep=0.75pt]  [font=\tiny]  {$y$};
\draw (510,204.4) node [anchor=north west][inner sep=0.75pt]  [font=\tiny,color={rgb, 255:red, 208; green, 2; blue, 27 }  ,opacity=1 ]  {$23$};
\draw (510,104.4) node [anchor=north west][inner sep=0.75pt]  [font=\tiny,color={rgb, 255:red, 208; green, 2; blue, 27 }  ,opacity=1 ]  {$19$};
\draw (530,154.4) node [anchor=north west][inner sep=0.75pt]  [font=\tiny,color={rgb, 255:red, 208; green, 2; blue, 27 }  ,opacity=1 ]  {$26$};
\draw (530,64.4) node [anchor=north west][inner sep=0.75pt]  [font=\tiny,color={rgb, 255:red, 208; green, 2; blue, 27 }  ,opacity=1 ]  {$21$};
\draw (397,205.4) node [anchor=north west][inner sep=0.75pt]  [font=\tiny,color={rgb, 255:red, 208; green, 2; blue, 27 }  ,opacity=1 ]  {$22$};
\draw (400,94.4) node [anchor=north west][inner sep=0.75pt]  [font=\tiny,color={rgb, 255:red, 208; green, 2; blue, 27 }  ,opacity=1 ]  {$25$};
\draw (420,164.4) node [anchor=north west][inner sep=0.75pt]  [font=\tiny,color={rgb, 255:red, 208; green, 2; blue, 27 }  ,opacity=1 ]  {$20$};
\draw (420,54.4) node [anchor=north west][inner sep=0.75pt]  [font=\tiny,color={rgb, 255:red, 208; green, 2; blue, 27 }  ,opacity=1 ]  {$24$};
\draw (251.8,134.8) node [anchor=north west][inner sep=0.75pt]  [font=\tiny]  {$1$};
\draw (204.6,138.8) node [anchor=north west][inner sep=0.75pt]  [font=\tiny]  {$0$};
\draw (196.2,76.4) node [anchor=north west][inner sep=0.75pt]  [font=\tiny]  {$2$};
\draw (153.4,140.8) node [anchor=north west][inner sep=0.75pt]  [font=\tiny]  {$3$};
\draw (189.8,176.8) node [anchor=north west][inner sep=0.75pt]  [font=\tiny]  {$4$};
\draw (251.4,79.6) node [anchor=north west][inner sep=0.75pt]  [font=\tiny]  {$5$};
\draw (141.4,76.4) node [anchor=north west][inner sep=0.75pt]  [font=\tiny]  {$6$};
\draw (138.6,192) node [anchor=north west][inner sep=0.75pt]  [font=\tiny]  {$7$};
\draw (253.8,187.6) node [anchor=north west][inner sep=0.75pt]  [font=\tiny]  {$8$};

\end{tikzpicture}

%% file: p1_amr.tex
\section{Adaptive Mesh Refinement}\label{sec:amr}

This section begins with a discussion on related work and presents an overview of the current methodology. We illustrate the access patterns used in the mesh adaptation procedures and solver routines, and then detail the refinement and coarsening algorithm. Finally, we provide the routines for data communication along coarse-fine interfaces.


    \subsection{Related Work}\label{sec:amr_work}

    Tree-based AMR has received much attention in the literature, especially in the context of scalable algorithms \cite{Octor, Dendro, Peano, Burstedde2013, p4est, EnzoP, Schornbaum2018, Schornbaum2018Thesis}, and algorithms that implement mesh data structures and adaptation routines on GPUs \cite{Giuliani2019, Pavlukhin2019, Menshov2020, Wang2024}. The particular case of quad/octree is the method of choice used in GPU grid-refined LBM codes (both static and dynamic) such as Palabos \cite{Palabos}, ESPResSo \cite{Lahnert2016,Mehl2019}, and waLBerla \cite{waLBerla}. Quad/octrees are data structures where each node can be subdivided into up to four/eight child nodes \cite{Sundar2008}. Nodes are denoted as leaf nodes if they do not possess children, and interior nodes otherwise. The root node is the unique node in the tree without a parent, while all other nodes have exactly one. A node's level is the number of parent-child links between it and the root node. The level of the root node is defined as zero.

    While octree algorithms and implementations have been extensively studied over several decades \cite{Sundar2008, Cornerstone}, their specific application to AMR on GPUs remains relatively unexplored in the current literature. Pointer-based tree implementations that enable recursive construction do not possess good data locality \cite{Wang2024}, so linear representations are typically used instead. The Morton Z-order SFC \cite{Morton1966} encoding is a well-known way to implicitly represent the leaf node data of an octree in linear memory. The SFC representation retains good data locality \cite{Zumbusch2003}, an important property for load balancing when distributing across multiple processes \cite{Schornbaum2018, Schornbaum2018Thesis}. It has been applied in AMR packages such as p4est \cite{p4est}, Peano \cite{Peano}, and waLBera \cite{waLBerla}. Menshov and Pavlukhin \cite{Pavlukhin2019, Menshov2020} employed this approach when implementing octree AMR natively on the GPU. Wang et al. \cite{Wang2024} developed algorithms for construction, 2:1 balancing, and nodal connectivity of Morton-encoded octrees for GPU-based discontinuous finite element methods with AMR. Tree data structures can alternatively be represented using integer lists that store the indices of mesh elements in the data arrays. This is more common in AMR codes on unstructured grids \cite{Luo2016, Giuliani2019} where a Morton encoding is unavailable. In the current work, we diverge from the usual Morton encoding and utilize an explicit representation via integer lists storing indices of block data in separate metadata arrays. In contrast to the classic linear tree \cite{Sundar2008}, the data of both interior and leaf nodes are stored to facilitate specific steps in the refinement and coarsening algorithm presented in Section \ref{sec:amr_ref} and data transfer between coarse and fine grids. It will be shown that the connectivity update step, which establishes/reverts neighbor links between inserted/removed nodes, can be performed with careful use of shared memory and arithmetic due to this explicit identification.

    A common reason for selecting a cell-based approach in AMR over a block-based approach is a finer granularity in which fewer resources are required for a given proscribed error tolerance \cite{Saetra2015, Giuliani2019}. However, an explicit index-based representation (a common approach in recent literature \cite{Giuliani2019, Wang2024}) presents some disadvantages: 1) the memory footprint of a cell-based scheme can become an issue depending on the number of neighbor links that need to be stored (this would be up to 27 if employed in a 3D LBM code), 2) contiguity in memory is reduced as new cells are inserted, which can significantly reduce the efficiency of global memory transactions on the GPU in grid advancement routines involving neighbor data (contiguous in groups of 32 bits are required for global memory transactions to coalesce on GPUs). For solvers with explicit time-stepping, padding of the fine grid with one or more layers of ghost cells is required, which increases the complexity of the mesh adaptation algorithm. In contrast, Fakhari and Lee \cite{Fakhari2014} observed that a block-based approach provides memory savings since metadata is shared among groups of cells. There is also no need for additional refinement near coarse-fine interfaces for smoothness, as this is taken care of automatically. Since quad-/octree data structures can be implicitly represented and linearized with space-filling curves, explicit storage of the neighbor links is not necessarily required and the memory footprint issue in the cell-based scheme can be remedied in principle. However, the lack of contiguity remains a matter of concern. The local structured-grid arrangement of cells in the block-based framework allows for the computation of neighbor cell IDs with a known formula, enabling a simplified, coalesced exchange of information within the block. It is common for GPU-parallel codes for the LBM to utilize block-based arrangements \cite{Palabos, waLBerla}, and a similar approach will be taken in this paper.

    Subdivision of leaf nodes in a linear octree stored with explicit representation is equivalent to the construction of new children in metadata arrays. Coarsening is equivalent to the removal of child blocks from these arrays. For example, if a block is being subdivided and the current total number of blocks is $N_{\text{curr.}}$, the data (e.g., node level, coordinates) corresponding to eight new blocks can be inserted at locations $[N_{\text{curr}}, N_{\text{curr}}+8)$ in the metadata arrays. Coarsening is straightforward in that the indices of blocks to be removed are deemed inactive. In an explicit representation, the indices of these blocks are removed from the list of active blocks. However, this results in gaps that no longer correspond to any point in the computational grid. Failure to eliminate these gaps could result in less efficient global load/store operations or premature memory exhaustion \cite{Pavlukhin2019}; this has been achieved in previous GPU-native AMR approaches. Luo et al. \cite{Luo2016} report an index recycling strategy where gaps were tracked (i.e., indices removed from the list were temporarily stored) and re-used during the next call to refinement. Giuliani and Krivodonova \cite{Giuliani2019} describe a hybrid approach, employing index recycling and, if an insufficient number of cells were added during refinement to cover existing gaps, stream compaction to close those left over. Menshov and Pavlukhin \cite{Menshov2020} describe a traversal of the octrees managing the grid where they apply a prefix-scan operation to fill the gaps and preserve memory locality. Shifting of data can become costly depending on the strategy undertaken and would require a complicated connectivity update due to modification of the block data locations. A recycling strategy is preferable, so this work uses an explicit octree representation. This requires tracking the available indices (i.e., locations in the metadata arrays where blocks have not yet been defined) to enable parallel construction without overwriting when many blocks are marked for refinement. Two sets of integer arrays accomplish this: 1) ID sets, which store the IDs of active blocks, and 2) the so-called gap set, which enumerates all available IDs that can be assigned to blocks inserted in the grid. The refinement and coarsening strategy presented in Section \ref{sec:amr_ref} revolves around a parallel manipulation of these arrays.

    \subsection{Overview}

    The mesh is organized as a forest of quad/octrees in which nodes represent square/cubic blocks of cells of size $M_b$. A node can be split into exactly $N_c$ child nodes in our implementation, where $N_c$ takes on a value of four in 2D and eight in 3D. The root nodes of these trees are arranged as a structured grid denoted by the root grid of the hierarchy. We will refer only to octrees for the remainder of the paper as the mesh management procedure outlined in this section is identical in 2D and 3D. A collection of nodes across all octrees on level $L$ is denoted grid level $L$, and the set of grid levels $0 \leq L < L_{\text{max.}}$ is referred to as the grid hierarchy. A 2:1 balance \cite{Sundar2008} is maintained across the whole forest at all times (so that the width of a cell on grid level $L+1$ is always half of that on level $L$). The forest is described altogether by a set of one-dimensional arrays that store flattened multidimensional octree solution data and mesh metadata in a structure of arrays format. 
    These are classified as cell data arrays with labels $\texttt{cells\_*}$ and cell-block data arrays with labels $\texttt{cblock\_*}$.
    \begin{figure}[t]
        \centering
        \begin{subfigure}[b]{0.47\textwidth}
            \centering
            \includegraphics[scale=0.5]{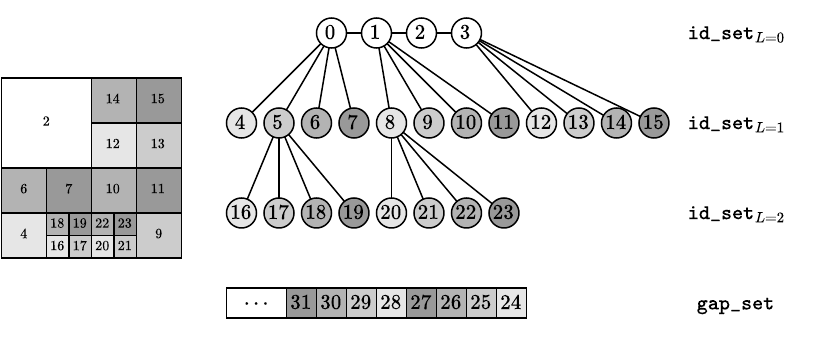}
            \label{fig:amr_tree_1}
            \caption{Sample grid and octree, along with ID sets and gap set.}
        \end{subfigure}
        \hspace{0.5cm}
        \begin{subfigure}[b]{0.47\textwidth}
            \centering
            \includegraphics[scale=0.8]{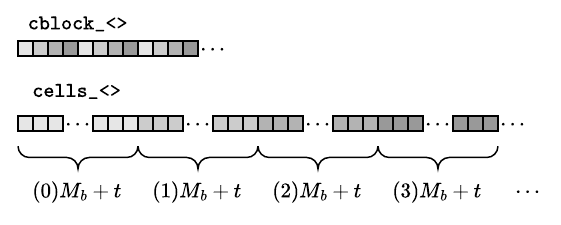}
            \label{fig:amr_tree_2}
            \caption{Locations in data arrays.}
        \end{subfigure}
        \caption{Illustration of the index list representation of the current octree implementation.}
        \label{fig:amr_tree}
    \end{figure}
    
\begin{figure}
        \centering
        \begin{subfigure}[b]{0.47\textwidth}
            \centering
            \includegraphics[scale=0.5]{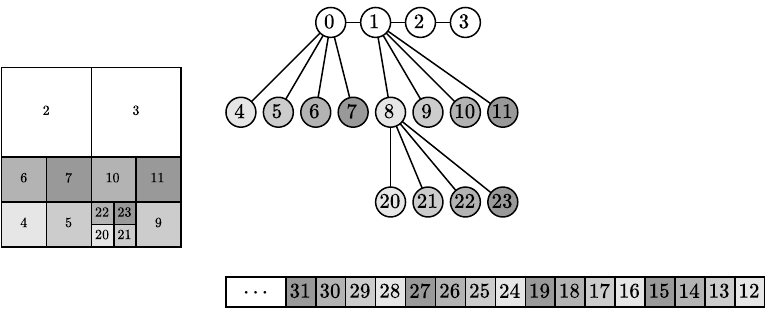}
            \label{fig:amr_tree_coarsen}
            \caption{Blocks 3 and 5 are coarsened.}
        \end{subfigure}
        \hspace{0.5cm}
        \begin{subfigure}[b]{0.47\textwidth}
            \centering
            \includegraphics[scale=0.5]{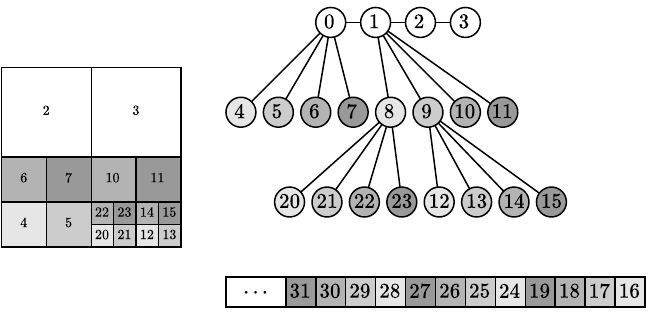}
            \label{fig:amr_tree_refine}
            \caption{Block 9 is refined afterwards.}
        \end{subfigure}
        \caption{Illustration of the refinement and coarsening operations.}
        \label{fig:amr_tree_refinement}
    \end{figure}

    Individual nodes are explicitly identified with an integer (a block ID) corresponding to the location of block data in these arrays. Two sets of arrays denoted the ID sets (\texttt{id\_set}\}) and the gap set (\texttt{gap\_set}) store active block IDs and IDs that are available for assignment to newly-inserted blocks during refinement, respectively. Each grid level has its own ID set $\texttt{id\_set}_L$ that stores both leaf and interior nodes. The gap set is enumerated at the beginning of the simulation, defined as all indices $N_{\text{blocks,init.}} \leq \kappa < N_{\text{blocks}}$, where $N_{\text{blocks,init.}}$ is the number of blocks in the root grid. When blocks are marked for refinement, an equal number of gaps spaced by an amount $N_c$ are assigned, representing the IDs of children to be created. The IDs are pulled from the gap set and inserted into the appropriate ID sets. To avoid re-arranging the \texttt{gap\_set} at runtime, it is enumerated in reverse (so that smaller IDs are used up first), and the required set of indices is split from the end. When an element is coarsened, child IDs are identified and pulled from the ID sets before re-inserting into the gap set. Figure \ref{fig:amr_tree} illustrates this representation for a sample 2D grid, and Figure \ref{fig:amr_tree_refinement} shows how the ID and gap sets are modified to coarsen and refine the grid. This controls grid traversal during temporal integration. By separating the ID sets by grid level, synchronization can be carefully carried out along the interface at different stages of the integration.

        \subsubsection{Data Structures}\label{sec:amr_term}
    
\begin{figure}[b]
        \centering
        \begin{subfigure}[b]{0.48\textwidth}
            \centering
            \includegraphics[scale=0.45,trim={0 1cm 0 0 }]{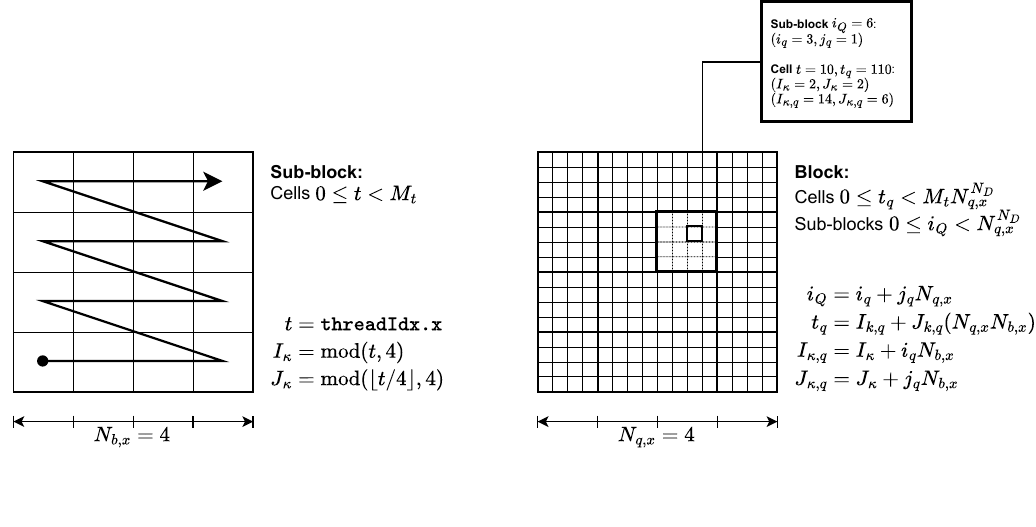}
            \caption{Cell-block arrangement.}
            \label{fig:amr_cellblock_arrangement}
        \end{subfigure}
        \begin{subfigure}[b]{0.48\textwidth}
            \centering
            \includegraphics[scale=0.45]{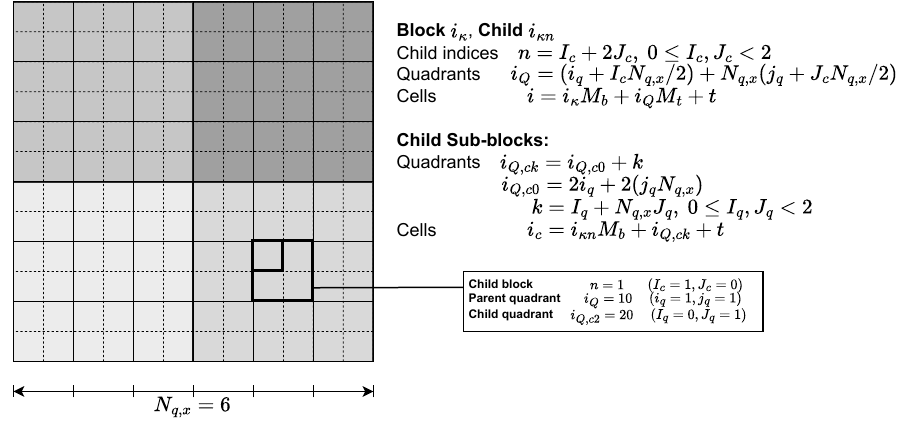}
            \caption{Cell-block traversal.}
            \label{fig:amr_cellblock_subblock_access}
        \end{subfigure}
        \caption{Left: arrangement of a cell-block decomposed into sub-blocks of size $M_b$. While $N_{b,x}=4$ is fixed in the current implementation, $N_{q,x}$ (shown here equal to 4) can be specified at compile time. Blocks are processed by looping over the sub-blocks for which the solver kernels introduced in Section \ref{sec:solver} are designated. Right: visualization of parent-child data transfer for a cell-block with $N_{q,x}=6$. Loops over the parent sub-blocks are split among the block's quadrants, each corresponding to a child cell block and its associated sub-blocks.}
        \label{fig:amr_access_arrangement}
        \end{figure}

        A single solution array stores the DDFs for all cells. Mesh metadata includes spatial coordinates of the blocks, cell and block masks that indicate participation in coarse-fine grid communication, and the block IDs of neighbors. These are summarized in Table \ref{tab:amr_data_struct}. Solution data is organized so that array index $i$ of a cell $t$ (which is always the local CUDA thread index \texttt{threadIdx.x}) in sub-block $i_Q$ of block $i_\kappa$ is obtained with $i = t + i_Q M_t + i_\kappa M_b$. The local coordinates of the cell are recovered with modulo and integer division operations. Figures \ref{fig:amr_cellblock_arrangement} and \ref{fig:amr_cellblock_subblock_access} visualize how cells and blocks are arranged and traversed, respectively, in the implementation. Metadata is accessed so that index $\kappa + d N$ for $N \in \{N_{\text{cells}},N_{\text{blocks}}\}$, where $N_{\text{cells}}$ and $N_{\text{blocks}}$ are the respective maximum number of cells and blocks, retrieves the $d^{\text{th}}$ component of cell/block $i_\kappa$’s data.
    
\begin{figure}
            \centering
            \begin{subfigure}[b]{0.45\textwidth}
                \centering
                \includegraphics[width=\textwidth,trim={0 0.5cm 0 0}]{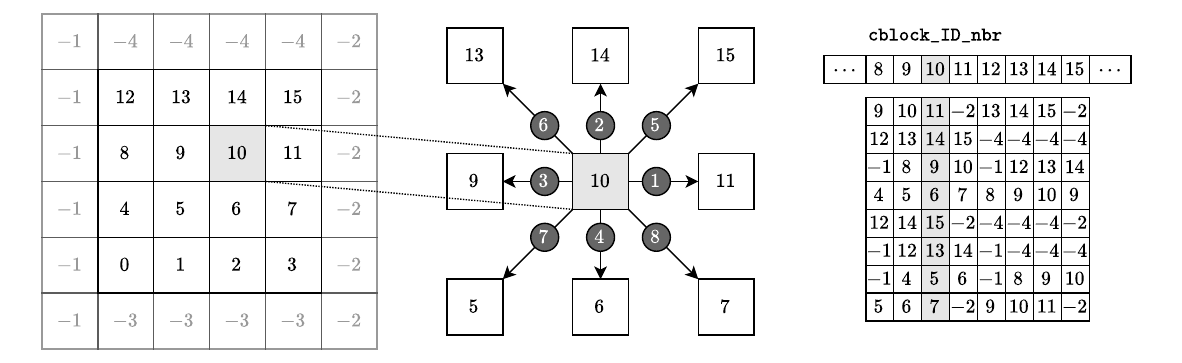}
                \label{fig:amr_conn_1}
                \caption{Neighbor index list.}
            \end{subfigure}
            \hspace{0.5cm}
            \begin{subfigure}[b]{0.45\textwidth}
                \centering
                \includegraphics[width=\textwidth,trim={0 0.5cm 0 0}]{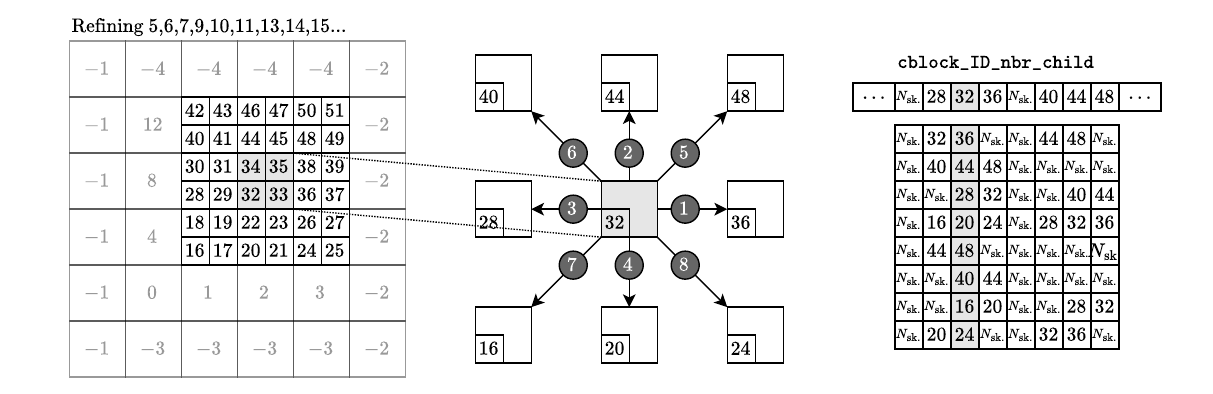}
                \label{fig:amr_conn_2}
                \caption{Neighbor-child index list.}
            \end{subfigure}
            \caption{Visualization of connectivity lists.}
            \label{fig:amr_conn}
        \end{figure}

        Neighbor links between nodes are stored explicitly since an encoding is not used. IDs of neighbor blocks, referred to in this paper simply as ‘neighbors,’ are indices such that $\texttt{cblock\_ID\_nbr}[i_\kappa + pN_{\text{blocks}}]$ is the $p^{\text{th}}$ neighbor block. A convention similar to the velocity sets introduced in the previous section is used to order neighbors. 9 and 27 neighbors are linked in 2D and 3D, giving each block access to a full surrounding halo. The indices of the first child block of every neighbor are also stored explicitly in \texttt{cblock\_ID\_nbr\_child}; these are denoted as ‘neighbor-children.’ These two arrays (Figure \ref{fig:amr_conn}) form the mesh connectivity. The values of the neighbor-children of a given block are used to determine the neighbor indices for its children. When blocks are refined, the corresponding neighbor-child values of its neighbors are modified to record the refinement. This represents a connectivity transmission from one grid to the next, beginning with the root grid. Since the neighbor indices of blocks on the root grid are all known, the neighbor indices of new child blocks added in separate octrees can be identified without resorting to searches. This representation of connectivity therefore enables parallel execution of the refinement and coarsening algorithm simultaneously over the whole forest of octrees.
    
        The cell-block size is designated by specifying the number of cells along an axis $N_{b,x} = 4k, \ k \in \mathbb{N}$ (so that $M_b = N_{b,x}^{N_d}$,  where $N_d \in \{2,3\}$ is the number of dimensions) and by choosing $k$ based on a suitable thread-block size. Fixed thread-block sizes $M_t, \overline{M}_t$ launch the CUDA kernels that update cell and cell-block data, respectively). Due to hardware limits, the maximum thread-block size needs to be smaller or equal to the cell-block size. For example, the maximum $M_t$ that can accommodate full occupancy on Nvidia GPUs with Compute Capability 3.0+ is 1024.
        However, large values of $M_t$ will increase register pressure and may lead to register spilling. $M_t$ and $\overline{M}_t$ should be multiples of 32 to ensure that all threads in the block are active. $M_t$ should also divide $(4k)^{N_d}$ to ensure that all threads participate in processing the cell block. The values $k=1$, $M_t = 4^{N_d}$ and $N_{b,x}=4$ are fixed in the current implementation, with an equal minimum cell-block size. Larger cell blocks can be constructed by arranging sub-blocks of size $M_t$ in a regular grid of size $N_{q,x}^{N_d}$, where $N_{q,x}$ is the number of sub-blocks \footnote{These are sometimes referred to in the code as quadrants/octants (especially during grid communication when a sub-block is identifying the sub-blocks in the main block's children).} along one axis. This parameter is controlled at compile time with the variable \texttt{Nqx}. In 2D, $M_t = 16$ is less than the GPU warp size, which results in a penalty for load and store instruction efficiency, though this is countered by the significantly smaller datasets being processed. The value of $\overline{M}_t$ can be chosen more freely; $\overline{M}_t=128$ is used in the current implementation.
    
        Initialization of the mesh comprises of memory allocation on the CPU and GPU and the construction of the root grid based on input parameters read from a text file. $\text{N}_{\text{cells}}$ is determined at runtime by surveying the available memory on the GPU and allocating a specified fraction \texttt{M\_frac} of it for all of the above arrays (some memory can be left free for other processes, e.g., on a personal computer). This number is then rounded to a number divisible by 32 to ensure proper alignment of the arrays for coalesced access.
        
        \begin{table}[t]
            \centering
            \small
            \begin{tabular}{l c c p{3.6 in}}
                \hline
                \hline
                Name & No. Elem. & Bytes & Description \\
                \hline
                \hline
                \texttt{cells\_f\_F} & $N_Q$ & $N_p$ & Stores the solution field of cells in the grid. For the LBM, these are the density distribution functions (DDFs). \\
                \texttt{cells\_ID\_mask} & 1 & 4 & Stores cell mask IDs which differentiate between interior cells of a fine mesh, interface cells (whose values are communicated to the coarse cell at a fine-grid boundary), and ghost cells which receive data from the coarse grid to facilitate time advancement on the fine grid. \\
                \texttt{cblock\_f\_X} & $N_d$ & $N_p$ & Stores block global coordinates of the (bottom) southwest corner. \\
                \texttt{cblock\_ID\_nbr} & $N_{Q,{\text{max.}}}$ & 4 & Stores the IDs of all neighbor blocks. \\
                \texttt{cblock\_ID\_nbr\_child} & $N_{Q,{\text{max.}}}$ & 4 & Stores the IDs of the first children of all neighboring blocks. \\
                \texttt{cblock\_ID\_ref} & 1 & 4 & Stores refinement IDs representing block status at runtime. Several values are represented: unrefined, refined, refined w/ child (i.e., a refined block with at least one refined child, for restricting coarsening), refined-permanent (do not consider for coarsening), marked for refinement, marked for coarsening, marked new, marked for removal, inactive. \\
                \texttt{cblock\_level} & 1 & 4 & Stores grid levels which blocks occupy. This is required during the evaluation of the refinement criterion and when computing new spatial locations of children added during refinement. \\
                \texttt{cblock\_ID\_mask} & 1 & 4 & Stores block mask IDs which identify whether or not blocks will be participating in coarse-fine grid communications. The indicator returns positive if at least one cell among child blocks has been designated as an interface or ghost cell. \\
                \texttt{cblock\_ID\_onb} & 1 & 4 & Stores indicators for blocks lying on the domain boundary. Used to determine whether to proceed with boundary condition imposition. \\
                \hline
            \end{tabular}
            \caption{List and descriptions of data structures utilized in implementing the mesh adaptation procedure and solver. $N_p$ is the number of bytes in a floating-point word, either four or eight in single- and double-precision, respectively.}
            \label{tab:amr_data_struct}
        \end{table}
    
        \begin{figure}[b]
         \centering
         \begin{subfigure}[b]{0.495\textwidth}
             \centering
             \includegraphics[width=0.99\textwidth]{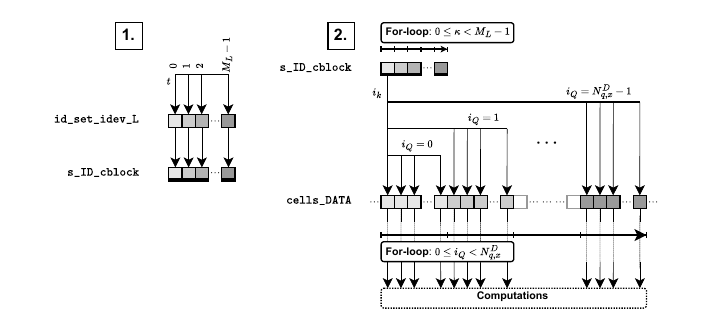}
             \caption{Primary mode of access.}
             \label{fig:amr_memaccess_A}
         \end{subfigure}
         \begin{subfigure}[b]{0.495\textwidth}
             \centering
             \includegraphics[width=0.99\textwidth]{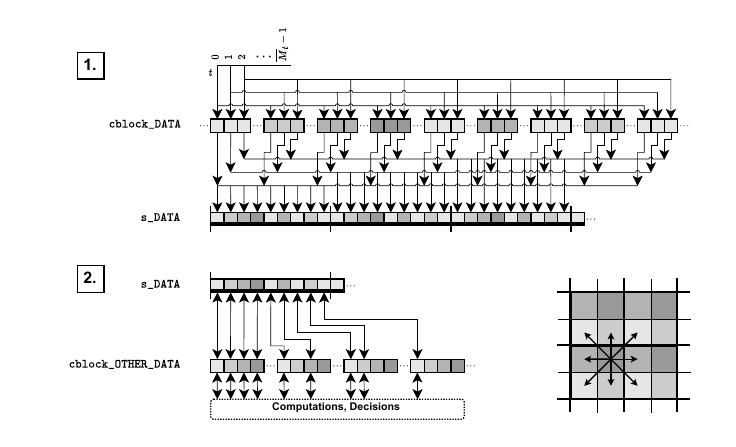}
             \caption{Secondary mode of access.}
             \label{fig:amr_memaccess_B}
         \end{subfigure}
            \caption{Memory access strategies used in the current implementation. An example of access to neighbor metadata is used to illustrate the secondary-mode pattern.}
            \label{fig:amr_memaccess}
        \end{figure}
        
        \subsubsection{Memory Access Strategies}\label{sec:amr_mem}

        Two access patterns update cell and cell-block data, respectively, denoted the primary and secondary modes of access (Figure \ref{fig:amr_memaccess}); these serve as representative templates for the CUDA kernels to dynamically adapt the mesh and temporally integrate the numerical solution.
        
        The primary mode of access, applied whenever cell data arrays $\texttt{cells\_*}$ need to be modified, is the template pattern used in all solver kernels for temporal integration and grid communication, and in modifying the mask IDs which determine participation in the kernels for communication between grid levels. CUDA kernels based on this template receive an ID set for a particular grid level as input, and a subset of IDs is extracted and traversed by each thread-block. Cell-blocks are processed by assigning one of these IDs at a time to all threads simultaneously, allowing them to access contiguous cell data.
        
        We introduce a parameter $M_L$ that determines the number of cell-blocks that are processed by each thread-block, controlled at compile time with the parameter \texttt{M\_LBLOCK}. A set of $M_L \leq M_t$ block IDs is first read from an appropriate ID set in global memory into shared memory. This is followed by a for-loop over the stored IDs, where the data of cells in the sub-block is processed simultaneously. If $M_t < M_b$, a loop over the various sub-blocks is performed within the outer loop over the IDs.

        The secondary mode of access is applied whenever cell-block data arrays $\texttt{cblocks\_*}$ need to be modified. It is the central template pattern used in the refinement and coarsening algorithm when an update to cell-block metadata depends on the data of its neighbors or children. It exploits the contiguity of sibling blocks to achieve a degree of coalescence. Threads access the metadata arrays directly instead of traversing an ID set as with the primary mode.
        
        Cell-block IDs are traversed first, and the IDs of  neighbors/children are copied into shared memory based on a specified criterion. A second traversal is performed over these stored IDs to load neighbor/child block data. The data can replace the IDs in memory since the latter are no longer needed at this point. Coalesced access of the neighbor/child data is guaranteed to a limited extent since children are always inserted contiguously. Finally, the original cell-block indices are traversed again, and tasks are performed based on the available neighbor/child data. An example of this strategy is shown for the case of neighbor-data access in Figure \ref{fig:amr_memaccess_B}.

    \subsection{Refinement and Coarsening} \label{sec:amr_ref}

\begin{figure}[h!]
        \centering
        \begin{subfigure}[b]{0.3\textwidth}
             \centering
             \includegraphics[scale=0.5]{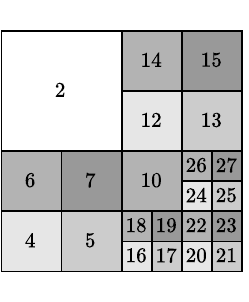}
             \caption{Sample grid.}
             \label{fig:amr_step_sample_A}
         \end{subfigure}
         \begin{subfigure}[b]{0.6\textwidth}
             \centering
             \includegraphics[scale=0.5]{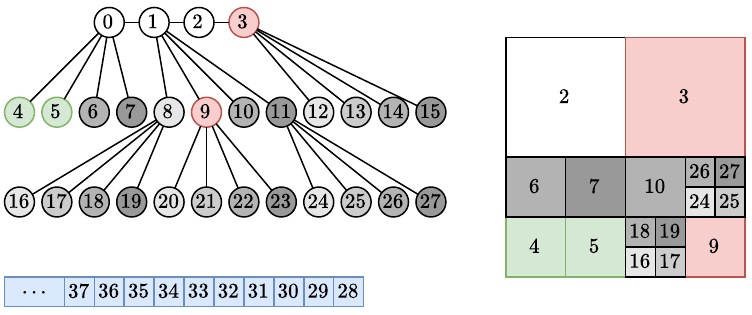}
             \caption{Corresponding ID and gap sets.}
             \label{fig:amr_step_sample_B}
         \end{subfigure}
        \caption{A sample 2D AMR grid to aid in visualizing the refinement and coarsening algorithm. Blocks 3 and 9 are marked (in red) for coarsening, while blocks 4 and 5 are marked (in green) for refinement. Gap set IDs are shown in blue to aid in later visualizations.}
        \label{fig:amr_step_sample}
    \end{figure}

    Refinement and coarsening are combined in a single routine and applied to all grid levels in the hierarchy simultaneously. This routine begins with a preparation step and is followed by an eight-step procedure, detailed below. A sample 2D grid (shown in Figure \ref{fig:amr_step_sample}) with two levels of refinement is used to visualize the various steps.

    \input{algos/Cu_AddRemoveBlocks}

        \begin{enumerate}
        \item {\bf Preparation}. The total number of blocks marked for refinement and coarsening ($N_{\text{marked,R}}$ and $N_{\text{marked,C}}$) are obtained with Thrust's \texttt{count\_if} over \texttt{cblock\_ID\_ref}. If both are zero, the remainder of the routine is skipped, otherwise all intermediate arrays are reset to negative values which will indicate invalidity or non-participation in future steps if unmodified.

        \item {\bf Minimal Map} The minimal map involves the fewest number of IDs participating in updates to connectivity and is essentially a list of all marked blocks and their neighbors. Blocks are traversed and, if marked, have their neighbors' IDs copied into an intermediate array. These IDs are arranged contiguously with \texttt{copy\_if}, filtered of duplicates with \texttt{count\_if} and \texttt{unique\_copy}, then finally sorted and scattered with \texttt{sort} and \texttt{scatter}. Now, the map value for a given block is checked before the execution of subsequent subroutines (and it is skipped if the value is negative). Figure \ref{fig:amr_step_1} shows the construction of the minimal map for the sample grid.
        \begin{figure}
            \centering
            \begin{subfigure}[b]{0.495\textwidth}
                 \centering
                 \includegraphics[scale=0.35]{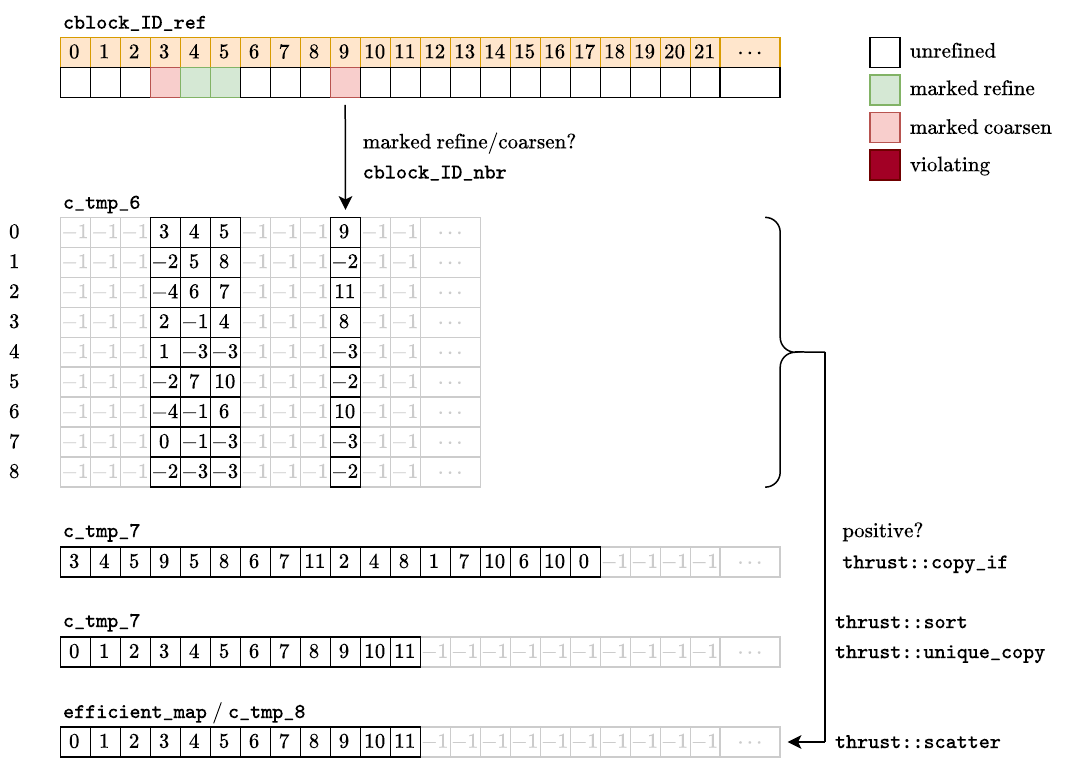}
                 \caption{}
                 \label{fig:amr_step_1}
             \end{subfigure}
             \begin{subfigure}[b]{0.495\textwidth}
                 \centering
                 \includegraphics[scale=0.35]{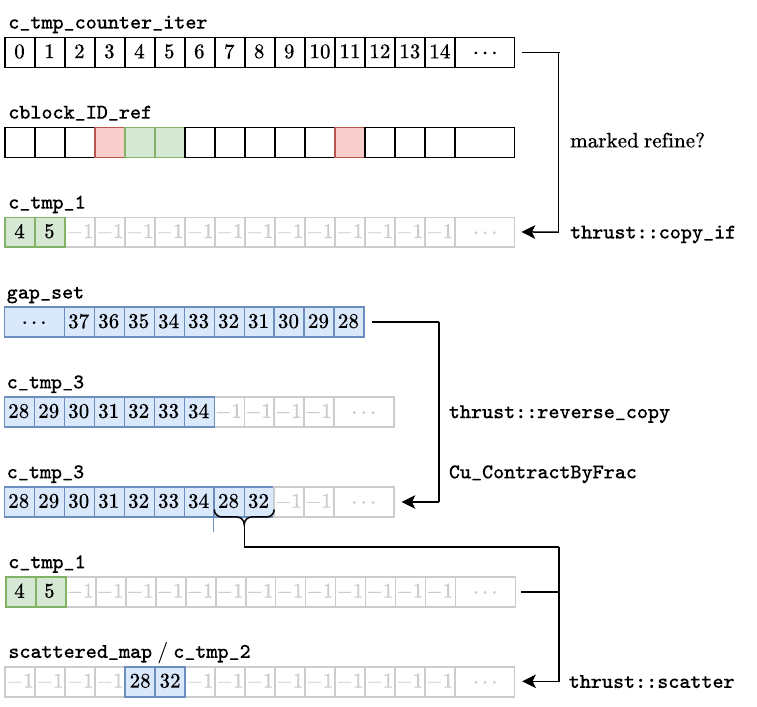}
                 \caption{}
                 \label{fig:amr_step_2a}
             \end{subfigure}
            \caption{Left: development of the minimal map (called 'efficient map' in the code). Right: assignment of new children to blocks marked for refinement. Yellow cells indicate array element indices.}
            \label{fig:amr_steps_12}
        \end{figure}

        \item {\bf Map Gaps to Marked Blocks} If $N_{\text{marked,R}} > 0$, the IDs of the marked blocks are retrieved with \texttt{copy\_if}. A \texttt{reverse\_copy} call is made to acquire $N_c N_{\text{marked}}$ available gaps. A separate `contracted' copy of these gaps takes only the first of each group of $N_c$ indices. These are assigned to the marked blocks with Thrust's \texttt{scatter} as the IDs of their first children and from which all other children are reconstructed. The assignment is completed in a custom routine that updates connectivity in \texttt{cblock\_ID\_nbr\_child}. This routine exploits the secondary mode of access to neighbors as follows: neighbor IDs are placed in shared memory in the first loop and traversed in a second loop where their block IDs are replaced with their corresponding refinement IDs; then, in a third loop, shared memory is accessed in the original order to determine whether or not a block possesses a refined neighbor. If so, the ID stored in \texttt{cblock\_ID\_nbr\_child} for that corresponding direction has its value updated to the new mapped value. Figure \ref{fig:amr_step_2a} illustrates how new child IDs are assigned to blocks marked for refinement from the gap set for the sample grid. Figure \ref{fig:amr_step_2b} shows how the secondary mode of access is used to update $\texttt{cblock\_ID\_nbr\_child}$.

        
        \item {\bf Enforcement of 2:1 Balancing} A 2:1 balancing of the underlying octrees is ensured by (a) prevention of refinement when the resulting children would be adjacent to an unrefined block and (b) reversion of blocks marked for coarsening if they are near a refined block. The former is performed during computation of the refinement criterion prior to invocation of the refinement and coarsening routine (this is discussed further in Section \ref{sec:solver_amr}). The latter is divided into two steps: 1) unrefined blocks are traversed and marked as violating if at least one neighbor-child is non-negative, and 2) blocks marked for coarsening are traversed in a secondary-access mode and reverted if at least one child was marked as violating. Modifying the unrefined blocks in this step does not affect subsequent refinement/coarsening algorithm subroutines. Now, $N_{\text{marked,C}}$ is updated to account for reversions. Suppose $N_{\text{marked,C}}$ remains positive. In that case, the second half of the connectivity update in the data arrays is performed in the same manner as that for mapped children where the values of \texttt{cblock\_ID\_nbr\_child} for neighbors marked for coarsening are changed to an arbitrary $N_{\text{skip}}$ (excluding the current block with neighbor index 0 since references to the children to be removed are still needed in the next step). A neighbor index $N_{\text{skip}}$ indicates that the block does not have a neighbor in that direction. Block 3 of the sample grid is marked for coarsening, but the mark is reverted since coarsening would result in its adjacency to two blocks with lengths four times smaller (shown in Figure \ref{fig:amr_qc}). Figure \ref{fig:amr_step_3a} illustrates how this is detected with the secondary access mode.
        \begin{figure}[b]
            \centering
            \includegraphics[scale=0.35]{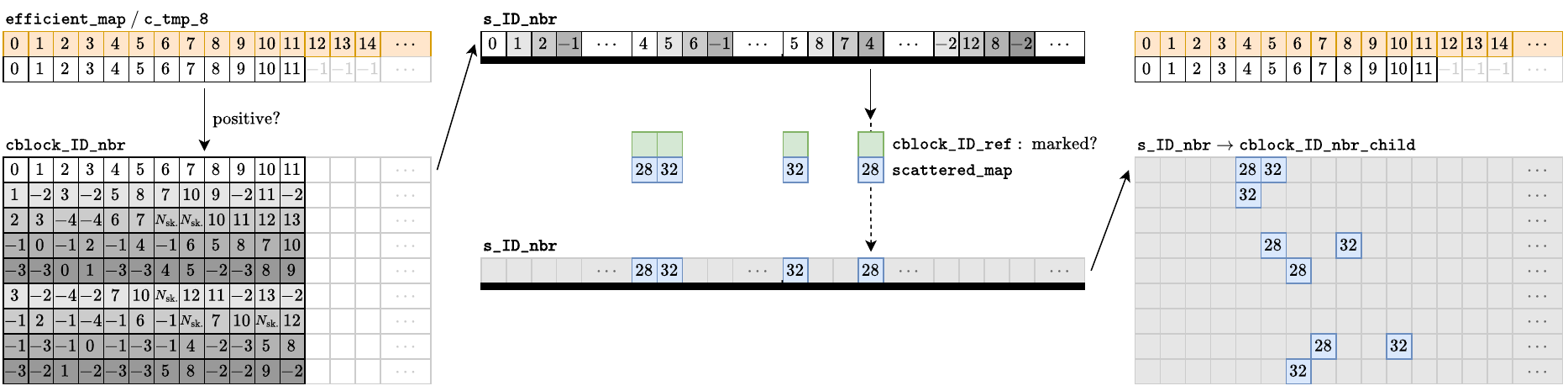}
            \caption{Visualization of the first update of $\texttt{cblock\_ID\_nbr\_child}$ with the secondary mode of access. Neighbor-children are shaded in gray to illustrate the change in order between the first and second traversals. Cells with a thick lower edge indicate shared memory.}
            \label{fig:amr_step_2b}
        \end{figure}

        \item {\bf Build / Reset Children} If $N_{\text{marked,R}} > 0$ or $N_{\text{mark,C}} > 0$, a custom kernel (summarized in Algorithm \ref{alg:ref_s1}) is called which builds the data of new children based on parent information and resets children to be removed in preparation for updates to the Id sets. The strategy to construct new children is to place mapped child IDs in shared memory (which perfectly coalesces the read) and to invoke a loop over it separately to rebuild the remaining children. Metadata is the same for all children for a single block - new children receive the increment to the grid level of the parent (from \texttt{cblock\_level}) and a refinement ID corresponding to `new' while children to be removed have their refinement ID set to `remove' (later written to \texttt{cblock\_ID\_ref}). The global coordinates of the new children can be calculated using reference parent coordinates (from \texttt{cblock\_f\_X}) incremented based on the thread index. Since $\overline{M}_t$ threads are processed in parallel, each child is mapped once at the kernel's beginning using integer division and modulo operators. Increments in space are found with:
        \begin{align}
            \Delta x_{\kappa} &= \text{mod}(t, 2) \Delta x_{\text{root}} / 2^{L_q}, \\
            \Delta y_{\kappa} &= \text{mod}(t / 2, 2) \Delta x_{\text{root}} / 2^{L_q}, \\
            \Delta z_{\kappa} &= \text{mod}(t / 4, 2) \Delta x_{\text{root}} / 2^{L_q},
        \end{align}
        \noindent where $\Delta x_{\text{root}}$ is the spatial step of the root grid and $L_q$ is the grid level that the child blocks occupy.

        \item {\bf Coarsen Blocks in ID Sets} Indices of blocks marked for removal are retrieved via a call to Thrusts’s \texttt{copy\_if} and appended to the same intermediate array used for collecting IDs of blocks marked for refinement. Retrieved as well are the IDs of the children themselves and their corresponding levels in the grid hierarchy (from \texttt{cblock\_level}). The child IDs to be removed are sorted by level using Thrust’s \texttt{sort\_by\_key}. Now, looping over the levels, \texttt{count\_if} is applied to count the IDs that need to be pulled from the ID set of that level. If positive, a custom routine is used that traverses the ID set and resets the IDs to $N_{\text{skip}}$ by individual comparison to all IDs stored in intermediate memory. This is followed by Thrust’s \texttt{remove\_if}, which strips the ID set of block IDs that had just been transformed. The number of removed blocks is decremented from the ID set size and added to the total gap set count. The removed IDs are concatenated to the gap set in reverse to preserve their order in groups of $N_c$, then discarded.
        \begin{figure}
            \centering
            \begin{subfigure}[b]{0.27\textwidth}
                 \centering
                 \includegraphics[scale=0.5]{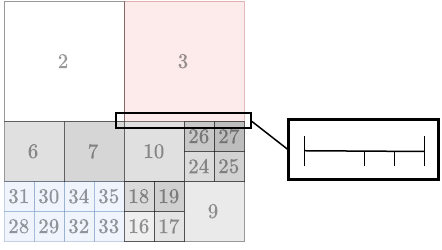}
                 \caption{Coarsening block three results in a violation of the 2:1 balance requirement.}
                 \label{fig:amr_qc}
             \end{subfigure}
             \hspace{0.2cm}
             \begin{subfigure}[b]{0.7\textwidth}
                 \centering
                 \includegraphics[scale=0.35]{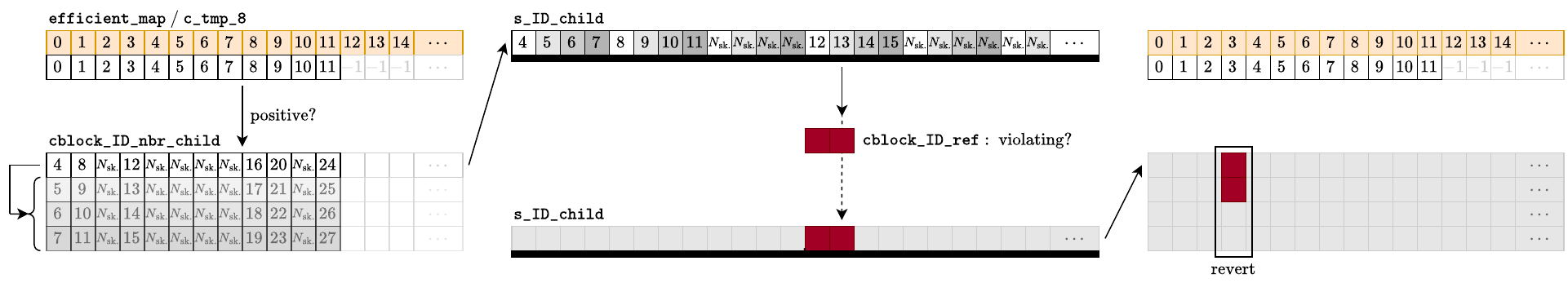}
                 \caption{Detection of violating child blocks with the secondary access mode. Indices read from $\texttt{cblock\_ID\_nbr\_child}$ are used to recover the remaining children.}
                 \label{fig:amr_step_3a}
             \end{subfigure}
            \caption{Visualizations of the enforcement of the 2:1 balancing requirement.}
            \label{fig:amr_step_3}
        \end{figure}

        \item {\bf Refine Blocks in ID Sets} Insertion of new children proceeds with the same process except that 1) concerned blocks are marked as `new,' and 2) the gap setting is updated in size only.
    
        \item {\bf Interpolate to Children, Finalize Ref. IDs} Neighbor-children are inspected to update block masks. Blocks are marked for coarse-fine communication if at least one neighbor-child is invalid (indicating proximity of at least one child block to an invalid neighbor). Next, data from refined blocks are interpolated to new children. Refinement IDs that were used to facilitate the present algorithm (i.e., marks for refinement/coarsening, ‘new,’ ‘remove’) are reverted to ‘unrefined’ if newly added, ‘refined’ if previously marked for coarsening, or ‘inactive’ if removed. Branch-leaf relations are also updated at this stage. If a refined block possesses at least one refined child, coarsening it before coarsening the child would lead to orphan blocks. A unique refinement ID (i.e., 'refined w/ child') records this to restrict coarsening during the evaluation of the refinement criterion. To do this, a custom routine that uses the secondary mode of access targets the refinement IDs of children. Blocks are declared branches if at least one child has been refined and leaves otherwise.

        \item {\bf Update Connectivity and Designate Ghosts} The last step is to update mesh connectivity and the cell masks. Connectivity is updated with a custom routine (Algorithm \ref{alg:conn_s2}) employing the secondary mode of access and code generation to retrieve neighbor-children of parent blocks, re-construct the neighbors of their children in order, and arrange them in shared memory so that global writes to \texttt{cblock\_ID\_nbr} at the child level retain a degree of coalescence. Negative neighbor-child IDs (not equal to $N_{\text{skip}}$) indicating adjacency to the domain boundary are also transmitted to children. Figure \ref{fig:amr_step_8a} illustrates the code-generated neighbor index formulas. Figure \ref{fig:amr_step_8b} shows how the secondary mode of access is used to update child cell neighbors. Once new neighbor IDs are established, a second routine converts the neighbor-children to the values of the parent blocks' negative neighbors in all directions. This transmits boundary information for future connectivity updates. Finally, cell masks indicating eligibility for grid communication are updated since new coarse-fine boundaries may have developed. This concludes the refinement and coarsening algorithm.
        \end{enumerate}
        \begin{figure}[b]
            \centering
            \begin{subfigure}[b]{0.495\textwidth}
                 \centering
                 \includegraphics[scale=0.35]{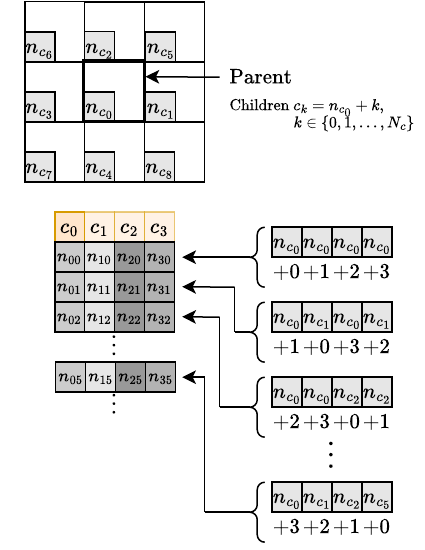}
                 \caption{Code-generated formulas based on neighbor-children.}
                 \label{fig:amr_step_8a}
             \end{subfigure}
             \begin{subfigure}[b]{0.495\textwidth}
                 \centering
                 \includegraphics[scale=0.35]{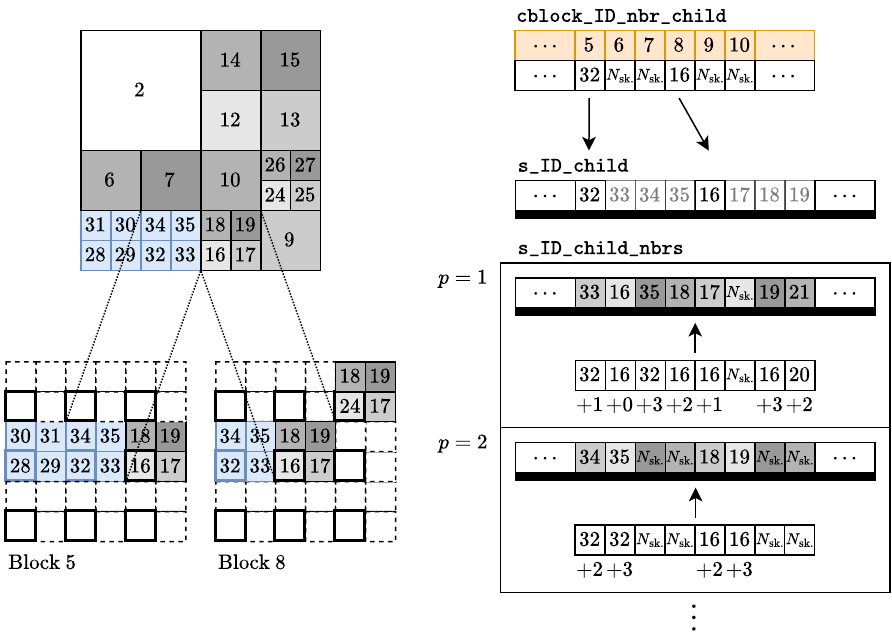}
                 \caption{Updating neighbors with the secondary mode of access.}
                 \label{fig:amr_step_8b}
             \end{subfigure}
            \caption{Left: index formulas of neighbors of child blocks based on neighbor-child values. Right: construction, arrangement, and global write of neighbors in shared memory for parent blocks 4 and 5 in the sample grid.}
            \label{fig:amr_step_8}
        \end{figure}

        \input{algos/Cu_UpdateConnectivity}

    \subsection{Grid Communication} \label{sec:amr_comm}

\begin{figure}
        \centering
        \begin{subfigure}[b]{0.48\textwidth}
            \centering
            \includegraphics[width=\textwidth]{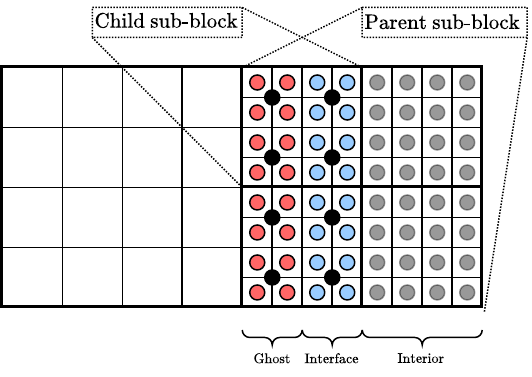}
            \caption{Linear interpolation.}
            \label{fig:amr_befill}
        \end{subfigure}
        \hspace{0.3cm}
        \begin{subfigure}[b]{0.48\textwidth}
            \centering
            \begin{subfigure}[b]{0.75\textwidth}
                \centering
                \includegraphics[width=\textwidth]{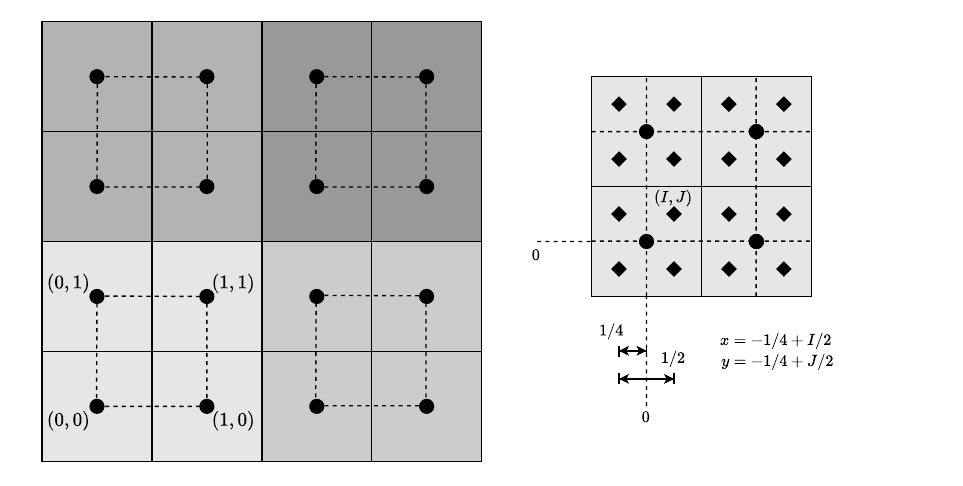}
                \caption{Linear interpolation.}
                \label{fig:amr_interp_linear}
            \end{subfigure}
            \begin{subfigure}[b]{0.75\textwidth}
                \centering
                \includegraphics[width=\textwidth]{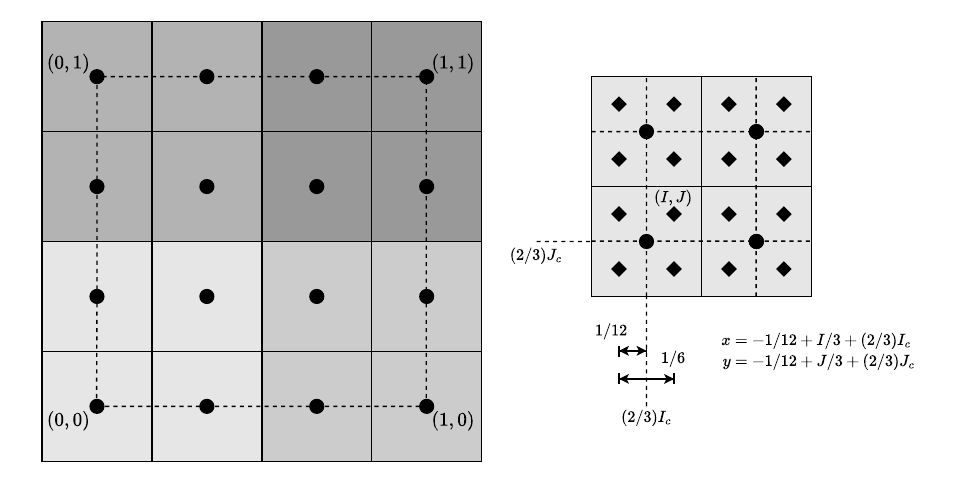}
                \caption{Cubic interpolation.}
                \label{fig:amr_interp_cubic}
            \end{subfigure}
            \label{fig:amr_interp}
        \end{subfigure}
        %
        %
        \caption{Left: ghost, interior, and interface cells along the refinement interface. Right: unit square locations for linear and cubic interpolation.}
        \label{fig:amr_interp}
    \end{figure}

    Data along refinement interfaces must be exchanged at each time step to couple the coarse and fine grids.
    A cell-block on the fine grid is adjacent to an interface if at least one of its neighbor links is set to $N_{\text{skip}}$. Interpolation is used to transfer data from the coarse grid to the fine grid, while averaging is used to transfer data in the opposite direction. The two layers of cells adjacent to an interface are designated ghost cells and receive interpolated DDFs from the coarse grid that are later discarded after advancement. Two further layers are designated interface cells. The parents of these cells receive transferred DDFs during averaging, completing the grid coupling. Fine grid cells that do not participate in grid communication are denoted interior cells. These cell layers are illustrated in Figure \ref{fig:amr_befill}.
    
    The arrangement of the grid and the order of interpolation accuracy can inform the time-integration strategy and impact the quality of the numerical solution. Grid refinement schemes based on a vertex-centered arrangement (e.g., implemented in Palabos \cite{Palabos}) require filtering or high-order polynomial interpolation to reduce artifacts at coarse-fine interfaces, while those based on a cell-centered arrangement (e.g., implemented in waLBerla \cite{waLBerla}) benefit from a staggering of the coarse and fine grid nodes so that linear interpolation suffices with regards to accuracy \cite{Palabos}. Lagrava et al. \cite{Lagrava2012} showed that linear interpolation with a vertex-centered grid refinement scheme produces a discontinuity in the pressure field at the coarse-fine interface, which they resolved with cubic interpolation. The discontinuity resulted from an incompatibility between the respective first- and second-order global spatial discretization errors of linear interpolation and the LBM.
    Fakhari and Lee \cite{Fakhari2014} observed this effect and attributed the discontinuity to a degradation in accuracy resulting from explicit computation of a velocity gradient term in their finite-difference LBM scheme. They resolved this with biquadratic interpolation.

    Filippova and Hanel \cite{Filippova1998} developed expressions connecting the coarse and fine grids based on analytical estimates for the non-equilibrium components of the DDFs, requiring a rescaling step during grid communication based on the estimate\footnote{This expression depends on how $\tau$ is defined. In other works, the definition $\nu = c_s^2(\tau-1/2)\Delta t$ implies an estimate $f_p^{\text{neq.}} = \mathcal{O}(\Delta t \tau)$} for the non-equilibrium distribution $f_p^{\text{neq.}} = f_p - f_p^{\text{eq.}} = \mathcal{O}(\tau)$. These expressions contain singularities later removed with the improved formulation of Dupuis and Chopard \cite{Dupuis2003}. Rohde et al. \cite{Rohde2006} introduced a mass-conservative approach based on a volumetric interpretation of the cell-centered arrangement. This method employs a homogeneous distribution of DDFs that eliminates the need for spatial/temporal interpolation or rescaling, thus providing generality to various collision models. While this approach offers a robust framework, subsequent work by Chen et al. \cite{Chen2006} demonstrated that incorporating mass- and momentum-conservative interpolation techniques can further enhance accuracy.
    
    More advanced collision operators such as the multi-relaxation-time (MRT) \cite{Lallemand2000} and cascaded \cite{Geier2006} operators enable collision in moment space, providing benefits in stability \cite{dHumieres2002}. T{\"o}lke et al. \cite{Tolke2006} used cubic spatial and linear temporal interpolations in moment space for a multiple-relaxation-time LBM. Cubic spline interpolation was needed to consistently transfer flux terms across interfaces, preventing the generation of spurious oscillations \cite{Yu2002, Yu2002Thesis}. Moment-space representation can also enable higher-order polynomial interpolations with greater locality. Geier et al. \cite{Geier2009} used a cascaded LBM to recover quadratic interpolation functions for momentum in a square spanned by four nodes (which generally provide enough information for linear interpolation in velocity space).
    
    This work employs a nodal interpretation of the cell-centered arrangement, the rescaling method of Dupuis and Chopard \cite{Dupuis2003}, and considers the BGK-LBM to restrict interpolation to velocity space rather than moment space. We perform high-order polynomial interpolation instead by exploiting the contiguity of the $M_t$ cells in a sub-block (the exact number of sampling points required for cubic interpolation in 2D and 3D) to enable parallel computation of the interpolation weights. Linear and cubic interpolation routines are implemented and compared in the test cases to follow.

        \subsubsection{Linear Interpolation} \label{sec:amr_interp}

        Linear interpolation is performed with the polynomial representations:
        \begin{align}
            \text{2D: } \quad & F(x,y) \approx a_0 + a_1x + a_2y + a_3y, \\
            \text{3D: } \quad & F(x,y,z) \approx a_0 + a_1x + a_2y + a_3z + a_4xy + a_5yz + a_6xz + a_7xyz,
        \end{align}
        
        \noindent where weights for sampling points $F_{ijk}=F(i,j,k)$ ($F_{ij}=F(i,j)$ in 2D) defined on the corners of a representative unit square/cube are defined by:
        \begin{align}
            \text{2D: } \quad & \begin{cases}
                a_0 &= F_{00} \\
                a_1 &= F_{10} - F_{00} \\
                a_2 &= F_{01} - F_{00} \\
                a_3 &= F_{11} - F_{01} - F_{10} + F_{00},
            \end{cases}, \quad \text{3D: } \quad & \begin{cases}
                a_0 &= F_{000} \\
                a_1 &= F_{100} - F_{000} \\
                a_2 &= F_{010} - F_{000} \\
                a_3 &= F_{001} - F_{000} \\
                a_4 &= F_{110} - F_{100} - F_{010} + F_{000} \\
                a_5 &= F_{101} - F_{100} - F_{001} + F_{000} \\
                a_6 &= F_{011} - F_{001} - F_{010} + F_{000} \\
                a_7 &= F_{111} - F_{100} - F_{010} + F_{001} - F_{011} - F_{101} - F_{011} + F_{000}.
            \end{cases}
        \end{align}
        The parent block is partitioned into quad/octants to enable four/eight-batch execution of the interpolation kernel, and cells in the parent block interpolate to children by drawing the required sampling points and computing the polynomials described above. A local coordinate calculation is performed using block indices $I_{d} = (I, J, K)$ (computed at the beginning of the kernel) so that, for a cell-block of size $4^{N_d}$ the coordinates of a child cell $l$ within the unit square/cube (as shown in Figure \ref{fig:amr_interp_linear}) are found by $x_{d,\text{local}} = (-1/4) + I_{d} (1/2)$. Local indices are computed from thread indices $0 \leq l=\texttt{threadIdx.x} < M_t$ only once for each thread-block and re-used for all processed cell-blocks according to:
        \begin{align}
            I &= \text{mod}(l,4), \quad J = \text{mod}(l/4,4), \quad K = l/4^2.
        \end{align} 
    

        The DDFs are rescaled prior to being sampled for interpolation. In the following equations, variables with superscripts $C$ and $F$ denote values on coarse and fine grids, respectively. By dividing the coarse and fine non-equilibrium DDFs with their respective relaxation rates and observing that macroscopic properties are identical in both grids (i.e., constants in $\mathcal{O}(\tau)$ are identical for both expressions, and that $f_p^{C,\text{eq.}}(\rho,\textbf{u}) = f_p^{F,\text{eq.}}(\rho,\textbf{u})$), a ratio can be taken to produce a formula for the rescaled interpolated DDFs on the fine grid in terms of coarse values:
        \begin{align}
            f_p^F \approx f_p^{C,\text{eq.}} + (f_p^C - f_p^{C,\text{eq.}})\frac{\tau^F}{\tau^C},
        \end{align}

        \noindent where values on the right-hand side are computed individually by each thread, stored in shared memory, and then sampled for interpolation.

        \subsubsection{Cubic Interpolation}

        A benefit of selecting $M_t = 4^{N_d}$ is that the exact number of sampling points required to fit a cubic polynomial
        \begin{align}
            F(x,y,z)  \approx \sum_{k=0}^{3^{N_d-2}} \sum_{j=0}^3 \sum_{i=0}^3 \alpha_{ijk} x^i y^j z^k,
        \end{align}
        \noindent where $\alpha_{ijk}$ are interpolation weights and $l = i + 4j + 16k$ are cell indices, are always available. A unit square/cube whose corners are the corner cell-centers of the sub-block (Figure \ref{fig:amr_interp_cubic}) is now used to define relative positions. A linear system is formed to determine the weights by requiring the interpolating function to take on the value of the sampling points $F(x_m,y_n,z_l)$ at their respective coordinates:
        \begin{align}
            A_{UV} \alpha_V \equiv \sum_{k=0}^{3^{N_d-2}} \sum_{j=0}^3 \sum_{i=0}^3 \alpha_{ijk} x_m^i y_n^j z_l^k = F(x_m,y_n,z_l) \equiv F_U, \quad\quad 0 \leq m,n,l < 4,
        \end{align}

        \noindent where $U = m + 4n + 16l$ and $V = i + 4j + 16k$. With this representation, the inverted matrix is multiplied by the sampled value to determine the weights. This matrix-vector multiplication can be represented as:
        \begin{align}
            \alpha_U = \sum_{V=0}^{3^{N_d}} A_{UV}^{-1} F_V.
        \end{align}

        Since all cell-blocks have a regular arrangement, the matrix $A$ is constructed beforehand and inverted to compute the interpolation weights. These are then used to unroll the sum explicitly via code generation. Multiplication takes place with sampled values stored in shared memory as with linear interpolation. Horner's method is used to efficiently compute the polynomials. Intermediate values $\beta_{jk}$ and $\gamma_k$ are introduced, defined by factorizing the polynomial:
        \begin{align}
            F(x,y,z) = \sum_{k=0}^{3^{N_d-2}} \sum_{j=0}^3 \sum_{i=0}^3 \alpha_{ijk} x^i y^j z^k &= \sum_{k=0}^{3^{N_d-2}} \sum_{j=0}^3 \underbrace{ \left( \sum_{i=0} \alpha_{ijk} x^i \right) }_{\beta_{jk}} y^j z^k = \sum_{k=0}^{3^{N_d-2}} \underbrace{ \left( \sum_{j=0}^3 \beta_{jk} y^j \right) }_{\gamma_k} z^k = \sum_{k=0}^{3^{N_d-2}} \gamma_k z^k,
        \end{align}

        \noindent to enable recursive calculations $F_k = \gamma_k + z_{\text{local}} F_{k+1}$, $(\gamma_{k})_j = \beta_{jk} + y_{\text{local}}(\gamma_k)_{j+1}$, and $(\beta_{jk})_i = \alpha_{ijk} + x_{\text{local}}(\beta_{jk})_{i+1}$, where $\textbf{x}_{\text{local}}$ are the child cell coordinates, $(\cdot)_{4}=0$, and $(\cdot)_0 = (\cdot)$ such that $F_0 = F(x,y,z)$. In 2D, there is only one value each of $\gamma_k, F_k$ such that $\gamma_0 = F_0 = F(x,y)$. To avoid recalculating the polynomial weights multiple times for each child sub-block, $N_c$ registers are allocated, and the results for the cells of child sub-blocks $c = I_c + 2J_c + 4K_c$ are computed with the spatial locations $x_{d,\text{local}} = -(1/12) + I_d(1/6) + I_{c,d}(2/3)$. Global memory writes are then performed sequentially to each child afterwards. The same principles for memory access, local coordinate calculation, and DDF rescaling are utilized as described for linear interpolation. 
    
        \subsubsection{Averaging} \label{sec:amr_average}

        Averages are computed with child data from:
        \begin{align}
            F(x_l,y_l,z_l) \approx \frac{1}{N_c}\sum_{i=0}^{N_c} F(x_{l,i}, y_{l,i}, z_{l,i}),
        \end{align}
    
        \noindent where $l$ is the index of the parent cell, and $\textbf{x}_{l,i}$ is the spatial location of the parent cell's $i^{\text{th}}$ child. For a cell with index $l$, the octant in which its children lie and their respective IDs are computed from:
        \begin{align}
            \text{octant}_l &= I + 2J + 4K, \quad \text{ID}_{l,c} = 2I' + 4(2J') + 4^2(2 K'), \\
            I &= \text{mod}(l,4)/2, \quad J = \text{mod}(l/4,4)/2, \quad K = (l/4^2)/2, \\
            I' &= \text{mod}(\text{mod}(l,4),2), \quad J' = \text{mod}(\text{mod}(l/4,4),2), \quad K' = \text{mod}((l/4^2)/2,2).
        \end{align}
        
        Values in 2D are identical except that the properties involving $K$ and $K'$ are set to zero. As with interpolation, quantities based on thread indices are only computed once for each thread block. Although it would be ideal for the parent DDFs to be written to global memory in one step, this is complicated by the memory requirements imposed during the rescaling of the child DDFs. It would require $N_c N_Q (N_p/4)$ registers to simultaneously store all the child data required to perform rescaling without repeated global memory reads ($N_c$ children, $N_Q$ DDFs, one/two 32-bit registers for single-/double-precision floating point numbers). This number is acceptable for small velocity sets such as D2Q9 in single-precision, but increases rapidly to 432 registers for the D3Q27 set in double-precision, exceeding the maximum allowable number of 255 registers per thread for devices with Compute Capability 3.5+. Child sub-blocks are consequently processed in batches as with interpolation but with updates restricted to parent cells possessing children in each batch, requiring a total of $N_c$ global writes. This could be alleviated by storing equilibrium distributions in global memory. However, memory traffic would be doubled as a result, and the number of global memory writes required to store these distributions would vastly exceed the number of writes introduced here.
        
        Rescaling when averaging from fine to coarse grids is provided by:
        \begin{align}
            f_p^C &\approx f_p^{F,\text{eq.}} + (f_p^F - f_p^{F,\text{eq.}})\frac{\tau^C}{\tau^F}.
        \end{align}

        Since averaging is a special case of second-order linear interpolation from fine-to-coarse grids, we tested a cubic interpolation procedure that is compatible with the transfer from coarse-to-fine grids in order of accuracy. The fine-to-coarse interpolation kernel utilizes the computational strategy of coarse-to-fine cubic interpolation and the memory access strategy of the averaging procedure. However, simulations were unstable even when the root grid was refined and the mesh adaptivity was disabled. The scripts implementing this algorithm \texttt{solver\_lbm\_average\_cubic*.cu} remain available in the repository for the interested user but were not used in any of the test cases reported in Section \ref{sec:tests}.

%% file: algos/Cu_AddRemoveBlocks.tex
    \begin{algorithm}[h!]
    \footnotesize
        \caption{Adding and Removing Blocks}\label{alg:ref_s1}
        \KwData {$\kappa_{\text{max.}}, N_{\text{blocks}}, \Delta x,$ \texttt{cblock\_ID\_ref}, \texttt{cblock\_level}, \texttt{cblock\_f\_X}, \texttt{cblock\_ID\_nbr\_child}, \texttt{scattered\_map}}
        Allocate shared memory arrays \texttt{s\_child\_ID}, \texttt{s\_ref}, \texttt{s\_level}, \texttt{s\_x}, \texttt{s\_y}, \texttt{s\_z} with size $\overline{M}_{t}$ \\
        \For(\tcp*[f]{The for-loop is divided among 1D blocks of threads of size M\_BLOCK.}){$\kappa=0 \to \kappa_{\text{max.}}$}
        {
            $t \gets$ \texttt{threadIdx.x} \\
            $t_{\text{i}} \gets \text{mod}(t,N_c)$ \\
            $t_{\text{b}} \gets t/N_c$ \\
            $\Delta x_t \gets \text{mod}(t,2) \Delta x$ \\
            $\Delta y_t \gets \text{mod}(t/2,2) \Delta x$ \\
            $\Delta z_t \gets \text{mod}(t/4,2) \Delta x$ \tcp*{Only applied if $N_d==3$.}
            \If(\tcp*[f]{Only build children if a child is assigned.}){$\texttt{scattered\_map}[\kappa] > -1$}
            {
                \tcp{Marked for refinement: prepare to build new children.}
                \If{$\texttt{cblock\_ID\_ref}[\kappa] = C_{\text{refine}}$}
                {
                    \texttt{s\_child\_ID}[$t$] $\gets$ $\texttt{scattered\_map}[\kappa]$ \\
                    \texttt{cblock\_ID\_nbr\_child}[$\kappa$] $\gets \texttt{scattered\_map}[\kappa]$ \\
                    \texttt{s\_level}[$t$] $\gets$ $\texttt{cblock\_level}[\kappa]$ \\
                    \texttt{s\_ref}[$t$] $\gets$ $C_{\text{new}}$ \\
                    \texttt{s\_x}[$t$] $\gets$ $\texttt{cblock\_f\_X}[\kappa + (0)N_{\text{blocks}}]$ \\
                    \texttt{s\_y}[$t$] $\gets$ $\texttt{cblock\_f\_X}[\kappa + (1)N_{\text{blocks}}]$ \\
                    \texttt{s\_z}[$t$] $\gets$ $\texttt{cblock\_f\_X}[\kappa + (2)N_{\text{blocks}}]$
                }
                \tcp{Marked for coarsening: prepare to remove children.}
                \If{$\texttt{cblock\_ID\_ref}[\kappa] = C_{\text{coarsen}}$}
                {
                    \texttt{s\_child\_ID}[$t$] $\gets$ $\texttt{scattered\_map}[\kappa]$ \\
                    s\_ref $\gets$ $C_{\text{remove}}$ \\
                    \texttt{cblock\_ID\_nbr\_child}[$\kappa$] $\gets$ $N_{\text{skip}}$
                }
            }
            \tcp{Up to $\overline{M}_{t}/N_c$ children may need to be written. New and removed children processed simultaneously.}
            \For{$q=0\to N_c$}
            {
                $i_q \gets$ \texttt{s\_child\_ID}[$t_{\text{b}} + q \overline{M}_{t} / N_c$] \\
                $i_{\text{ref.},q} \gets$ \texttt{s\_ref}[$t_{\text{b}} + q \overline{M}_{t} / N_c$] \\
                \If{$\text{Id}_{\text{ref.},q} = C_{\text{new}}$}
                {
                    $L_q \gets \texttt{s\_level}[t_{\text{b}} + q \overline{M}_{t} / N_c]+1$ \\
                    $\texttt{cblock\_level}[i_q + t_{\text{i}}] \gets L_q$ \\
                    $\texttt{cblock\_ref}[i_q + t_{\text{i}}] \gets C_{\text{new}}$ \\
                    $\texttt{cblock\_f\_X}[i_q + t_{\text{i}} + (0)N_{\text{blocks}}] \gets \texttt{s\_x}[t_{\text{b}} + q \overline{M}_{t} / N_c] + \Delta x_t/2^{L_q}$ \\
                    $\texttt{cblock\_f\_X}[i_q + t_{\text{i}} + (1)N_{\text{blocks}}] \gets \texttt{s\_y}[t_{\text{b}} + q \overline{M}_{t} / N_c] + \Delta y_t/2^{L_q}$ \\
                    $\texttt{cblock\_f\_X}[i_q + t_{\text{i}} + (2)N_{\text{blocks}}] \gets \texttt{s\_z}[t_{\text{b}} + q \overline{M}_{t} / N_c] + \Delta z_t/2^{L_q}$
                }
                \If{$i_{\text{ref.},q} = C_{\text{remove}}$}
                {
                    \texttt{cblock\_ref}[$i_q + t_{\text{b}}$] $\gets$ $C_{\text{remove}}$
                }
            }
        }
    \end{algorithm}

%% file: algos/Cu_UpdateConnectivity.tex
    \begin{algorithm}[h!]
    \footnotesize
        \caption{Updating Connectivity}\label{alg:conn_s2}
        \KwData {$\kappa_{\text{max.}}, N_{\text{blocks}}$, \texttt{cblock\_ID\_ref}, \texttt{cblock\_level}, \texttt{cblock\_f\_X}, \texttt{cblock\_ID\_nbr\_child, minimal\_map}}
        \For(\tcp*[f]{The for-loop is divided among 1D blocks of threads of size M\_BLOCK.}){$\kappa=0 \to \kappa_{\text{max.}}$}
        {
            $t$ $\gets$ \texttt{threadIdx.x} \tcp*{$\kappa = \texttt{blockIdx.x*blockDim.x + threadIdx.x}$}
            \If(\tcp*[f]{Proceed only if the current block requires an update.}){minimal\_map$[\kappa]>-1$}
            {
                \texttt{s\_child\_ID}[$t$] $\gets$ \texttt{cblock\_ID\_nbr\_child}$[\kappa]$ \tcp*{Load the IDs of each block's first child into shared memory.}
                \For{$p=0 \to N_Q$}
                {
                    \tcp{Load Ids of each neighbor's first child into register memory.}
                    $i_{\text{nbr,child},p} \gets \texttt{cblock\_ID\_nbr\_child}[\kappa + p N_{\text{blocks}}]$
                }
            }
            \For{$p=0 \to N_{Q,\text{max.}}$}
            {
                \If{\texttt{s\_child\_ID}$[t] > -1$}
                {
                    \For{$c=0 \to N_c$}
                    {
                        \tcp{The outer loop is unrolled, and code generation is used to generate arithmetic.}
                        Compute nbr Id $i_{\text{nbr},p,c}$ of child $c$ in direction $p$ based on stored $i_{\text{nbr,child}}$ \\
                        \texttt{s\_ID\_child\_nbrs}[$c + t N_c$] $\gets$ $i_{\text{nbr},p,c}$
                    }
                }
                \For{$q=0 \to N_c$}
                {
                    $i_q \gets$ \texttt{s\_child\_ID}$[ t/N_c + q M_b/N_c ]$ \\
                    \If{$i_q > -1$}
                    {
                        \texttt{cblock\_ID\_nbr}$[i_q + \text{mod}(t,N_c) + N_{\text{blocks}}] \gets \texttt{s\_ID\_child\_nbrs}[t + q M_b]$
                    }
                }
            }
        }
    \end{algorithm}

%% file: p2_solver.tex
\section{Lattice Boltzmann Solver} \label{sec:solver}

Our recursive time-stepping algorithm draws motivation from Schornbaum's work on extreme-scale parallel AMR-LBM \cite{Schornbaum2018Thesis} and includes in-place streaming and collision procedures based on a modification to the Esoteric Twist strategy of Geier and Schönherr \cite{Geier2017} suitable for block-based parallelization with reduced locality in memory. The following sections describe and illustrate these schemes for a cell-block size $M_b=M_t$. When $M_b>M_t$, these processes apply to the sub-blocks individually.

    \subsection{Grid Advancement} \label{sec:solver_adv}

\begin{figure}[h!]
        \centering
        \includegraphics[width=0.85\textwidth]{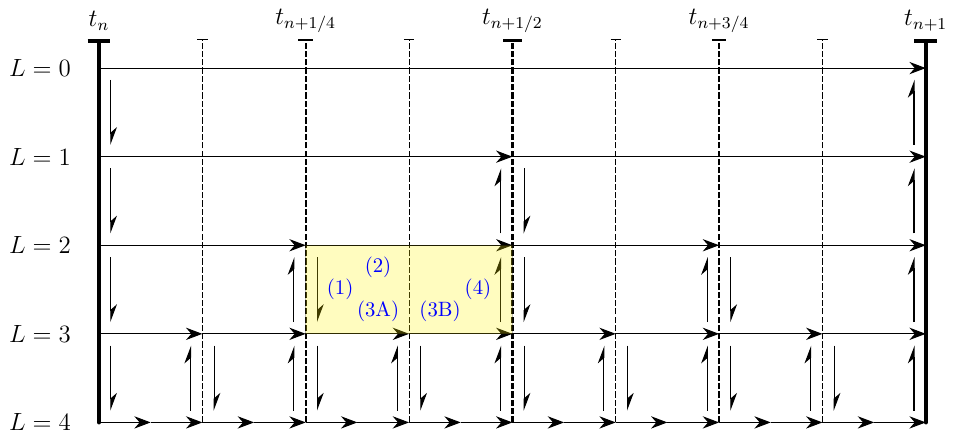}
        \caption{Visualization of the recursive advancement procedure for a grid with four levels of refinement. Highlighted is a single instance with four steps enumerated: 1) interpolation from $L$ to $L+1$, 2) stepping in time on $L$, 3A) first step in time on $L+1$, 3B) second step on fine on $L+1$ and 4) averaging from $L+1$ to $L$. Steps 1) and 4) are omitted for advancements on the finest level.}
        \label{fig:solver_adv}
    \end{figure}

\begin{figure}[h!]
        \centering
        \includegraphics[scale=0.65]{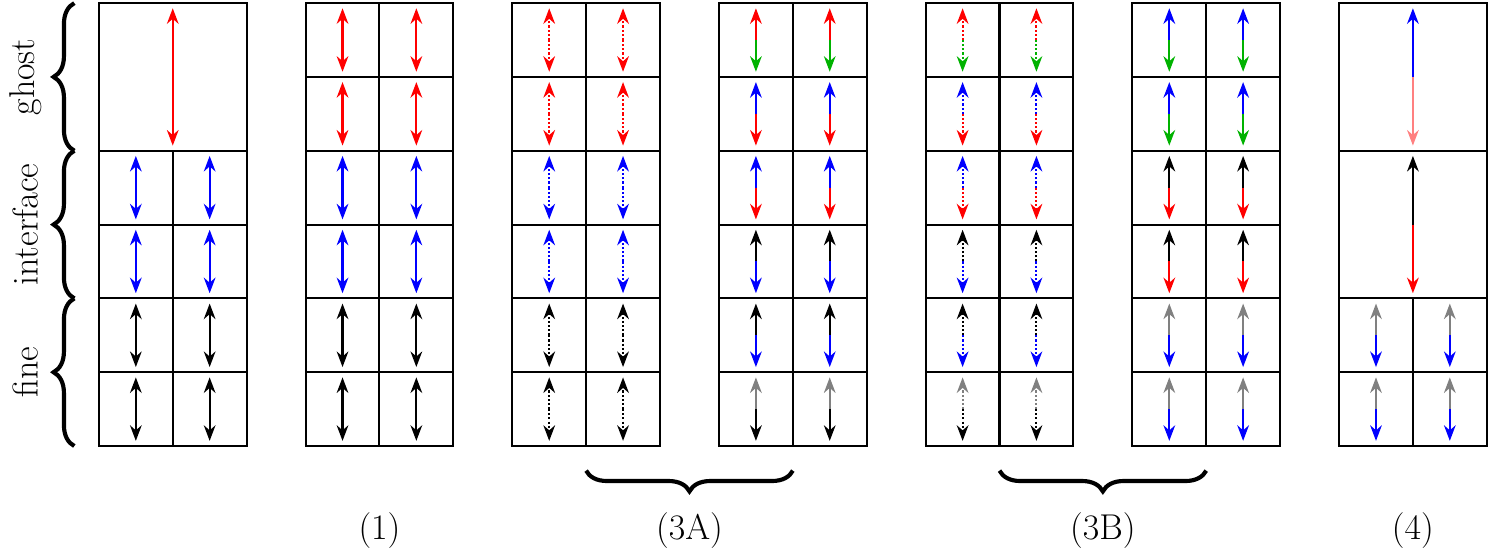}
        \caption{Visualization of coarse-fine communication via ghost and interface cells with only two sets of DDFs drawn for clarity. Numbering is consistent with the recursive instance defined by yellow shading in Figure \ref{fig:solver_adv}. Data is interpolated from a coarse cell to two layers of ghost cells first. Then, collision and streaming proceed twice on the fine grid, including ghost and interface cells. Finally, averaging is performed only over the interface layer. Green arrows indicate spoiled data that accumulates in the ghost layers and is discarded afterward.}
        \label{fig:p2-advance}
    \end{figure}

    The grid is advanced in a recursive breadth-first traversal of the underlying forest of octrees. The nodes of each level in the tree are enumerated by separate ID sets that are supplied individually to a set of solver and grid communication kernels and processed according to the primary mode of access described in Section \ref{sec:amr_mem} such that every cell is updated before the subsequent level can be processed. The routine is described with representative coarse and fine grid levels $L$ and $L+1$, where it is called recursively on the latter (except on level $L_{\text{max.}}-1$) until a final synchronization with the root grid is performed. This is shown in Figure \ref{fig:solver_adv} for a sample forest of octrees with four levels of refinement. The representative routine is decomposed into four steps: 1) interpolation of data from the coarse grid to ghost cells on the fine grid, 2) collision and streaming on the coarse grid with time step $\Delta t / 2^L$, 3) two calls to the routine on the fine grid if $L < L_{\text{max.}}-1$ or two calls to collision and streaming with time steps $\Delta t / 2^{L+1}$ otherwise, and 4) averaging of data on interface cells on the fine grid to their parents on the coarse grid. This ordering ensures that ghost and interface cell data are replenished on finer grids before synchronization with the data on their coarser parents. Figure \ref{fig:p2-advance} shows the step-by-step procedure for one step in time on an arbitrary level.
    
    After one round of collision and streaming on the fine grid, the outer layer of ghost cells on the fine grid contains invalid data, as they do not participate in streaming in directions originating from the nearby coarse grid. However, the inner ghost cell layer receives correctly collided and streamed data. After the second round, both ghost layers become invalid, but the interface cells similarly receive correct data. The data at the interface cells is then averaged to the coarse grid, concluding the advancement. 
    
    Ghost and interface cells in the fine grid both participate in collision and streaming, unlike the method of Rohde et al. \cite{Rohde2006}, where collision is restricted to DDFs distributed to the fine grid or streamed into it. In Schornbaum~\cite{Schornbaum2018}, it is mentioned that only DDFs streaming into a coarse block need to be communicated from the fine grid to the coarse grid\footnote{Interpolation and averaging in the current work play equivalent roles to the explosion and coalescence functions described by Schornbaum.}. The exact specification of the DDFs streaming into the coarse grid in different situations requires additional conditionals regarding the DDF arrangement in the streaming kernel, which would require more complex code generation. By including ghost and interface cells during collision and streaming, transfers between the coarse and fine grids can be achieved using just the fine grid so that averaging can safely be done for all DDFs rather than just those entering the coarse grid.
    
    \subsection{Streaming} \label{sec:solver_streaming}

\begin{figure}[h]
     \centering
     \begin{subfigure}[b]{0.45\textwidth}
         \centering
         \captionsetup{width=0.8\textwidth}
         \includegraphics[width=0.7\textwidth]{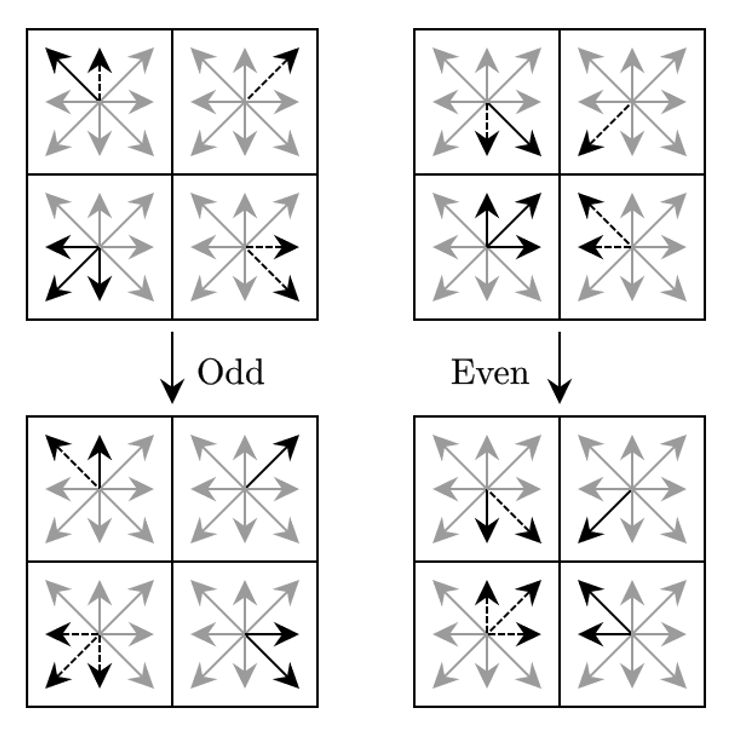}
         \caption{Esoteric Twist scheme distinguishing between odd and even time steps.}
         \label{fig:solver_streaming_cell}
     \end{subfigure}
     \begin{subfigure}[b]{0.45\textwidth}
         \centering
         \captionsetup{width=0.8\textwidth}
         \includegraphics[width=0.7\textwidth]{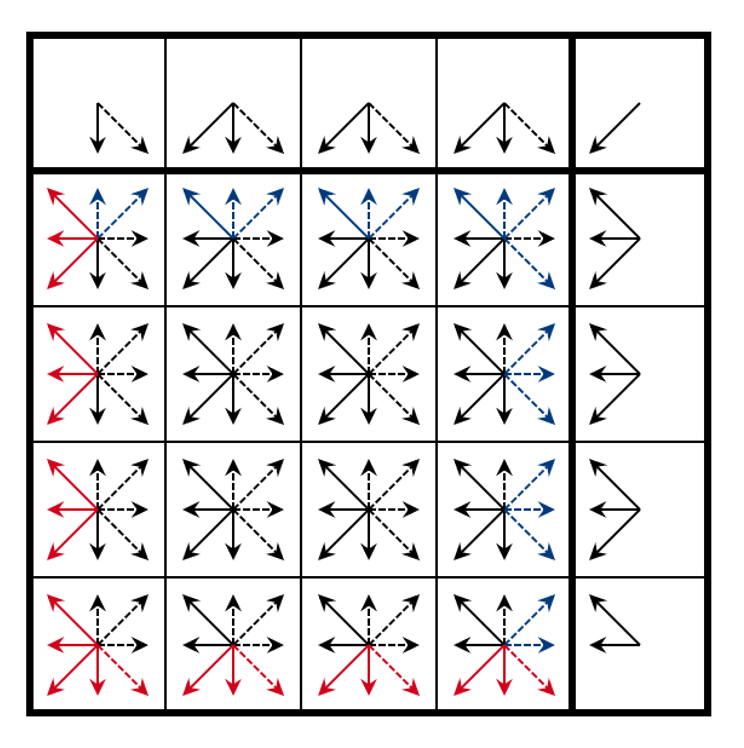}
         \caption{Block-based adaptation of Esoteric Twist applies to odd and even time steps.}
         \label{fig:solver_streaming_block}
     \end{subfigure}
        \caption{Visualizations of the Esoteric Twist streaming scheme based on the description of Geier and Schönherr  \cite{Geier2017} (left) and the block-based approach utilized in the current work (right). Red DDFs are ignored when streaming in the interior. Red and blue DDFs may participate in boundary condition imposition if applicable.}
        \label{fig:solver_streaming_1}
    \end{figure}

    Geier and Sch{\"o}nherr \cite{Geier2017} describe a streaming scheme that exploits the observation that DDFs can be 'twisted' and made to exchange locations during streaming to avoid introducing an intermediate grid for temporary storage (as shown in Figure \ref{fig:solver_streaming_cell}). That is, $f_p$ of a cell at location $\textbf{x}$ can be made to take the place of $f_{\overline{p}}$ at $\textbf{x}+\textbf{c}_p \Delta t$ in memory and vice-versa. This enables the implementation of a combined collision and streaming step or possibly distinct ones for odd and even time steps depending on the chosen arrangement of data.

    This approach is modified in a few ways for the current AMR scheme. One kernel implementation is maintained to ensure that DDFs are always in the same location when interpolating/averaging between time steps. Pointer swapping is impossible as one array stores the solution field for all grid levels simultaneously, so exchanges must be performed in register memory. According to Geier and Sch{\"o}nherr \cite{Geier2017}, block-based implementations of their scheme require additional ghost cells for exchanges between blocks, decreasing its economic efficiency in memory. However, shared memory can facilitate these exchanges locally on the GPU to avoid additional allocations. 
    
\begin{figure}[h]
        \centering
        \includegraphics[scale=0.3]{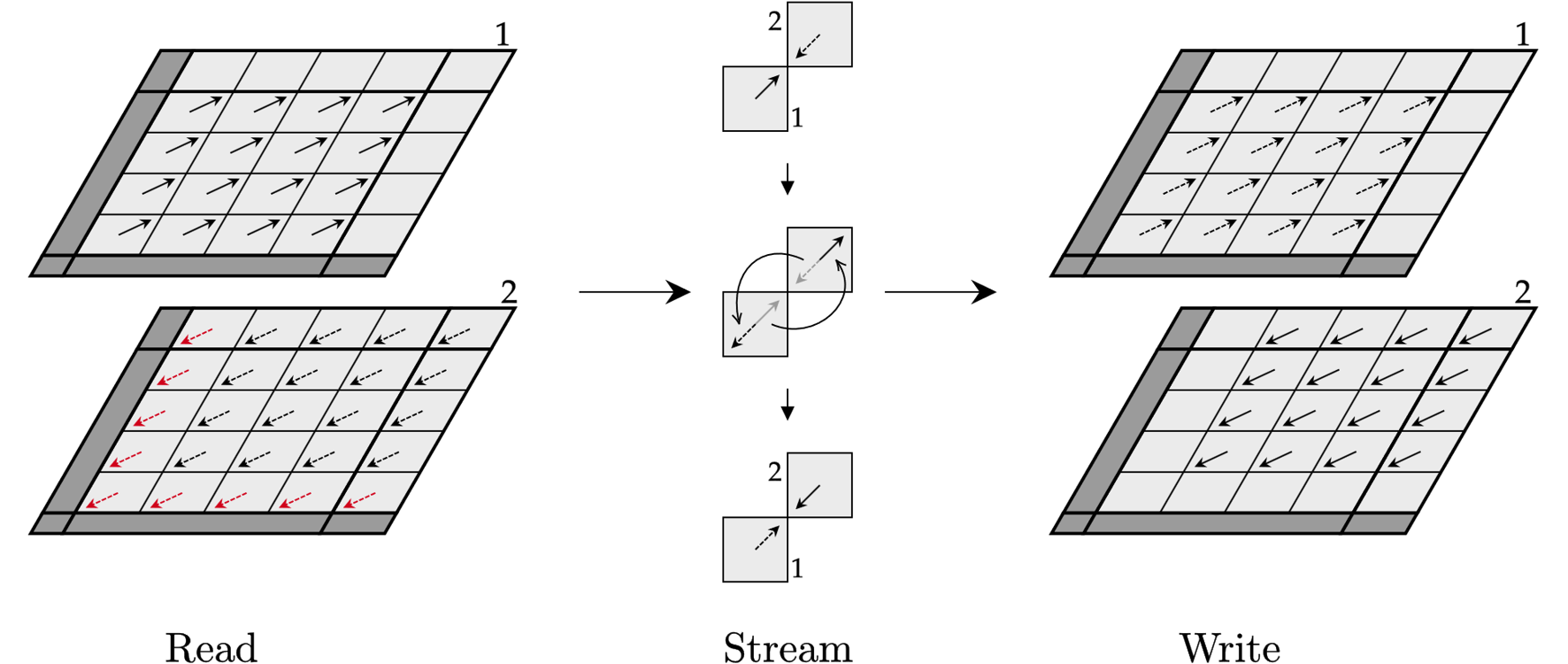}
        \caption{Illustration of streaming performed in shared memory (shown as a thin grid of cells). DDFs of opposing directions $p, \overline{p}$ (shown as solid and dashed arrows, respectively) are loaded in pairs into two separate arrays. The streaming exchange is performed between these arrays, resulting in DDFs of direction $p$ taking the place of those in $\overline{p}$ and vice-versa. Red arrows do not participate in streaming within the current block but will do so when called from an appropriate neighbor. A '-1' value in the halo is shown in dark gray and acts as a guard during the global write step.}
        \label{fig:solver_streaming_mem}
    \end{figure}

    Two shared memory arrays are initialized with sizes of $(4+2)^{N_d}$, which is necessary to store a block of cells encircled by an extra halo layer one cell thick. DDFs are traversed in pairs $p,\overline{p}$ where the post-collision DDFs $f_p^*$ and $f_{\overline{p}}^*$ are loaded from global memory and placed in the two arrays, respectively. Next, specific neighbors are processed depending on where the DDFs are expected to stream. For example, DDF 13 with particle velocity vector $\textbf{c}_{13} = (1,1,0)^T$ will require data to be streamed from block neighbors 2, 3, and 13 across the north, east, and north-east faces of the block, respectively. This is illustrated in Figure \ref{fig:solver_streaming_block}. When a neighbor is valid (i.e., its ID is non-negative), data is loaded into the halo region of the shared memory array based on conditionals applied to the local indices. For example, at the interface between a block and its neighbor 1, cells with index $(0,J,K)$ are loaded into the array via $0 + 4(J+1) + 4^2(K+1)$, where $0 \leq J, K \leq  3$. Although only specific threads will be active during these global loads, coalescence is guaranteed by the structured nature of the neighbor blocks, so only one transaction is required (assuming single precision storage). Consequently, collision and streaming can no longer be combined in one kernel. 
    Streaming is performed with $f_{\overline{p}}(t + \Delta t, \textbf{x}) = f_p^*(t + \Delta t, \textbf{x} - \Delta t \textbf{c}_p)$ and $f_{p}(t + \Delta t, \textbf{x}) = f_{\overline{p}}^*(t + \Delta t, \textbf{x} - \Delta t \textbf{c}_{\overline{p}})$. Global writes of the streamed $f_p^*, f_{\overline{p}}^*$ in $\overline{p}, p$ establish the twist as visualized in Figure \ref{fig:solver_streaming_mem}. These are corrected later in the following collision step.

    It should be noted that the GPU implementation of the streaming routine employs local communication between blocks on the same grid level via shared memory, such that a separate routine is not required. Explicit ‘explosion’ and ‘coalescence’ steps (i.e., conversion of the coarse cells into a fine one and vice versa) are avoided by storing both ghost cells and their parents and ignoring the latter during advancement. The inclusion of parents with their children in the data arrays reduces overhead in conversion at such interfaces.

    \subsection{Collision} \label{sec:solver_collision}

    Collision comprises a local evaluation of the collision operator with values at the current time step. For the BGK model, all DDFs are relaxed towards equilibrium using the macroscopic properties at a relaxation rate $\tau$ determined by viscosity $\nu = c_s^2(\tau - \Delta t/2)$. The relaxation rate must be modified on different grid levels in the context of grid refinement to recover the correct Reynolds number. If the lattice speed of sound $c_s$ is to be maintained constant across the hierarchy, then $\Delta x_L = \Delta t_L$ must also be maintained, implying that the time-step ratio will always be equal to the refinement ratio between grids, fixed equal to 2 in the current work. The relaxation rate on grid level $L$ is therefore recomputed with $\Delta t_L = \Delta t / 2^L$. This is done once during initialization and supplied as input to the CUDA kernels. 
    
    Boundary conditions are applied as a pre-streaming operation at the end of the collision routine after the collided DDF values and the computed macroscopic properties become available. The Esoteric Twist streaming strategy enables local imposition of boundary conditions without overwriting for no-slip and pressure boundary conditions. A post-collision DDF $f_p^*$ that would stream into a boundary has its direction reversed in the bounce-back condition and lands in the location of DDF $\overline{p}$. However, post-stream DDFs have their direction reversed, so the final location of a bounced-back DDF is its original location. If the velocity at the boundary is zero, the DDF value remains unchanged, and the boundary condition is automatically satisfied without doing anything. This phenomenon is called implicit bounce-back \cite{Geier2017, Lehmann2022}, and its main advantage is that specific neighbor IDs do not need to be checked.
    
    The implementation of collision is summarized with the following steps: 
    \begin{enumerate}
    \item 
     Load DDFs into register memory with alternating directions swapped,
     \item  Compute $\omega = \Delta t / \tau$ and $\omega' = 1 - \omega$, 
     \item Compute post-collision DDF values for all directions according to:
        \begin{align}
            f_p^*(t + \Delta t, \textbf{x}) = \omega' f_p(t, \textbf{x}) + \omega f_p^{\text{eq.}}(t, \textbf{x}),
        \end{align}
    \item Check neighbor block IDs if the current block lies on the boundary and apply boundary conditions,
    \item Write DDFs into \texttt{cells\_f\_F} in the correct order.
    \end{enumerate}

    \subsection{Adaptive Mesh Refinement} \label{sec:solver_amr}

    As the simulation progresses, regions of interest may require locally increased resolution. Such regions are identified using an arbitrary refinement criterion computed from physical properties of the system. In the current work, vorticity magnitude is chosen for this purpose as it captures the coherent structures that develop in the case studies to be considered for solver validation.

    The mesh requires a pre-computation step prior to invocation of the refinement and coarsening algorithm to ensure that any data transferred to newly-generated children is valid for the current time step. First, a global averaging procedure is performed over the entire grid hierarchy starting with the second-finest level and ascending to $L=0$. This ensures that any new fine-grid boundaries that have formed after coarsening can safely interpolate to freshly assigned ghost cells. Next, interpolation is performed to ghost cells on all levels in the opposite direction, starting from $L=0$ and descending onward. This ensures that the ghosts also possess valid data for the current time step. Macroscopic properties can be safely computed at this stage and placed in shared memory in preparation for the computation of the refinement criterion. In the case of vorticity magnitude, first-order finite differences are used.
    
    Once the refinement criterion has been found for all cells, the block-wise maximum value must be identified. Criterion values are loaded into shared memory following the primary mode of access and a reduction is performed. With the maximum value determined, the desired grid level can be calculated and compared with current levels to determine the course of action. When desired level $L_{\text{des.}} > L$, the block requires more accuracy and is marked for refinement. If $L_{\text{des.}} < L$, the block needs to be coarsened and is marked accordingly. The desired level is based on the logarithm of the criterion in base 2 according to the procedure of Algorithm \ref{alg:criterion}, where $\epsilon$ caps the log$_2$ of the maximum criterion value by 1 (a value selected by tuning during validation), and $N_{\text{start}}, N_{\text{inc.}}$ are tunable 'start' and 'incremental' parameters. To ensure that the octree remains 2:1 balanced, neighbor-children are checked and refinement is canceled if at least one neighbor-child is set equal to $N_{\text{skip}}$ (otherwise, at least one of the resulting children will be adjacent to a block that has not been refined such that the size ratio between them is greater than two).
    \input{algos/Criterion}

    An additional routine has been implemented for refinement according to distance from the nearest wall, providing additional resolution near walls of interest. Distance is computed based on block coordinates and domain boundaries. For successive near-wall refinements, the distance required is divided so that $\text{min}\{d_i\} < d_{\text{spec.}}/2^L$ specifies the refinement condition, where $d_i$ are the distances between the current block and the domain boundaries, $d_{\text{spec.}}$ is the specified distance, and $L$ is the level occupied by the current block. An option to freeze blocks added in this manner is enabled to prevent premature coarsening.

%% file: algos/Criterion.tex
\begin{algorithm}[h!]
\footnotesize
    \caption{Refinement/Coarsening Criterion}\label{alg:criterion}
    $\epsilon \gets \text{min}(1,\log_2|\omega|)$ \\
    $L_{\text{des.}} \gets L_{\text{max.}}-1$ \\
    \For{$p = 1 \to L_{\text{max.}}-1$}
    {
        \If{$\epsilon < N_{\text{start}} - p N_{\text{inc.}}$}
        {
            $L_{\text{des.}} \gets (L_{\text{max.}}-1)-p$
        }
    }
\end{algorithm}

%% file: p3_tests.tex
\section{Validation and Case Studies} \label{sec:tests}

The performance and accuracy of the current implementation are assessed using two benchmark problems: the lid-driven cavity (LDC) and the flow past a square cylinder (FPSC). Several tests are performed with variation in select parameters (Reynolds number Re, number of bytes for the chosen floating-point precision $N_p$, number of dimensions $N_d$, velocity set size $N_Q$, root grid size $N_{\text{coarse}}^{N_d}$, and maximum number of grid levels $L_{\text{max.}}$) to investigate how performance is affected and to identify possible bottlenecks in specific subroutines that can be targeted for later optimization. Execution times for all interpolation, collision, streaming, and averaging calls are recorded per grid level and used to compute the fraction of time taken in coarse-fine grid communication and lattice node updates per second. These serve as metrics for solver efficiency and solver performance, respectively. Total simulation times are also reported to demonstrate the utility of the implementation even on older hardware. Execution times for the steps taken in refinement and coarsening are recorded to identify the fraction of simulation time needed to adapt the mesh as a measure of AMR efficiency. A small fraction implies that using AMR does not detract from the time required to solve the governing equations. All execution times are estimated by computing the difference in wall time at instances before and after routine calls with device stream synchronization to ensure full completion between calls. A selection $N_{q,x}=1, M_L = M_t$ is used for all test cases (other than studying performance against cell-block size). Linear interpolation is used for the lid-driven cavity tests.

The results of the LDC simulations are presented first. The implementation of the single-relaxation-time LBM scheme with AMR is validated by comparing velocity profiles extracted from the interior of the cavity with profiles reported in the literature. 
For a fixed Reynolds number $\text{Re}=1000$, the execution time distributions required for time-stepping and grid refinement are tabulated and visualized across various combinations of velocity sets and GPU configurations. Performance over time concerning the latter is juxtaposed in separate figures by velocity set to highlight the relative speedup. The arguments supporting the use of a block-based approach to AMR are also verified by comparison of the average execution time for cell- and block-based grids with a single grid level. It is shown that by shuffling the cell/block indices representing the grid, a significant performance penalty is incurred by the cell-based approach, while performance remains consistent in the block-based approach even with reduced locality in memory. Performance is then assessed with respect to cell-block size and thread-block workload by varying $N_{q,x}$ and $M_L$, respectively.

Results of the FPSC simulation are reported next. Time-averaged drag coefficient and Strouhal number values are reported and compared with those in the literature. Total simulation times are reported with AMR and with corresponding uniform grids with effective resolutions $\Delta x / 2^{L_{\text{max.}}-1}$ extended to the whole domain to demonstrate considerable speedup of the former. CPU times for similar effective resolutions in the literature are also reported to demonstrate the effectiveness of the GPU-native implementation. It is verified that the pressure field obtained with linear interpolation is continuous across coarse-fine interfaces. Dimensionless quantities are computed and compared with both linear and cubic interpolation.

    \subsection{Solver Performance}
    
    A commonly used metric for assessing solver performance is MLUPS (Million/Mega Lattice node Updates Per Second). This is usually calculated for a grid with a fixed resolution and averaged for all time steps; however, with AMR, the metric must account for the repeated updates of nodes on higher grid levels for a single time step on the root grid. This is expressed as:
    \begin{align}
        \text{MLUPS}_{\text{inst.},k} &\approx \frac{1}{\Delta t_{\text{exec.},k}} \sum_{L=0}^{L_{\text{max.}}-1} 2^L \bigg(N_{\text{nodes},L} - \frac{N_{\text{nodes},L+1}}{N_c}(1-\delta_{L,L_{\text{max.}}-1}) \bigg), \\
        \overline{\text{MLUPS}} &\approx \sum_{k=t_f/N_x-1024}^{t_f/N_x-1} \text{MLUPS}_{\text{inst.},k}, \label{eq:MLUPS_ave}
    \end{align}
    
    \noindent where $2^L (N_{\text{nodes},L} - N_{\text{nodes},L+1}/N_c)$ is the number of updates resulting from $2^L$ advancements in time for active nodes on grid level $L$ (nodes with children that do not participate in advancement are subtracted from the total grid size), $\text{MLUPS}_{\text{inst.},k}$ represents an instantaneous calculation of the metric in discrete time $t_k = k \Delta t$ and $\overline{\text{MLUPS}}$ represents a steady-state average that is estimated with the last 1024 samples. These definitions include the time taken to communicate data between blocks on the same level (with the shared memory strategy described in Section \ref{sec:solver_streaming}) and between grids along the coarse-fine interface (with interpolation and averaging as described in Section \ref{sec:amr_comm}).

    \subsection{Lid-Driven Cavity} \label{sec:tests_ldc}

    Three sets of simulations are performed on the 970M, V100, and A100 GPU models to validate the numerical scheme described in Section \ref{sec:solver} (labeled V/A-S$_k$), compare performance when more precision and detail (i.e. increases in $N_Q, N_p$) are sought (labeled G/V/A$_k$), and establish the efficiency of the GPU-native AMR by showing that a minimal amount of time is spent in refinement throughout the simulations on all hardware. Table \ref{tab:tests_sim_params_ldc} summarizes the parameters employed in these simulations. For all simulations considered in this section, the mesh is adapted every 32 iterations at the coarsest level, and $N_{\text{start}}, N_{\text{inc.}}$ are set to -2 and 1, respectively.

    \small
    \begin{table}[h!]
        \centering
        \begin{tabular}{cc|ccccccc|c}
            \hline
            \hline
            \multicolumn{10}{c}{\textbf{Lid-Driven Cavity Simulations}} \\
            \hline
            \hline
            Sim. Label &flags& Re & $N_{\text{coarse.}}^{N_d}$ & $N_d$ & $L_{\text{max.}}$ & $t_f$ (s) & $N_p$ (bytes) & $N_Q$ & GPU \\
            \hline
            \hline
            G$_1$ &mar& 1000 & 128 & 2 & 4 & 1000 & 4 & 9 & GTX 970M \\
            G$_2$ &mar& 1000 & 128 & 2 & 4 & 1000 & 8 & 9 & GTX 970M \\
            G$_3$ &mar& 1000 & 64 & 3 & 3 & 1000 & 4 & 19 & GTX 970M \\
            G$_4$ &mar& 1000 & 64 & 3 & 3 & 1000 & 8 & 19 & GTX 970M \\
            G$_5$ &mar& 1000 & 64 & 3 & 3 & 1000 & 4 & 27 & GTX 970M \\
            G$_6$ &mar& 1000 & 64 & 3 & 3 & 1000 & 8 & 27 & GTX 970M \\
            V$_1$ &mvar& 1000 & 128 & 2 & 4 & 1000 & 4 & 9 & V100 \\
            V$_2$ &mar& 1000 & 128 & 2 & 4 & 1000 & 8 & 9 & V100 \\
            V$_3$ &mvar& 1000 & 64 & 3 & 4 & 1000 & 4 & 19 & V100 \\
            V$_4$ &mar& 1000 & 64 & 3 & 4 & 1000 & 8 & 19 & V100 \\
            V$_5$ &mar& 1000 & 64 & 3 & 4 & 1000 & 4 & 27 & V100 \\
            V$_6$ &mar& 1000 & 64 & 3 & 4 & 1000 & 8 & 27 & V100 \\
            A$_1$ &mar& 1000 & 128 & 2 & 4 & 1000 & 4 & 9 & A100 \\
            A$_2$ &mar& 1000 & 128 & 2 & 4 & 1000 & 8 & 9 & A100 \\
            A$_3$ &mar& 1000 & 64 & 3 & 4 & 1000 & 4 & 19 & A100 \\
            A$_4$ &mar& 1000 & 64 & 3 & 4 & 1000 & 8 & 19 & A100 \\
            A$_5$ &mar& 1000 & 64 & 3 & 4 & 1000 & 4 & 27 & A100 \\
            A$_6$ &mar& 1000 & 64 & 3 & 4 & 1000 & 8 & 27 & A100 \\
            \hline
            V-S$_1$ &v& 3200 & 64 & 2 & 4 & 1000 & 4 & 9 & V100 \\
            V-S$_2$ &v& 5000 & 64 & 2 & 4 & 2000 & 4 & 9 & V100 \\
            V-S$_3$ &v& 10000 & 512 & 2 & 3 & 5000 & 4 & 9 & V100 \\
            V-S$_4$ &v& 3200 & 128 & 3 & 4 & 2000 & 4 & 19 & V100 \\
            \hline
        \end{tabular}
        \caption{List of simulations considered for assessment of performance and validation of the current implementation, along with their parameters (flags indicate usage in this section, v: validation, m: MLUPS comparison, a: advancement time, r: mesh refinement/coarsening time).}
        \label{tab:tests_sim_params_ldc}
    \end{table}
    \normalsize

    \begin{figure}[h!]
        \centering
        \begin{subfigure}[b]{0.45\textwidth}
             \centering
             \includegraphics[width=\textwidth]{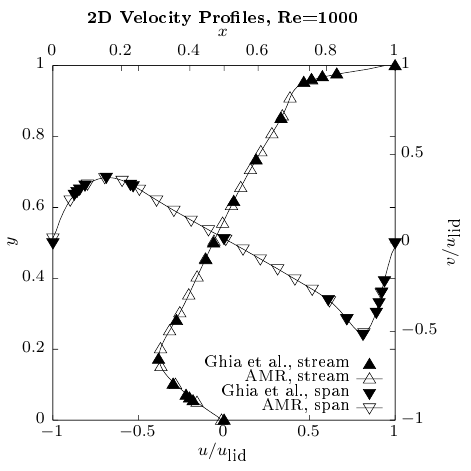}
             \caption{Profiles for $\text{Re}=1000$, 2D.}
         \end{subfigure}
         \hspace{0.5cm}
         \begin{subfigure}[b]{0.45\textwidth}
             \centering
             \includegraphics[width=\textwidth]{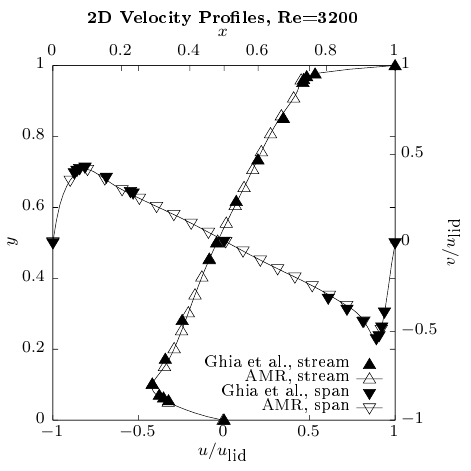}
             \caption{Profiles for $\text{Re}=3200$, 2D.}
         \end{subfigure}
         \begin{subfigure}[b]{0.45\textwidth}
             \centering
             \includegraphics[width=\textwidth]{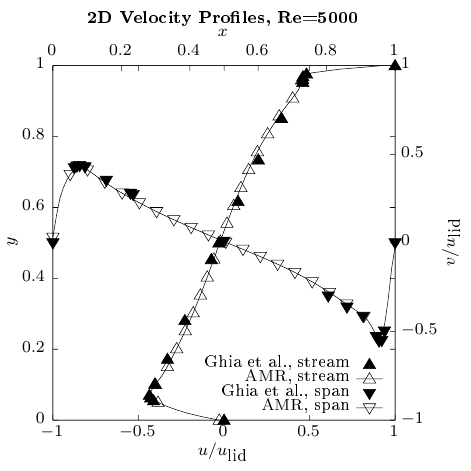}
             \caption{Profiles for $\text{Re}=5000$, 2D.}
         \end{subfigure}
         \hspace{0.5cm}
         \begin{subfigure}[b]{0.45\textwidth}
             \centering
             \includegraphics[width=\textwidth]{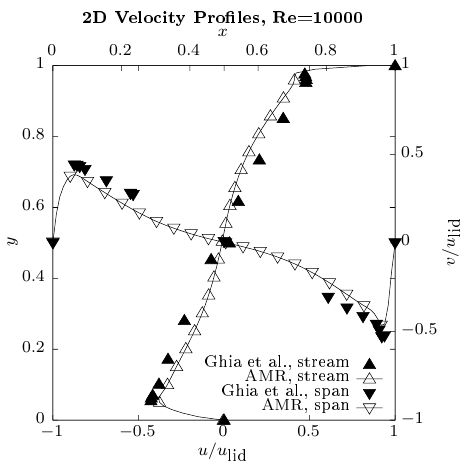}
             \caption{Profiles for $\text{Re}=10000$, 2D.}
         \end{subfigure}
        \caption{Plots of stream- and spanwise velocity profiles for simulations C$_1$, D, E and F corresponding to Re = 1000, 3200, 5000 and 10000. Profiles produced by the present solver are titled 'AMR'.}
        \label{fig:tests_ldc_valid_2d}
    \end{figure}

        \subsubsection{Velocity Profiles, Streamlines}

\begin{figure}[h!]
            \centering
            \begin{subfigure}[b]{0.4\textwidth}
                 \centering
                 \includegraphics[width=\textwidth]{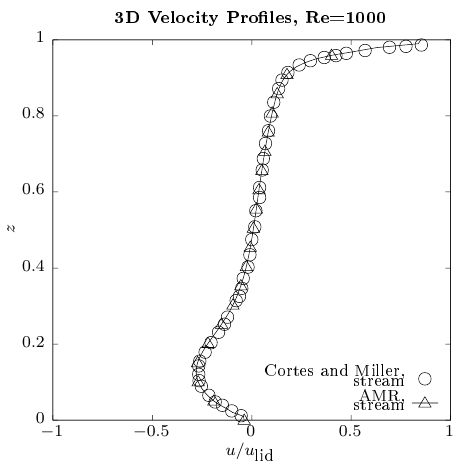}
                 \caption{Profiles for $\text{Re}=1000$.}
                 \label{fig:tests_ldc_valid_3d_A}
             \end{subfigure}
             \hspace{0.5cm}
             \begin{subfigure}[b]{0.4\textwidth}
                 \centering
                 \includegraphics[width=\textwidth]{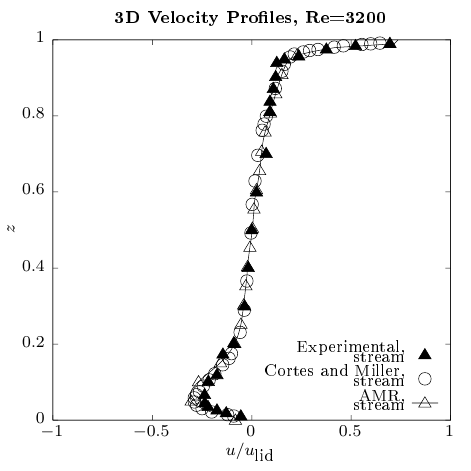}
                 \caption{Profiles for $\text{Re}=3200$.}
                 \label{fig:tests_ldc_valid_3d_B}
             \end{subfigure}
            \caption{Plots of normalized streamwise velocity profiles for simulations V$_3$, V-S$_4$ corresponding to Re = 1000, 3200.}
            \label{fig:tests_ldc_valid_3d}
        \end{figure}

        Solutions are validated by comparing velocity profiles obtained using the current implementation with those recorded by Ghia et al. \cite{Ghia1982} for 2D problems, and Cortes and Miller \cite{Cortes1994} for 3D problems across a range of Re. Reynolds numbers of 1000, 3200, 5000 and 10000 are considered in both 2D, and the former two in 3D. The experimental results of Prasad and Koseff \cite{Prasad1989} used by Cortes and Miller \cite{Cortes1994} for their validation of the Re=3200 case are also included for the 1000 and 3200 cases. The Reynolds number is modified by varying the fluid viscosity while keeping characteristic length and lid velocity fixed at unity and 0.05 m/s, respectively. The Mach number $\text{Ma} = u_{\text{lid}} / c_s$ is thus fixed at $\approx 0.1$. With the characteristic length and density set to unity, viscosity is determined via $\nu = 0.05/\text{Re}$. The normalized steam- and spanwise velocity profiles $u(y)/u_{\text{lid}}$, $v(x)/u_{\text{lid}}$ extracted at $x=0.5$ and $y=0.5$, respectively, for the 2D cases V$_1$, V-S$_k|_{1\leq k\leq 3}$ are displayed in Figure \ref{fig:tests_ldc_valid_2d}. The profiles $u(z)/u_{\text{lid}}$ of cases V$_3$ and V-S$_4$ are displayed in Figure \ref{fig:tests_ldc_valid_3d}.

        There is a general agreement between reference data and simulations with the present AMR solver. Deviation is mostly found at the curve inflection near the bottom of the profile in the region $0 \leq y \leq 0.2$ where low vorticity magnitude leads to less grid refinement. For Reynolds numbers $5000$ and $10000$, the simulation is run for $t_{f,\text{Re}=5000}=2000$ s and $t_{f,\text{Re}=10000} = 5000$ s, respectively, to achieve profiles close to equilibrium within job walltime limits. These two profiles match the data of Ghia et al. \cite{Ghia1982} at inflection points near the domain boundary in both stream- and spanwise plots and begin to straighten out in the cavity interior. 3D simulation results are also in general agreement with the profiles of Cortes and Miller \cite{Cortes1994}.
        
\begin{sidewaysfigure}[p]
            \centering
            \begin{subfigure}[b]{0.24\textwidth}
                 \centering
                 \includegraphics[width=\textwidth]{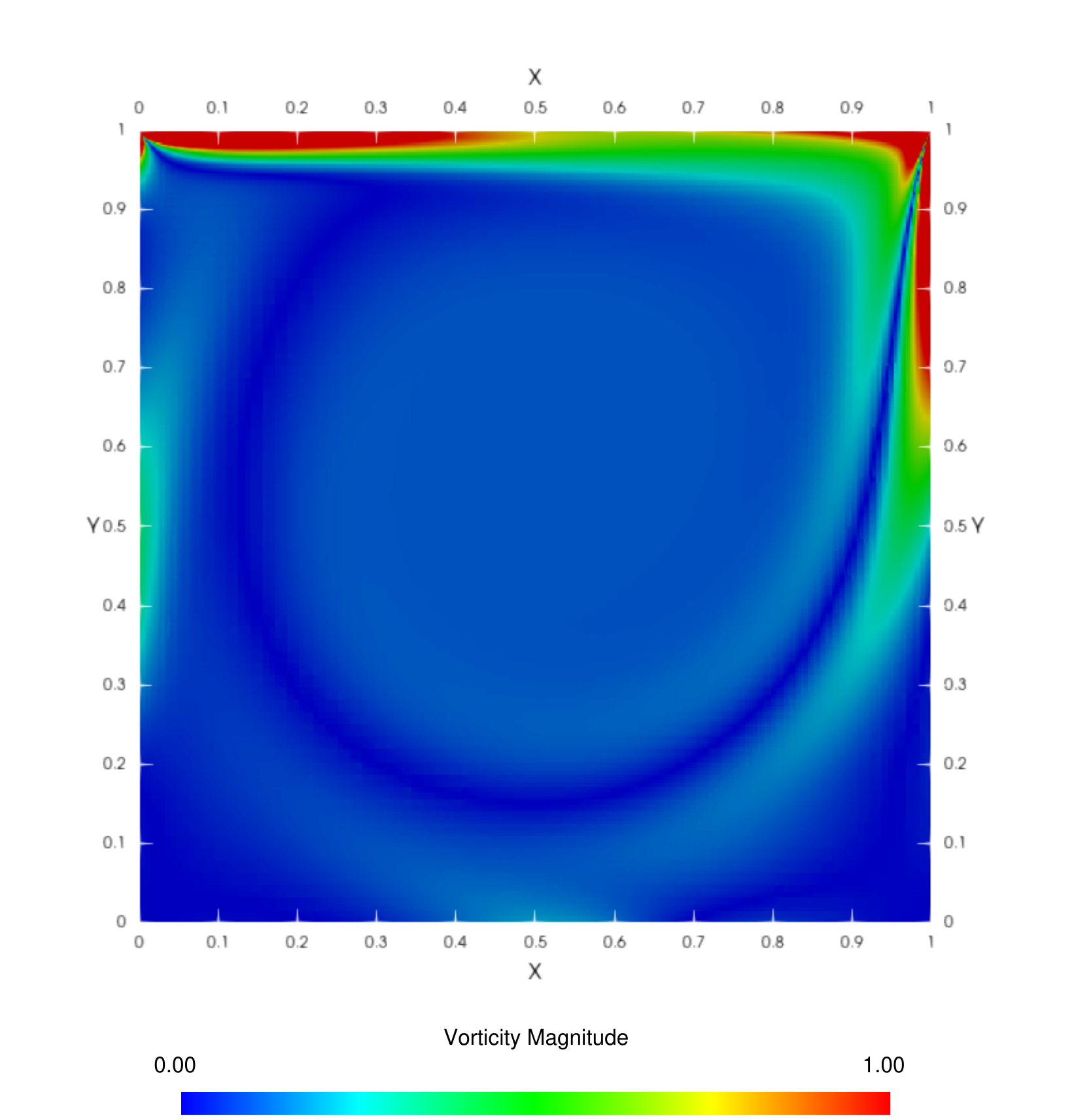}
                 \includegraphics[width=\textwidth]{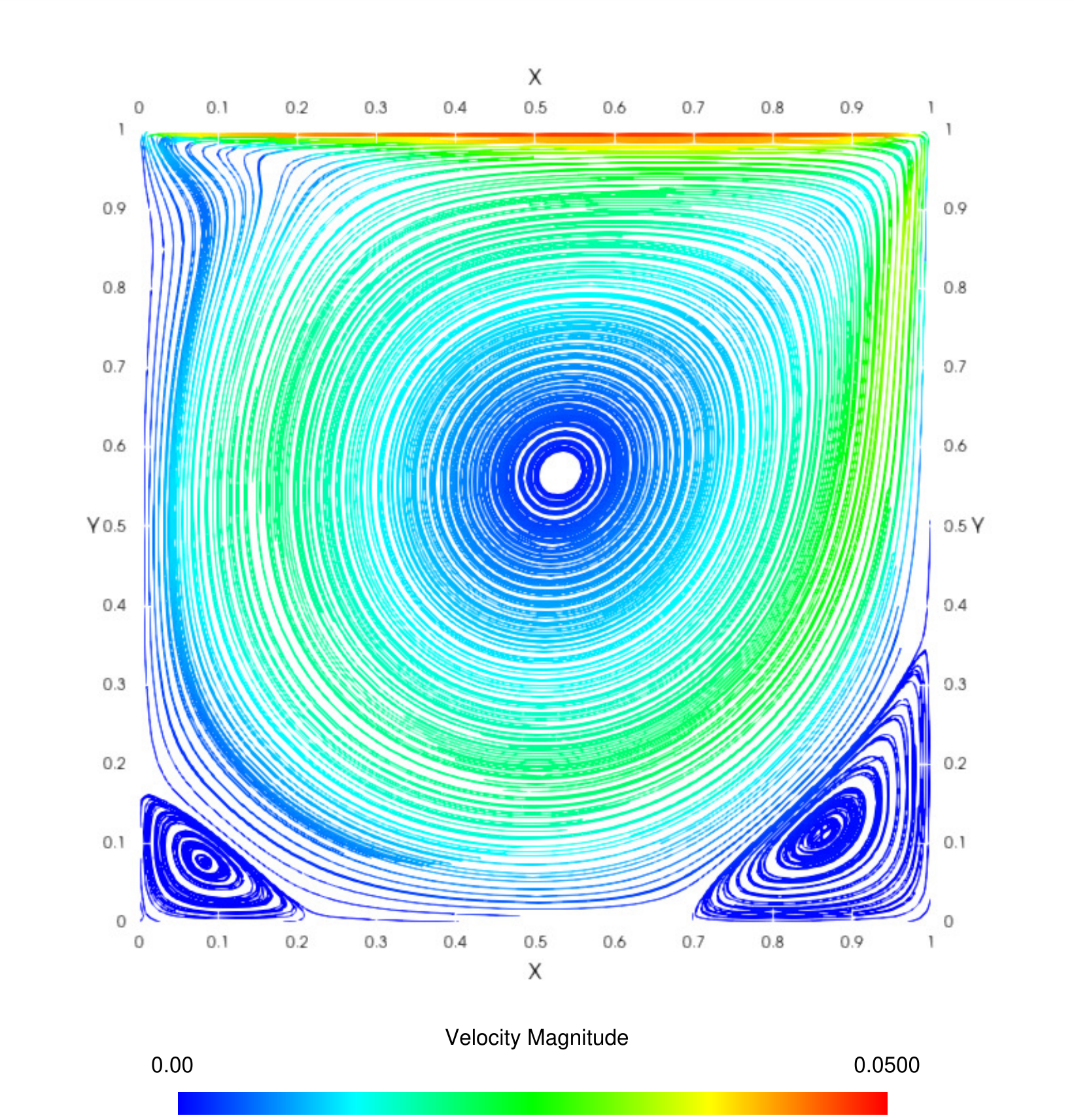}
                 \includegraphics[width=\textwidth]{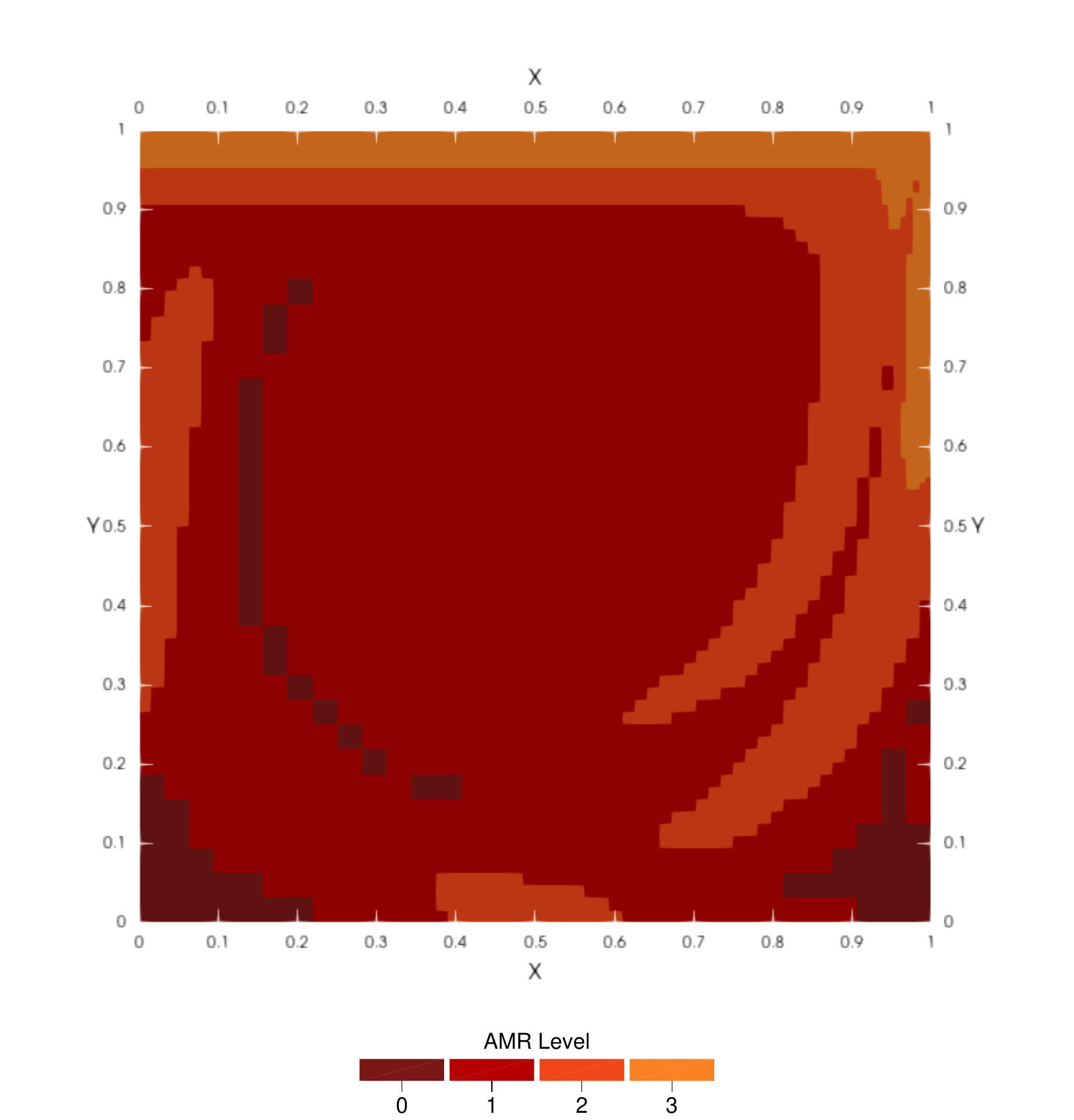}
                 \caption{$\text{Re}=1000$.}
                 \label{fig:y equals x}
            \end{subfigure}
            \begin{subfigure}[b]{0.24\textwidth}
                 \centering
                 \includegraphics[width=\textwidth]{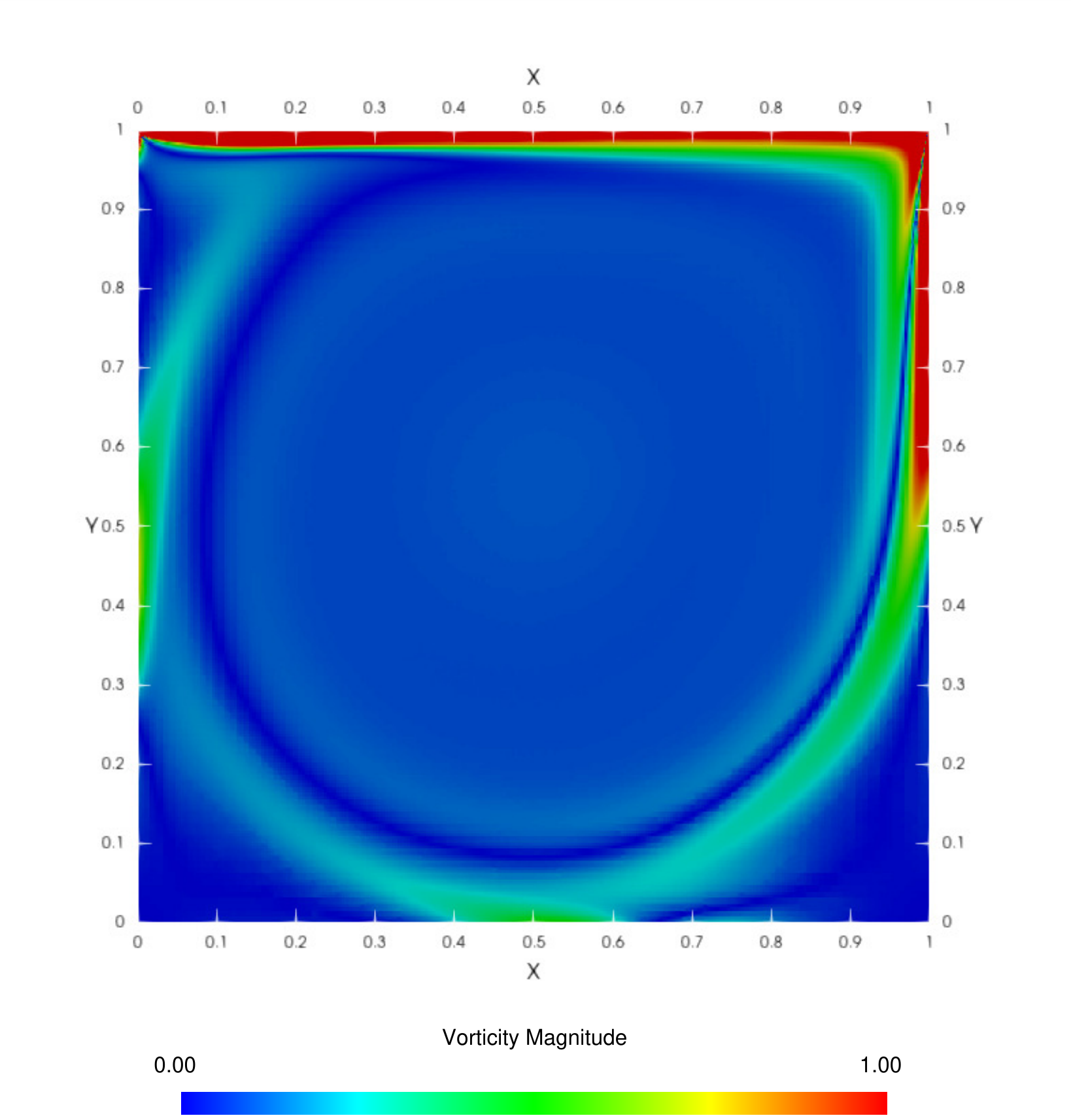}
                 \includegraphics[width=\textwidth]{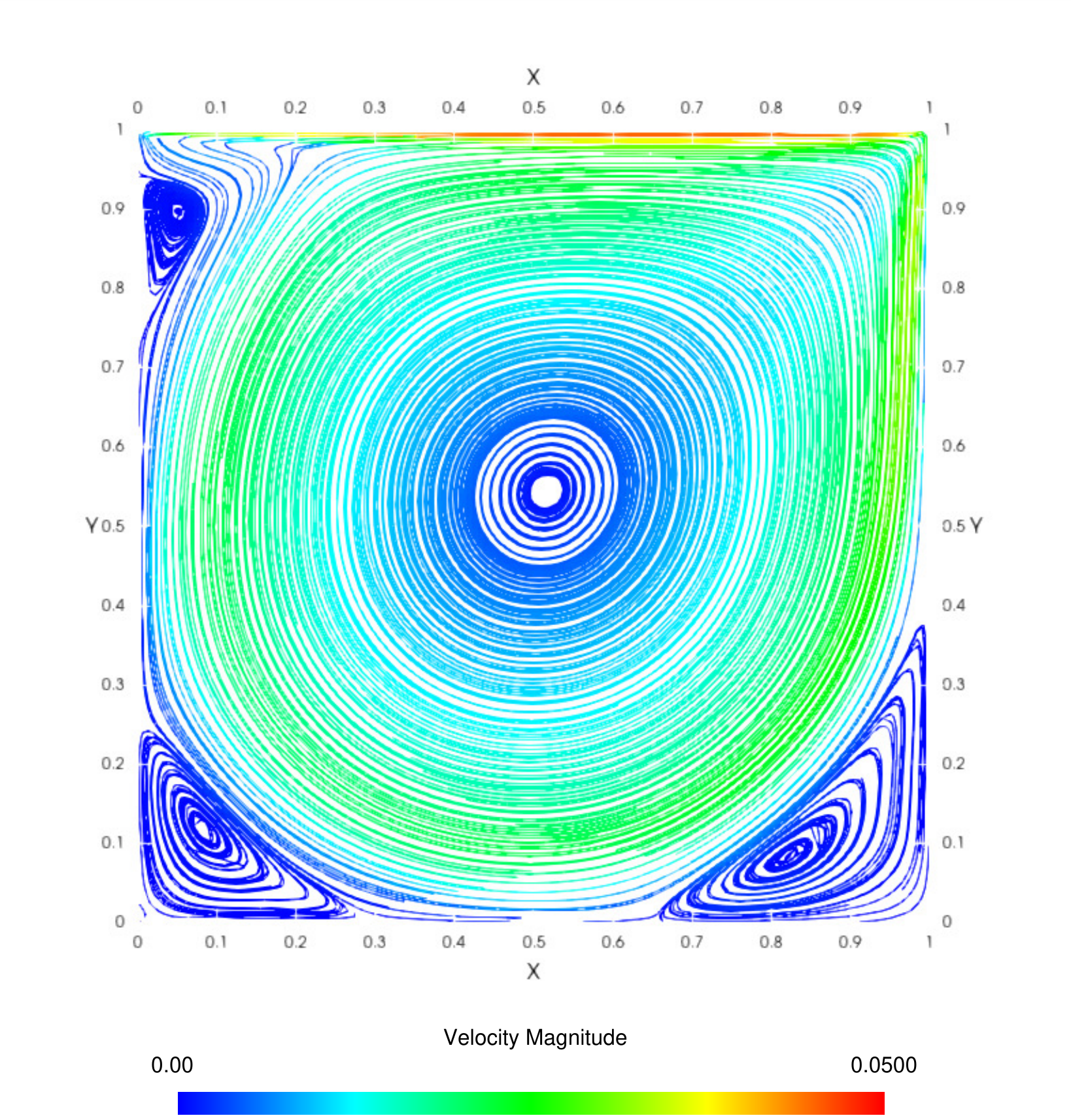}
                 \includegraphics[width=\textwidth]{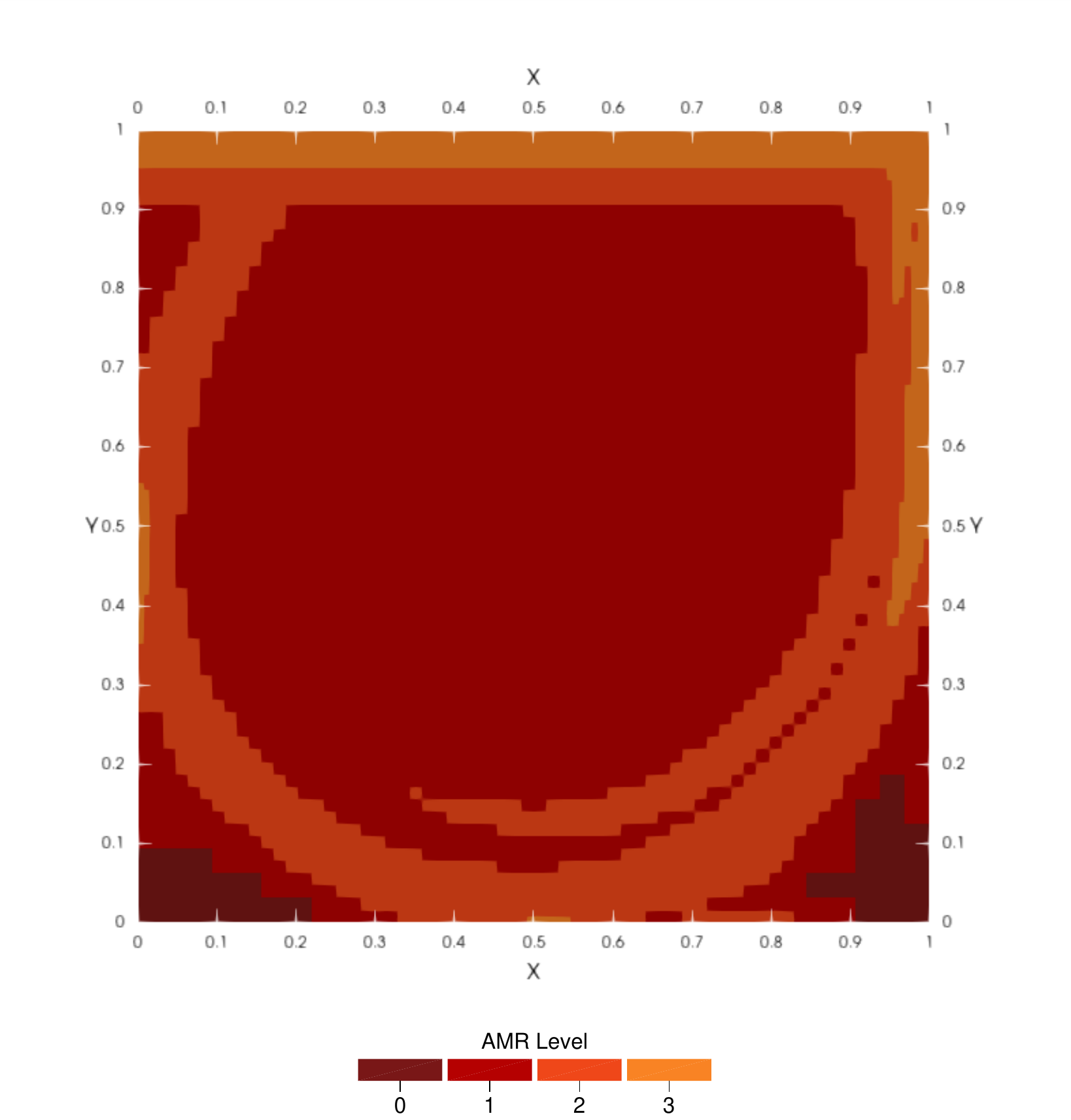}
                 \caption{$\text{Re}=3200$.}
                 \label{fig:y equals x}
            \end{subfigure}
            \begin{subfigure}[b]{0.24\textwidth}
                 \centering
                 \includegraphics[width=\textwidth]{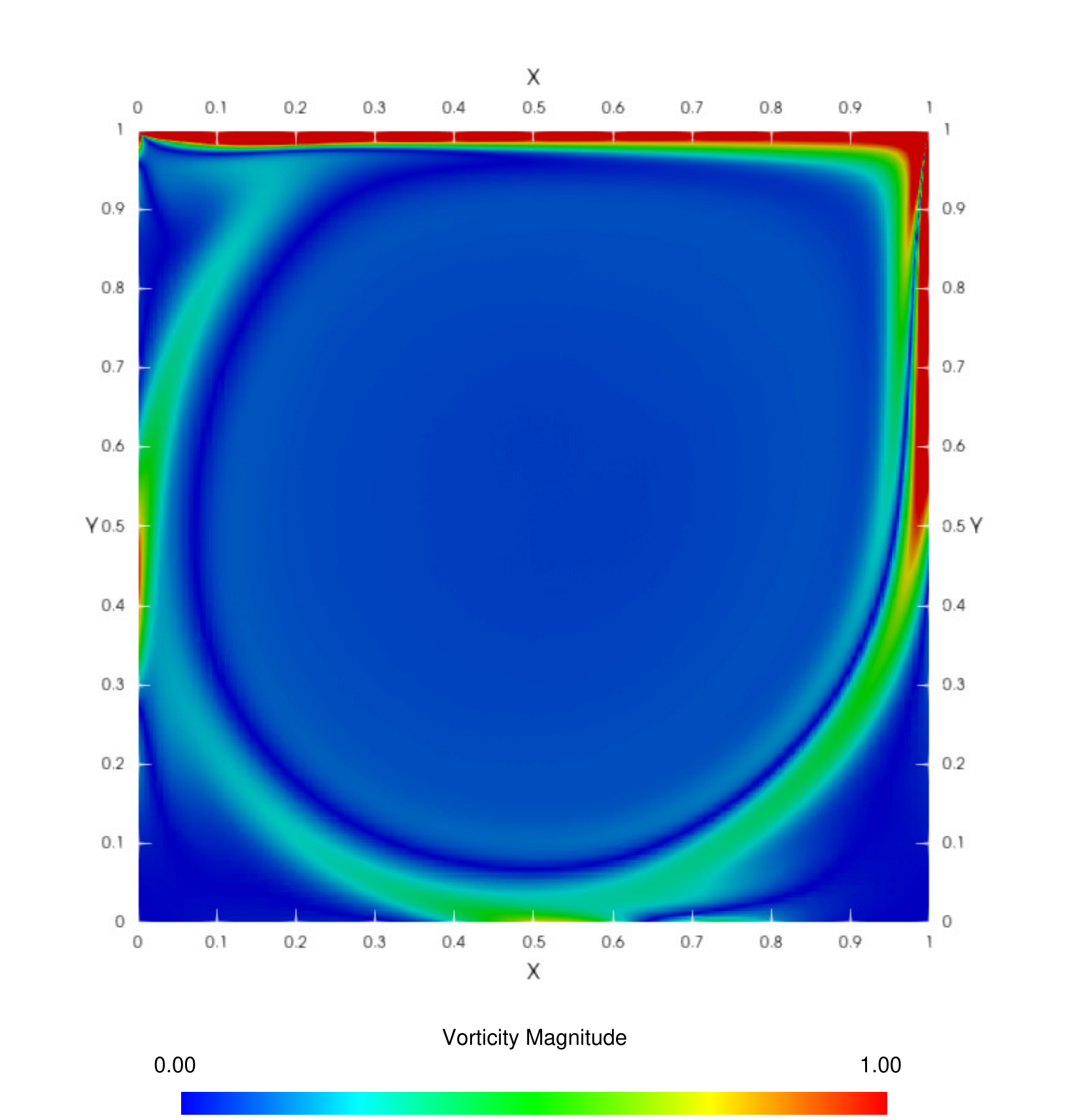}
                 \includegraphics[width=\textwidth]{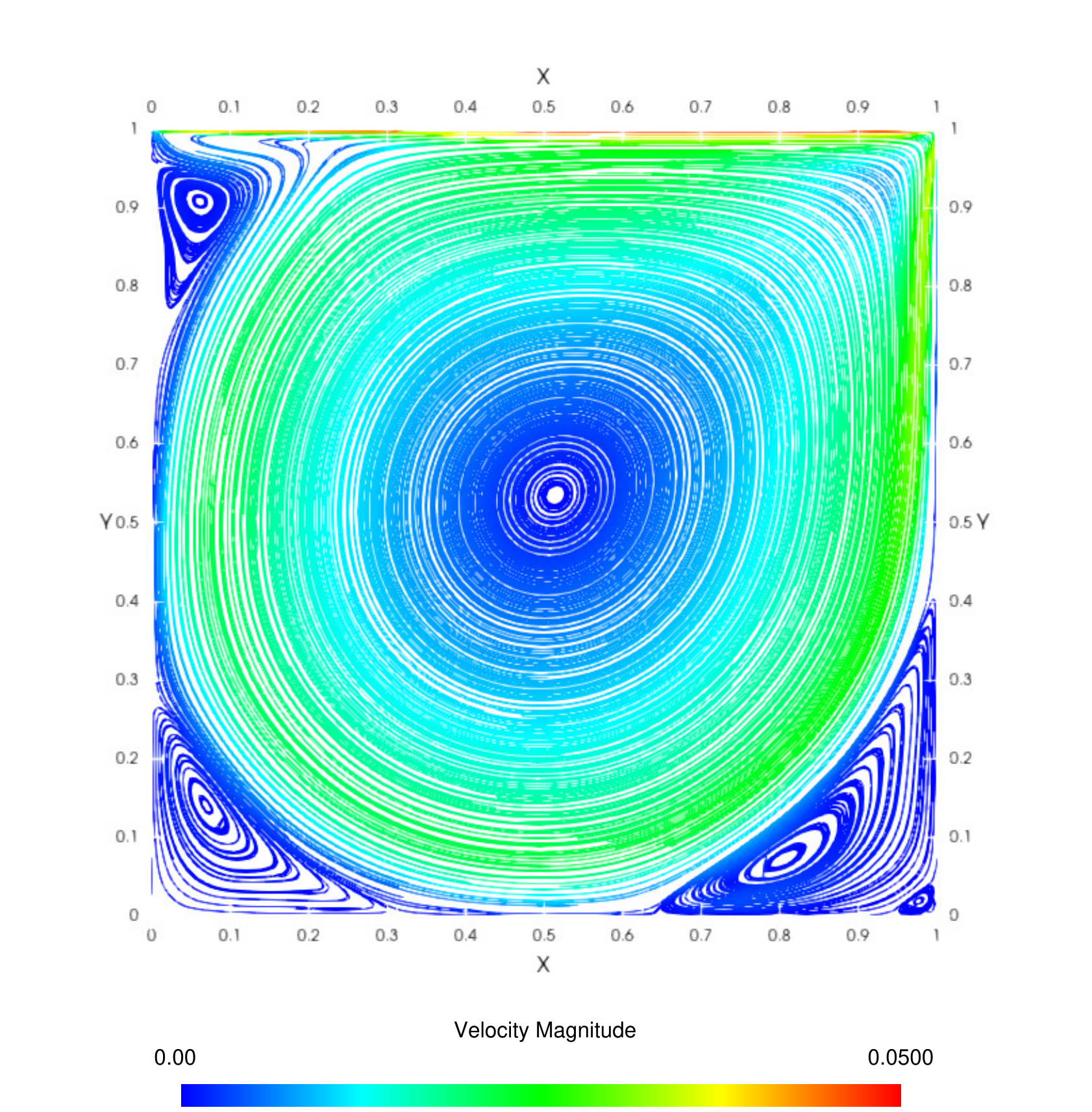}
                 \includegraphics[width=\textwidth]{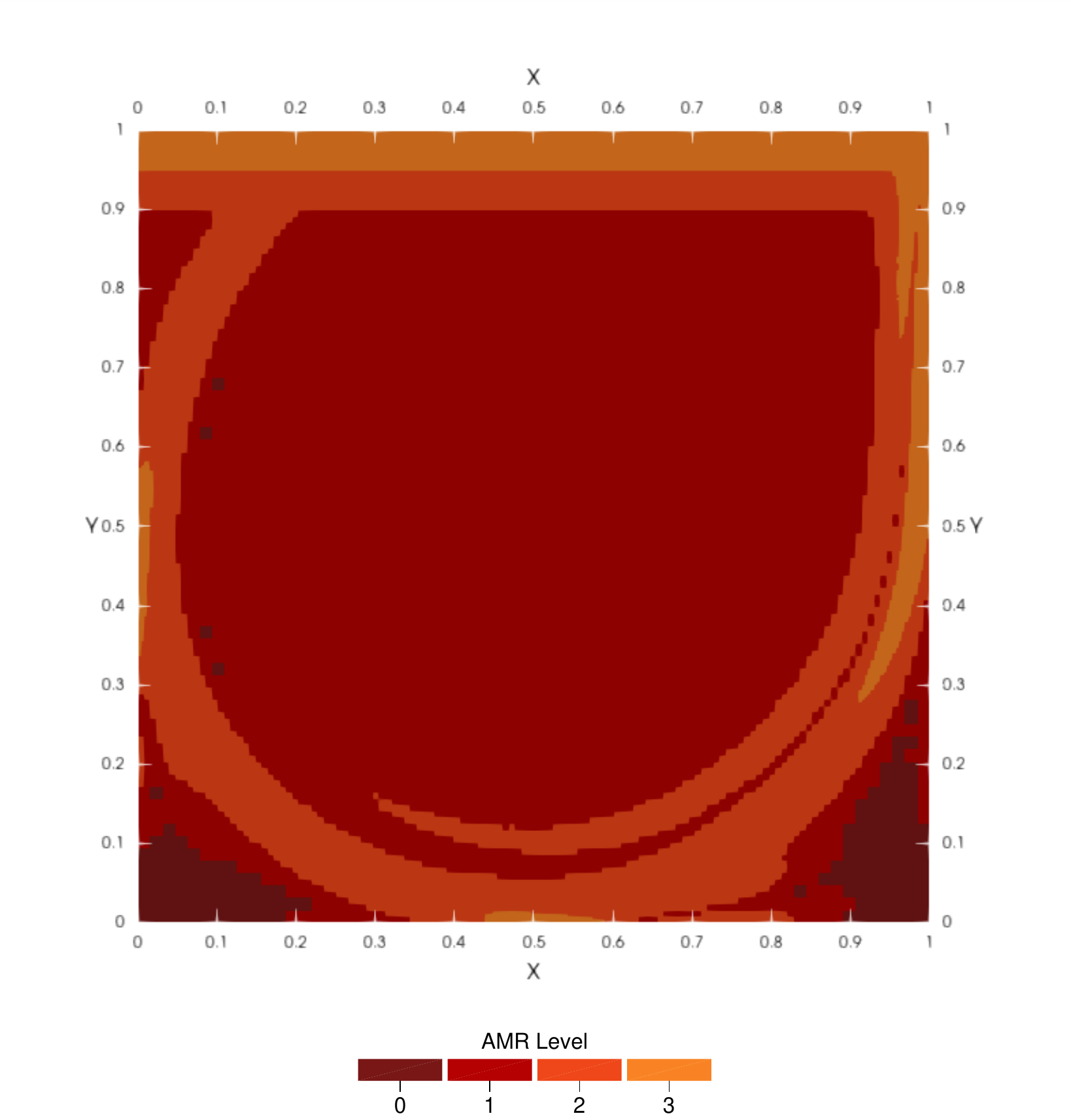}
                 \caption{$\text{Re}=5000$.}
                 \label{fig:y equals x}
            \end{subfigure}
            \begin{subfigure}[b]{0.24\textwidth}
                 \centering
                 \includegraphics[width=\textwidth]{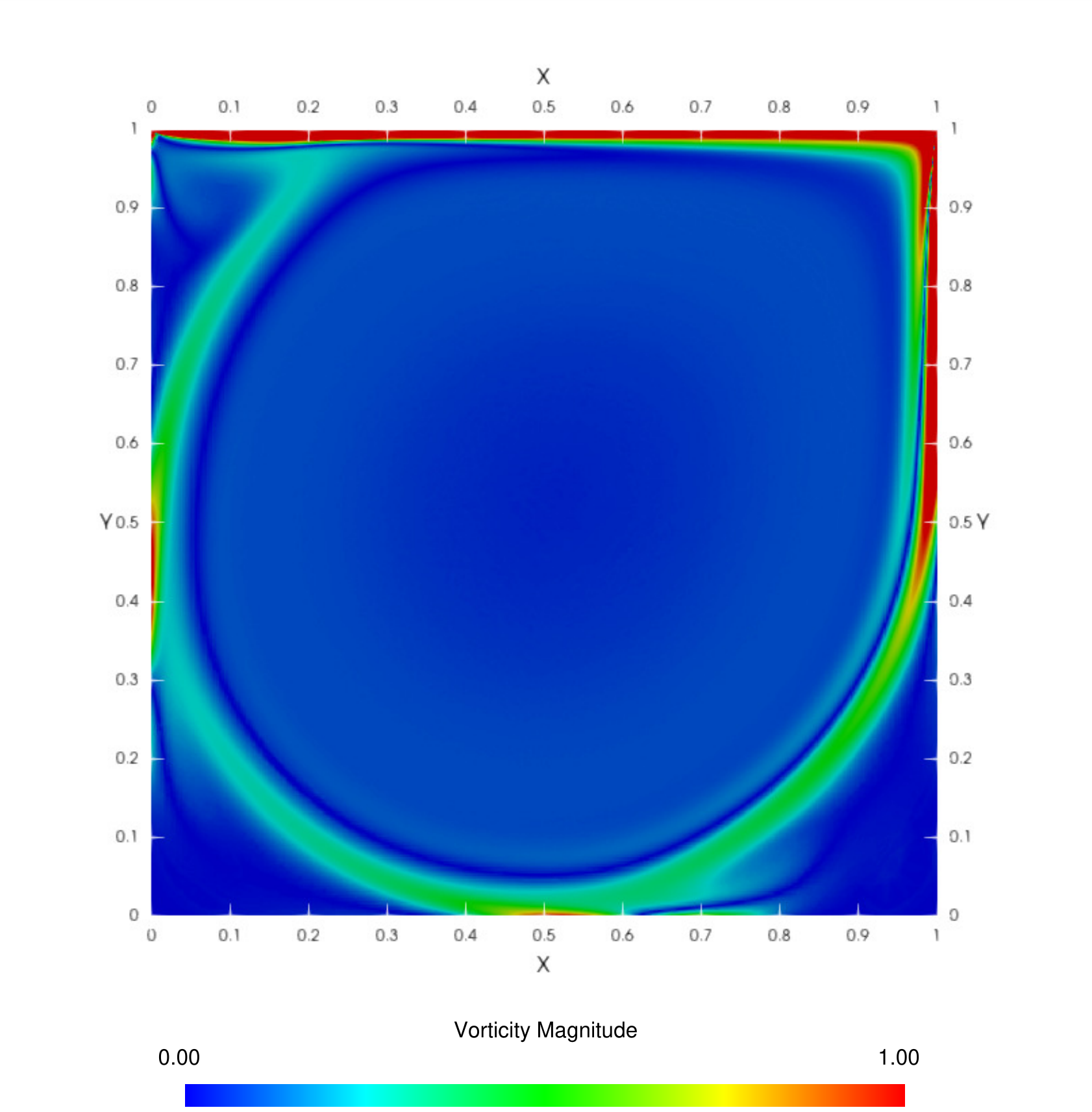}
                 \includegraphics[width=\textwidth]{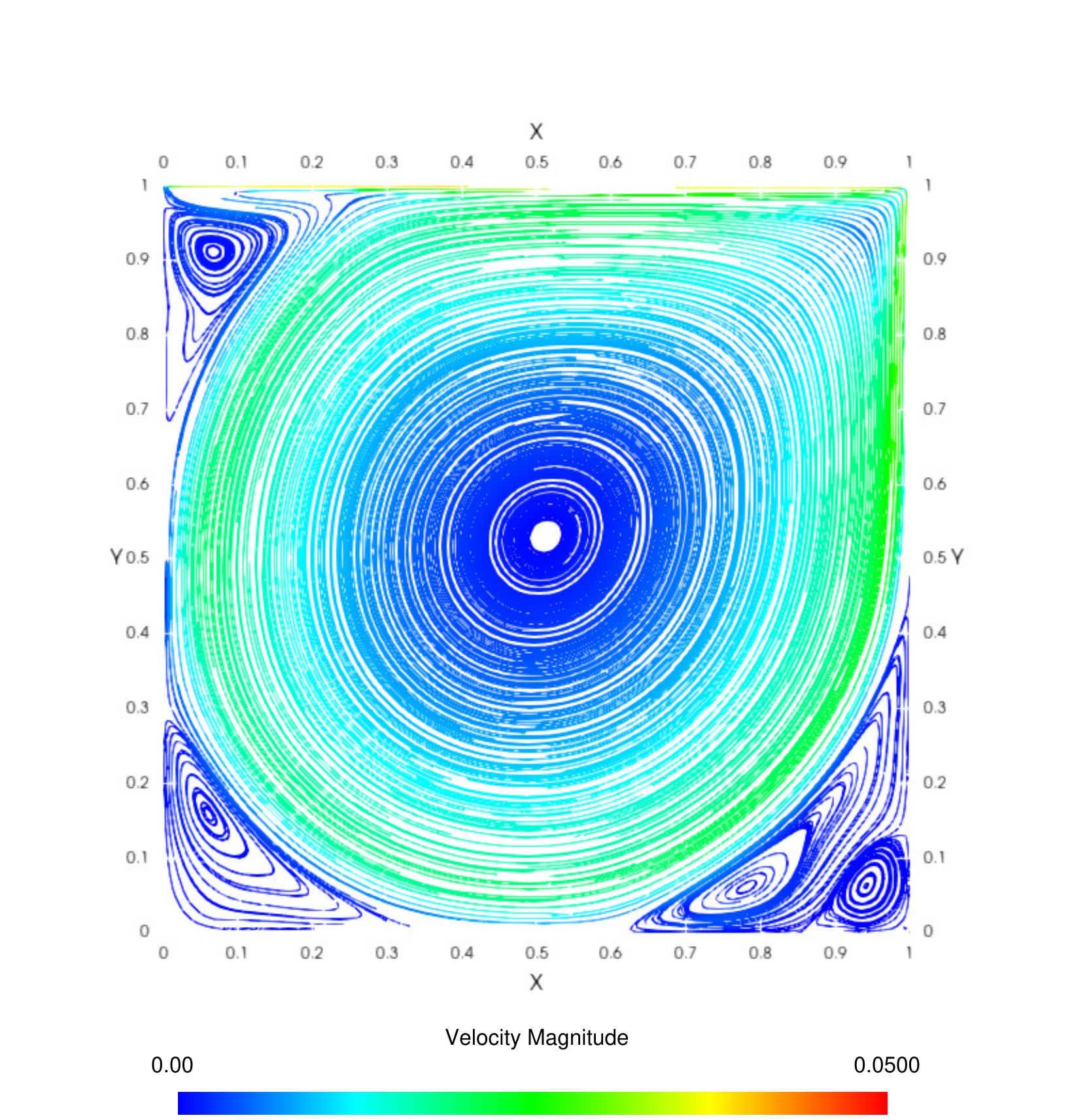}
                 \includegraphics[width=\textwidth]{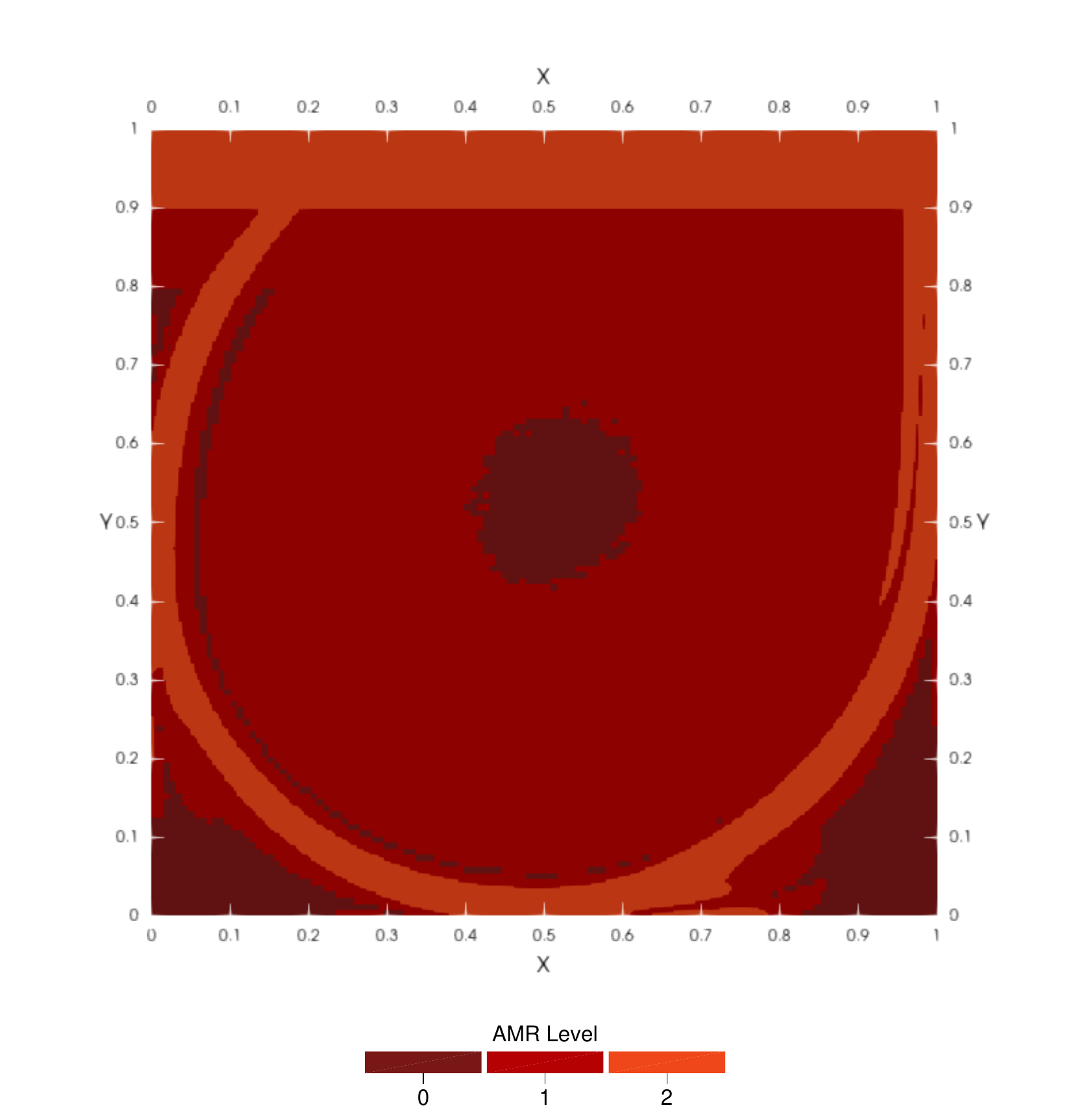}
                 \caption{$\text{Re}=10000$.}
                 \label{fig:y equals x}
            \end{subfigure}
               \caption{Plots of vorticity magnitude (top), streamlines (middle) and AMR grid levels (bottom) for simulations V$_1$ and V-S$_{k}|_{1\leq k\leq 3}$.}
               \label{fig:tests_ldc_valid_slices}
        \end{sidewaysfigure}

        Streamline plots for the 2D simulations colored by velocity magnitude are displayed in Figure \ref{fig:tests_ldc_valid_slices} and compared with surface plots of the cavity colored by grid level to indicate the regions of high vorticity magnitude. The shapes and locations of secondary corner vortices for different Re agree with the results of Ghia et al. \cite{Ghia1982}. As Re increases, the region of high vorticity magnitude reduces in accordance with the outline of the primary vortex. Low activity is detected in the cavity interior and over the corner vortices, which are of low magnitude and hence do not receive attention during refinement but nonetheless are captured adequately by the streamline plots even when the starting mesh is of low resolution.

        \subsubsection{Performance versus $N_p, N_Q$ and GPU}

\begin{figure}[h!]
            \centering
            \begin{subfigure}[b]{0.33\textwidth}
                \centering
                \includegraphics[width=\textwidth]{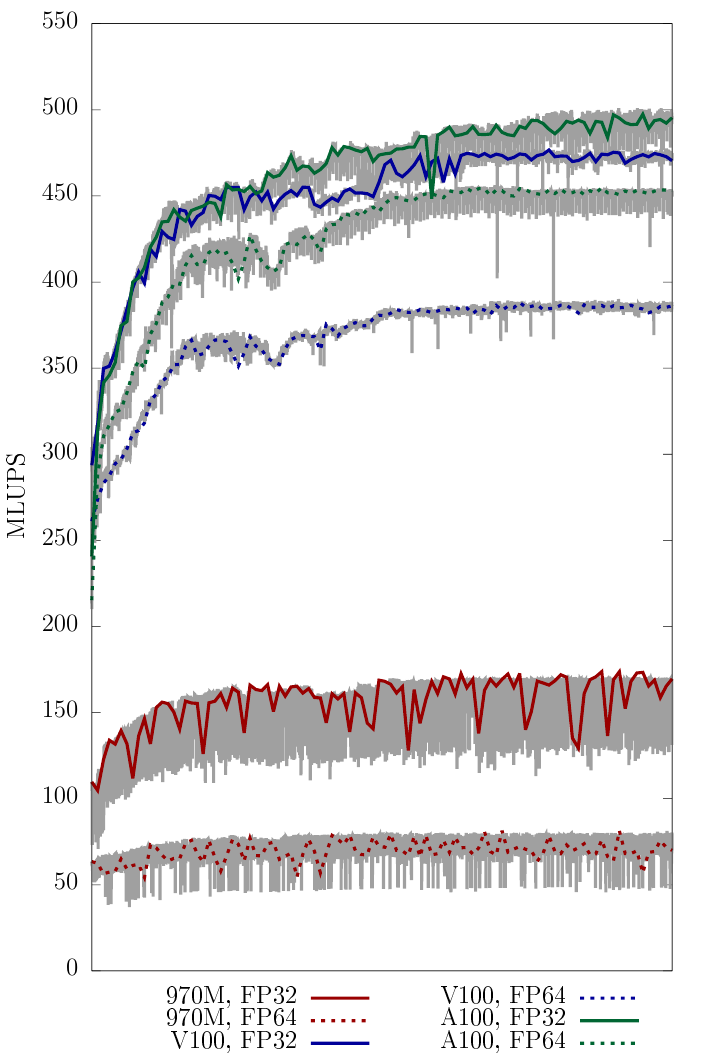}
                \caption{D2Q9}
                \label{fig:res4_3a}
            \end{subfigure}
            \begin{subfigure}[b]{0.33\textwidth}
                \centering
                \includegraphics[width=\textwidth]{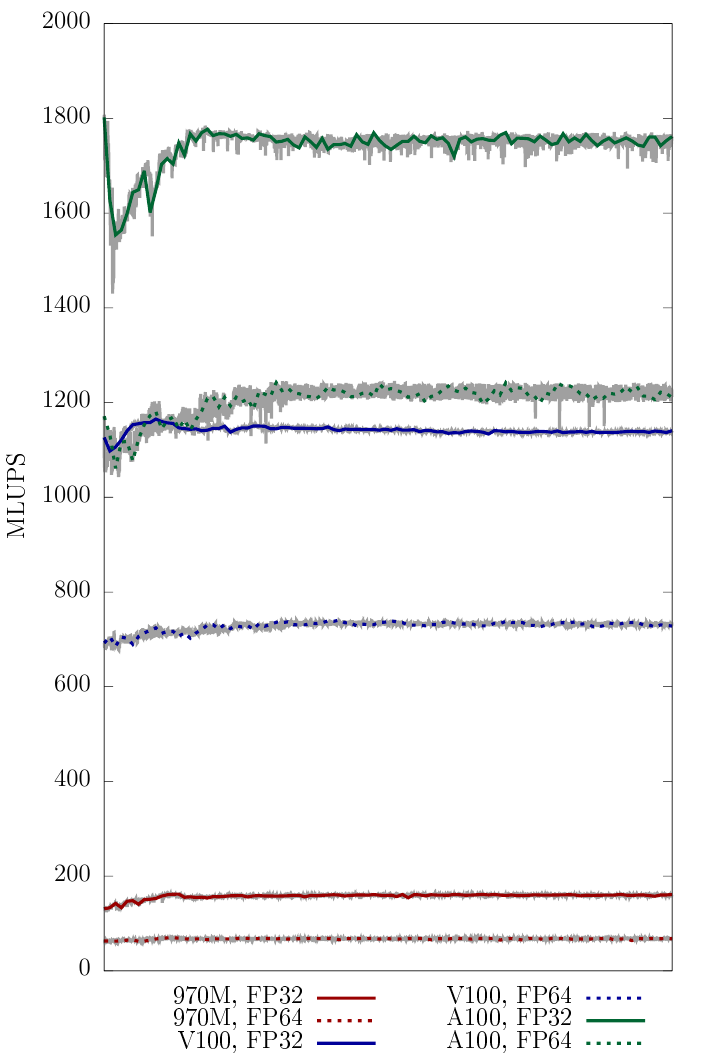}
                \caption{D3Q19}
                \label{fig:res4_3b}
            \end{subfigure}
            \begin{subfigure}[b]{0.33\textwidth}
                \centering
                \includegraphics[width=\textwidth]{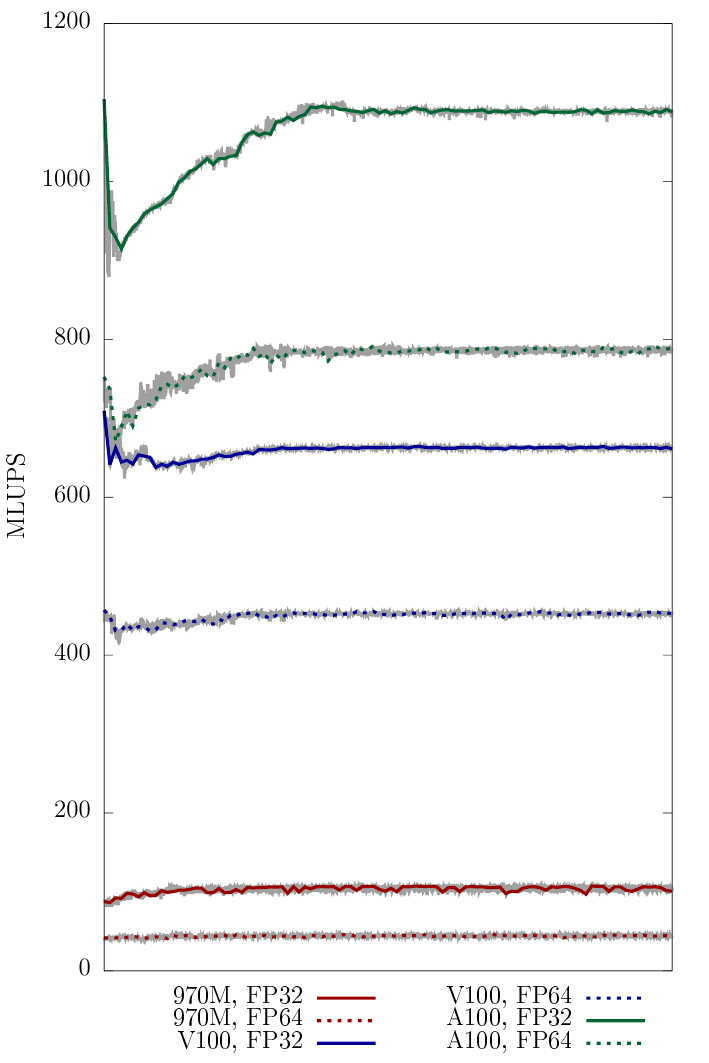}
                \caption{D3Q27}
                \label{fig:res4_3c}
            \end{subfigure}
            \caption{Plots of MLUPS over time for the three velocity sets on the 970M, V100, and A100 GPUs with both choices of floating-point precision.}
            \label{fig:tests_MLUPS}
        \end{figure}

        The $\text{MLUPS}_{\text{inst.},k}$ metric is evaluated per iteration at the coarsest level for simulations G/V/A$_k|_{1 \leq k \leq 6}$. Curves obtained with varied combinations of $N_p$ and GPU are displayed in Figure \ref{fig:tests_MLUPS} and compared for the three velocity sets. The total time spent executing interpolation, collision, streaming, and averaging subroutines is tabulated for these simulations in Table \ref{tab:tests_sim_times_ldc} to reveal the percentage of time spent in grid-communication and total simulation time.
        
        Results obtained with the A100 originally feature infrequent and random reductions in performance. These are filtered out before plotting and tabulation by detecting indices $k$ where $\text{MLUPS}_{\text{inst.},k}$ varies by more than 10\% with respect to $\text{MLUPS}_{\text{inst.},k-1}$ and replacing the corresponding value with an average computed from \{$\text{MLUPS}_{\text{inst.},k-i}\}_{5\leq i \leq 10}$. This smoothing operation is performed twice, which removes the dips in the curve without impacting earlier trends.

        Curves for all simulations increase from an initial value as blocks are added due to an increase in the size of the primary vortex, and eventually oscillate around a steady-state value after the numerical solution has converged in time. At this steady state, the performance increases achieved by selecting single-precision accuracy are similar for $N_Q = 19,27$, with rates of approximately 2.4, 1.5, and 1.4 times on the 970M, V100, and A100, respectively. For $N_Q=9$, the relatively smaller grids supply less work to the GPUs to balance out memory loads/stores, resulting in slightly lower respective rates of increase of 2.3, 1.2, and 1.1. The best performance is achieved by simulations on the A100, with $\overline{\text{MLUPS}}$ values 495, 1753 and 1089 corresponding to labels A$_1$, A$_3$ and A$_5$. For $N_Q=19,27$, the A100 simulations in double precision perform better than those in single-precision on the V100, with respective comparative $\overline{\text{MLUPS}}$ values 1219 vs. 1138 and 786 vs. 452. Single-precision simulations with $N_Q=9$ are similar between the V100 and A100, and unlike the 3D cases, the former perform better than the double-precision counterpart on the A100. $\overline{\text{MLUPS}}$ values for the 970M simulations with $N_Q=9,19$ are similar in both single and double precision, with respective comparative values 161 vs. 160 and 71 vs. 68. This decreases by an approximate factor of 1.5 for $N_Q=27$ to 105 vs. 44. The global memory load/store cost for accessing the numerous DDFs likely hinders these simulations from achieving higher performance. However, this is less pronounced on newer hardware where the speedups provided by a switch from single- to double-precision are around 1.5 compared to a theoretical value of 2.

        The percentage of time spent in grid communication ranges from 17\% (V$_4$) to 45\% (A$_1$) and is smallest for simulations performed on the V100 in 3D with double-precision. Single-precision simulations spend more (or nearly equal) time in communication than their double-precision counterparts on the V100 and A100, regardless of the velocity set. Still, they take longer when the 970M is used for the 3D simulations due to a smaller $L_{\text{max.}}$. Since calculations on more recent hardware are faster, the cost of retrieving and replacing DDFs from global memory during communication catches up, and this is more pronounced on the A100 where the fraction of time in communication for simulations in 3D is about 10\% higher than the V100 counterparts. 2D simulations are consistently around 40\% regardless of the choice of floating-point precision and GPU due to a lower amount of supplied work as with MLUPS vs. iteration.

\begin{figure}[h!]
        \centering
        \begin{subfigure}[b]{0.33\textwidth}
             \centering
             \includegraphics[width=\textwidth]{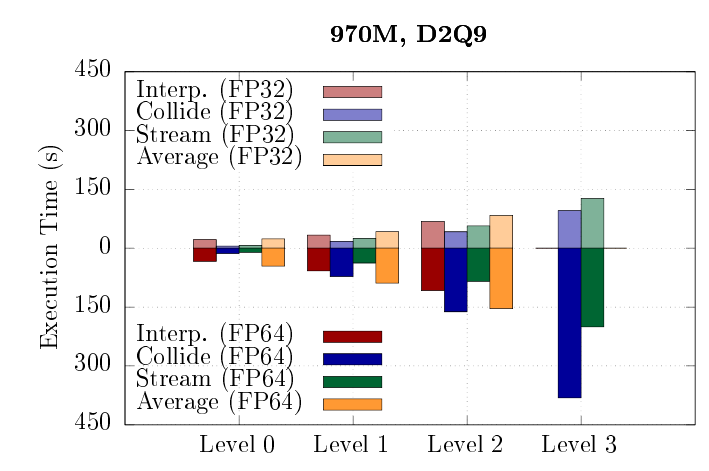}
             \includegraphics[width=\textwidth]{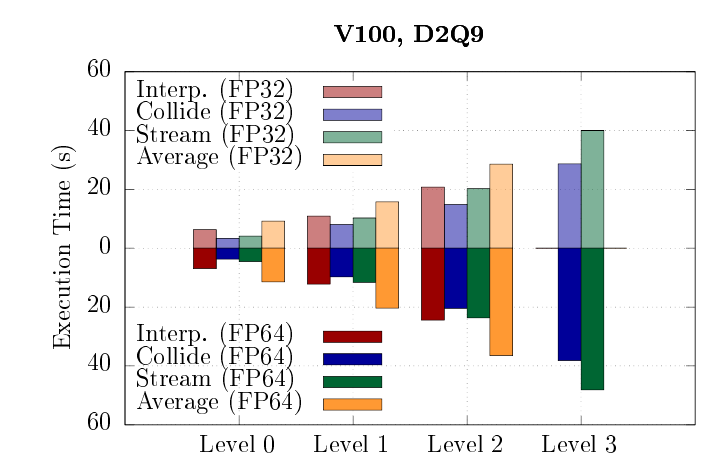}
             \includegraphics[width=\textwidth]{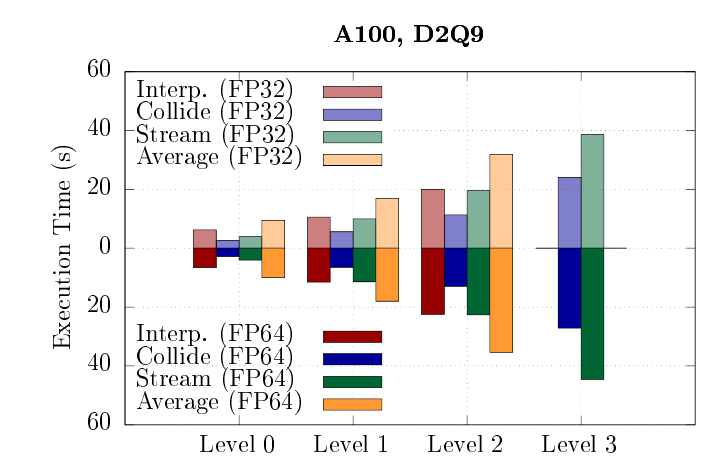}
             \caption{2D, $N_Q=9$}
         \end{subfigure}
         \begin{subfigure}[b]{0.33\textwidth}
             \centering
             \includegraphics[width=\textwidth]{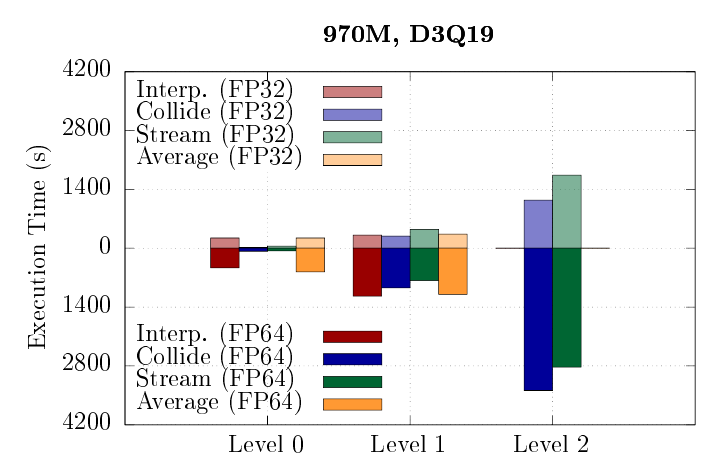}
             \includegraphics[width=\textwidth]{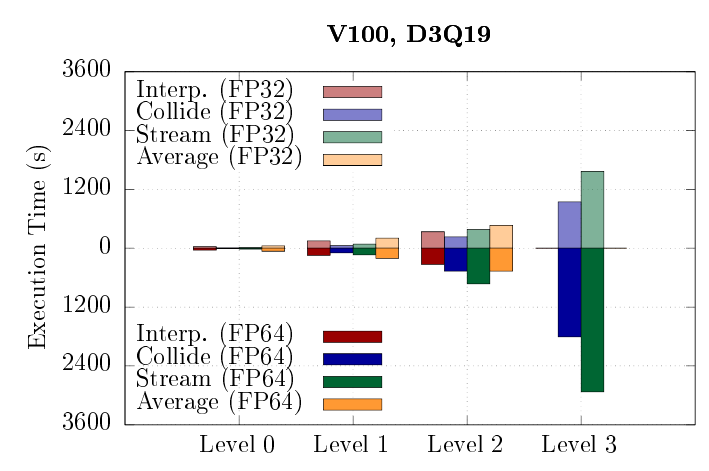}
             \includegraphics[width=\textwidth]{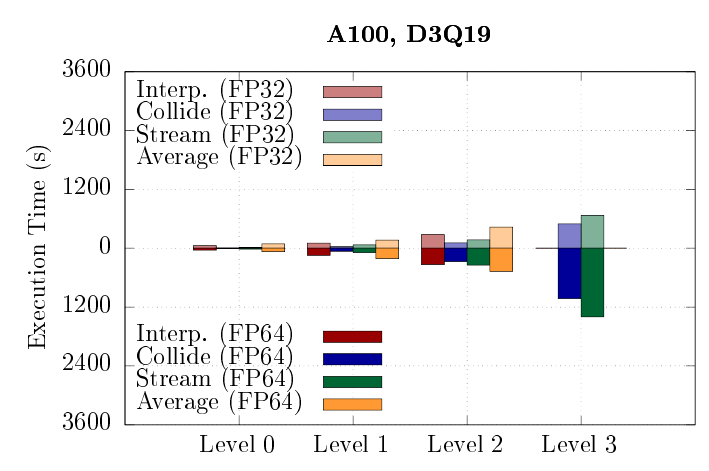}
             \caption{3D, $N_Q=19$}
         \end{subfigure}
         \begin{subfigure}[b]{0.33\textwidth}
             \centering
             \includegraphics[width=\textwidth]{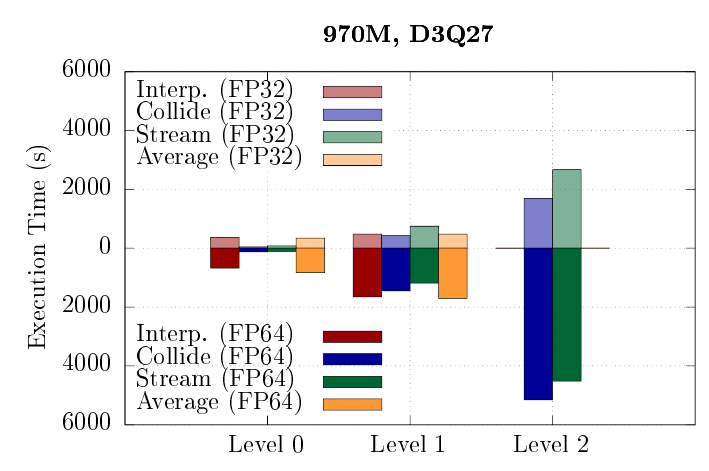}
             \includegraphics[width=\textwidth]{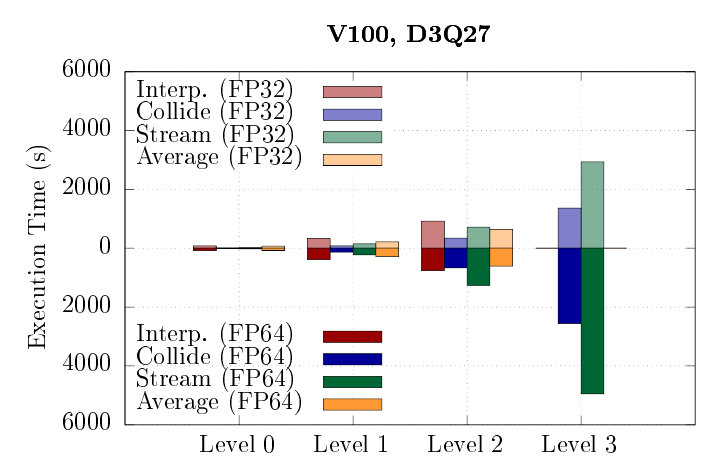}
             \includegraphics[width=\textwidth]{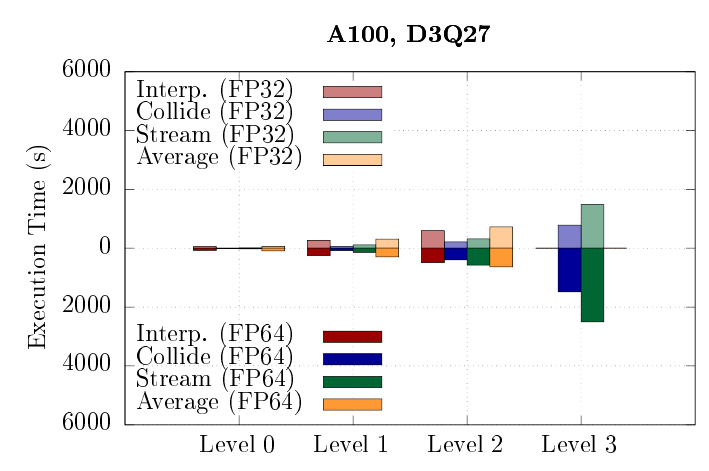}
             \caption{3D, $N_Q=27$}
         \end{subfigure}
            \caption{Comparison of execution time distributions by level and sub-process for simulations on the 970M (top), V100 (middle) and A100 (bottom). Velocity sets considered are $N_Q = 9, 19$ and 27 from left to right.}
            \label{fig:tests_ICSA}
        \end{figure}

        The distributions of time spent executing the individual advancement routines are displayed in Figure \ref{fig:tests_ICSA} for pairs of GPU and velocity set across grid level for simulations G/V/A$_k|_{1\leq k \leq 6}$. Single- and double-precision counterparts are shown side-by-side to visualize the cost of imposing higher floating-point accuracy. Most time is spent on the finest level in the 3D simulations owing to the larger fine-grid sizes and repeated advancements needed to advance an equivalent step on the coarsest level. Streaming is more expensive than collision except when double-precision is selected on the 970M. Similar execution times are observed for all simulations' communication and advancement subroutines on the second-finest level. Although interpolation and averaging are not as computation-heavy as collision, the numerous memory accesses to child cells render these operations nearly as expensive as collision and streaming. The 2D simulations feature execution times of the same order of magnitude across all grid levels due to the smaller grid sizes. They are nearly equal between single- and double-precision counterparts on the V100 and A100.

        \begin{sidewaystable}
            \centering
            \begin{tabular}{c|cccc|cccc|cc}
                \hline
                \hline
                \multirow{2}{*}{Label} & \multicolumn{10}{c}{Total Execution Times (s)} \\ \cline{2-11}
                 & Interp. & Collide & Stream & Average & Comm. & (\%) & Solv. & (\%) & Total & $\overline{\text{MLUPS}}$ \\
                \hline
                G$_{1}$ & 124.3 & 161.98 & 217.19 & 150.39 & 274.68 & (0.42) & 379.17 & (0.58) & 653.86 & 161 \\
                G$_{2}$ & 199.2 & 629.5 & 332.76 & 288.67 & 487.87 & (0.34) & 962.26 & (0.66) & 1450.13 & 71 \\
                G$_{3}$ & 557.37 & 1456.47 & 2241.14 & 584.01 & 1141.38 & (0.24) & 3697.61 & (0.76) & 4838.99 & 160 \\
                G$_{4}$ & 1604.61 & 4396 & 3659.53 & 1663.45 & 3268.06 & (0.29) & 8055.53 & (0.71) & 11323.59 & 68 \\
                G$_{5}$ & 855.9 & 2178.95 & 3508.29 & 823.43 & 1679.32 & (0.23) & 5687.24 & (0.77) & 7366.57 & 105 \\
                G$_{6}$ & 2329.01 & 6722.48 & 5814.9 & 2527.66 & 4856.67 & (0.28) & 12537.38 & (0.72) & 17394.05 & 44 \\
                V$_{1}$ & 38.09 & 55.01 & 74.72 & 53.68 & 91.77 & (0.41) & 129.73 & (0.59) & 221.51 & 474 \\
                V$_{2}$ & 43.61 & 72.05 & 87.94 & 68.32 & 111.93 & (0.41) & 159.98 & (0.59) & 271.91 & 386 \\
                V$_{3}$ & 518.35 & 1239.76 & 2054.71 & 726.08 & 1244.43 & (0.27) & 3294.47 & (0.73) & 4538.9 & 1138 \\
                V$_{4}$ & 509.51 & 2375.53 & 3806.9 & 733.79 & 1243.3 & (0.17) & 6182.42 & (0.83) & 7425.73 & 732 \\
                V$_{5}$ & 1338.21 & 1803.9 & 3824.53 & 934.92 & 2273.13 & (0.29) & 5628.43 & (0.71) & 7901.55 & 663 \\
                V$_{6}$ & 1230.89 & 3370.42 & 6461.17 & 961.06 & 2191.95 & (0.18) & 9831.59 & (0.82) & 12023.54 & 452 \\
                A$_{1}$ & 36.84 & 43.79 & 72.11 & 58.36 & 95.2 & (0.45) & 115.91 & (0.55) & 211.11 & 495 \\
                A$_{2}$ & 40.62 & 49.43 & 82.68 & 63.35 & 103.97 & (0.44) & 132.11 & (0.56) & 236.08 & 452 \\
                A$_{3}$ & 435.39 & 643.45 & 930.94 & 691.27 & 1126.66 & (0.42) & 1574.39 & (0.58) & 2701.05 & 1753 \\
                A$_{4}$ & 520.52 & 1365.13 & 1846.36 & 754.63 & 1275.15 & (0.28) & 3211.48 & (0.72) & 4486.63 & 1219 \\
                A$_{5}$ & 929.11 & 1059.91 & 1929.37 & 1102.98 & 2032.09 & (0.4) & 2989.28 & (0.6) & 5021.37 & 1089 \\
                A$_{6}$ & 811.07 & 1963.48 & 3234.55 & 1017.97 & 1829.04 & (0.26) & 5198.03 & (0.74) & 7027.06 & 786 \\
                \hline
            \end{tabular}
            \caption{Total execution times for the sub-processes in the advancement routine for the simulations defined in Table \ref{tab:tests_sim_params_ldc}.}
            \label{tab:tests_sim_times_ldc}
        \end{sidewaystable}

        \subsubsection{Refinement Performance}

\begin{figure}[t]
            \centering
            \begin{subfigure}[b]{0.475\textwidth}
                 \centering
                 \includegraphics[width=\textwidth]{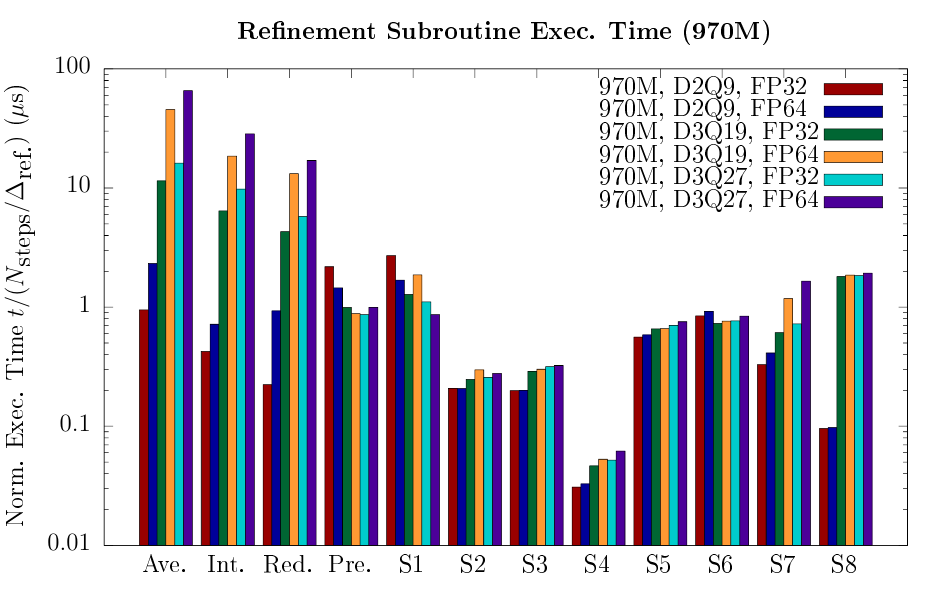}
                 \caption{Distribution of execution times by step on the 970M.}
                 \label{fig:res3_a}
             \end{subfigure}
             \begin{subfigure}[b]{0.475\textwidth}
                 \centering
                 \includegraphics[width=\textwidth]{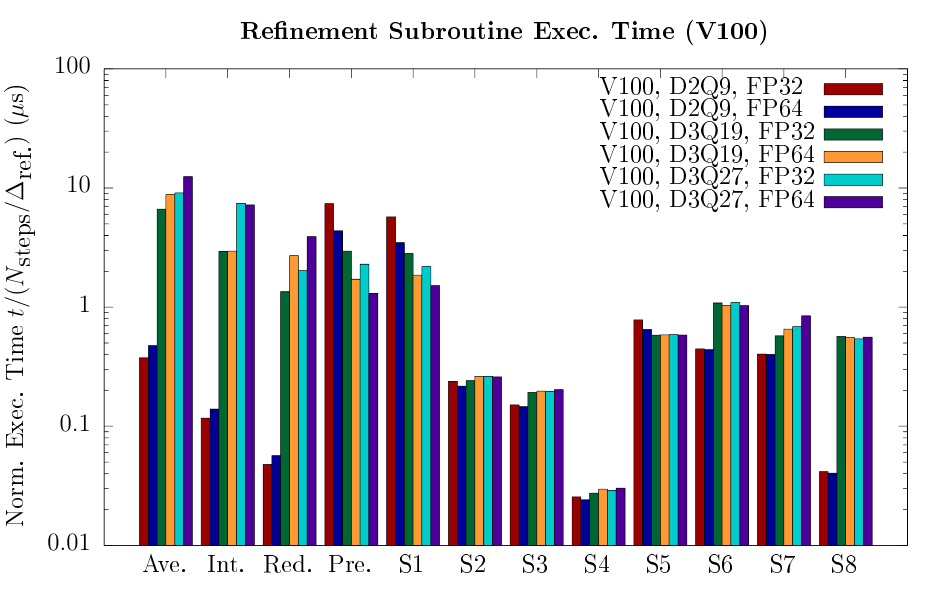}
                 \caption{Distribution of execution times by step on the V100.}
                 \label{fig:res3_b}
             \end{subfigure}
            \caption{Performance of the refinement/coarsening procedure for various simulations.}
            \label{fig:tests_ref_dist}
        \end{figure}

        The refinement and coarsening procedure is studied with respect to execution time, which is tabulated by step, and AMR efficiency where time spent in refinement is compared to total simulation time. Execution times are displayed in Table \ref{tab:tests_sim_ref_ldc} along with the fractions of time in refinement with respect to the totals computed from values shown in Table \ref{tab:tests_sim_times_ldc}. Figure \ref{fig:tests_ref_dist} displays the orders of magnitude for steps in the refinement/coarsening process after normalizing by the total number of calls for simulations on the 970M and V100 (results for the A100 were similar to the latter).

        The fraction of time adapting the mesh is always under 2\% of the total simulation time for the 3D simulations, attaining a maximum of 1.34\% in simulation A$_2$. The two largest fractions are 22.14\% and 18.74\% for simulations V$_1$ and A$_1$, respectively, where total grid sizes are among the smallest and computations among the quickest. The constant cost imposed by resets of intermediate arrays in the Preparation step results in a total execution time an order of magnitude higher than all other steps. For all other simulations, the two most expensive operations are averaging and interpolation, which are performed prior to mesh adaptation. The interpolation procedure in Step 7 does not incur a significant cost for any of the simulations even though the routine is generally as expensive as collision and streaming on the same level. This is due to the relatively high frequency in calls to refinement, limiting the number of children added at any point in time. Figure \ref{fig:tests_ref_dist} reveals that most steps incur a constant cost across floating-point precision, velocity set, and choice of GPU. The exceptions are the averaging, interpolation, and reduction steps performed prior to the mesh-adaptation call, where computations involving DDFs depend strongly on these parameters and where total times are 10-100 times higher in comparison to other steps. Step 4 pertaining to insertion/removal of refined/coarsened child blocks incurs the lowest cost, followed by Steps 2 and 3 where assignment of data to children is prepared and neighbor-child connectivity is updated. Step 8 is more costly in 3D simulations as more neighbors are checked in order to identify ghost and interface cells.

        \newpage
        \begin{sidewaystable}
            \centering
            \begin{tabular}{c|ccccccccccccc|c}
                \hline
                \hline
                \multirow{2}{*}{Label} & \multicolumn{13}{c|}{Total Execution Times (s)} & \multirow{2}{*}{\%} \\ \cline{2-14}
                 & Ave. & Int. & Red. & Pre & S1 & S2 & S3 & S4 & S5 & S6 & S7 & S8 & Total & \\
                \hline
                G$_{1}$ & 3.792 & 1.699 & 0.895 & 8.73 & 10.823 & 0.833 & 0.795 & 0.124 & 2.244 & 3.381 & 1.315 & 0.383 & 35.014 & 5.08 \\
                G$_{2}$ & 9.322 & 2.876 & 3.731 & 5.805 & 6.732 & 0.83 & 0.802 & 0.132 & 2.345 & 3.691 & 1.65 & 0.39 & 38.306 & 2.57 \\
                G$_{3}$ & 23.012 & 12.856 & 8.607 & 1.992 & 2.558 & 0.494 & 0.578 & 0.093 & 1.314 & 1.462 & 1.221 & 3.614 & 57.801 & 1.18 \\
                G$_{4}$ & 91.171 & 36.982 & 26.385 & 1.763 & 3.728 & 0.597 & 0.603 & 0.106 & 1.321 & 1.525 & 2.366 & 3.716 & 170.263 & 1.48 \\
                G$_{5}$ & 32.355 & 19.62 & 11.539 & 1.729 & 2.21 & 0.516 & 0.633 & 0.104 & 1.408 & 1.534 & 1.445 & 3.678 & 76.771 & 1.03 \\
                G$_{6}$ & 131.197 & 56.935 & 34.088 & 1.996 & 1.732 & 0.556 & 0.648 & 0.124 & 1.509 & 1.68 & 3.305 & 3.862 & 237.632 & 1.35 \\
                V$_{1}$ & 1.503 & 0.467 & 0.191 & 29.528 & 22.936 & 0.953 & 0.605 & 0.102 & 3.127 & 1.783 & 1.617 & 0.166 & 62.979 & 22.14 \\
                V$_{2}$ & 1.902 & 0.558 & 0.227 & 17.487 & 13.919 & 0.868 & 0.583 & 0.097 & 2.598 & 1.758 & 1.601 & 0.162 & 41.76 & 13.31 \\
                V$_{3}$ & 13.3 & 5.891 & 2.693 & 5.909 & 5.658 & 0.484 & 0.385 & 0.055 & 1.163 & 2.167 & 1.15 & 1.137 & 39.99 & 0.87 \\
                V$_{4}$ & 17.645 & 5.895 & 5.407 & 3.431 & 3.713 & 0.524 & 0.396 & 0.059 & 1.168 & 2.071 & 1.308 & 1.109 & 42.727 & 0.57 \\
                V$_{5}$ & 18.206 & 14.882 & 4.05 & 4.582 & 4.39 & 0.525 & 0.393 & 0.058 & 1.176 & 2.187 & 1.372 & 1.084 & 52.906 & 0.67 \\
                V$_{6}$ & 24.908 & 14.455 & 7.803 & 2.612 & 3.028 & 0.52 & 0.406 & 0.06 & 1.165 & 2.059 & 1.691 & 1.121 & 59.829 & 0.5 \\
                A$_{1}$ & 1.568 & 0.489 & 0.133 & 19.81 & 21.373 & 0.577 & 0.421 & 0.092 & 1.941 & 1.171 & 0.941 & 0.16 & 48.677 & 18.74 \\
                A$_{2}$ & 1.809 & 0.524 & 0.155 & 11.767 & 11.589 & 0.601 & 0.416 & 0.093 & 1.699 & 1.201 & 0.95 & 0.159 & 30.964 & 11.6 \\
                A$_{3}$ & 15.678 & 5.519 & 1.628 & 4.048 & 4.727 & 0.315 & 0.302 & 0.027 & 0.828 & 2.19 & 0.769 & 0.64 & 36.669 & 1.34 \\
                A$_{4}$ & 18.102 & 6.18 & 3.426 & 2.437 & 3.277 & 0.395 & 0.324 & 0.061 & 0.891 & 2.128 & 0.983 & 0.813 & 39.017 & 0.86 \\
                A$_{5}$ & 22.863 & 10.612 & 2.755 & 3.112 & 3.966 & 0.35 & 0.306 & 0.056 & 0.778 & 1.856 & 0.921 & 0.782 & 48.357 & 0.95 \\
                A$_{6}$ & 25.896 & 9.628 & 4.844 & 1.833 & 2.632 & 0.352 & 0.297 & 0.052 & 0.867 & 1.921 & 1.175 & 0.788 & 50.284 & 0.71 \\
                \hline
            \end{tabular}
            \caption{Total execution times of the various steps in the mesh refinement algorithm for the simulations defined in Table \ref{tab:tests_sim_params_ldc}. The fractions of time with respect to total solving time are provided.}
            \label{tab:tests_sim_ref_ldc}
        \end{sidewaystable}
        \newpage

\begin{figure}[h!]
            \centering
            \includegraphics[width=0.6\textwidth]{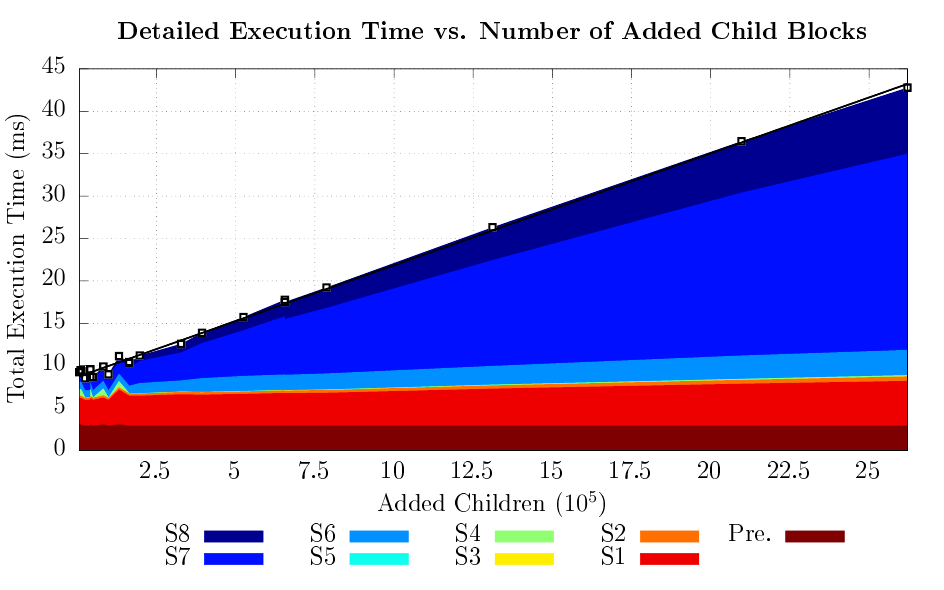}
            \caption{Scaling of the refinement procedure as the number of inserted blocks is increased.}
            \label{fig:tests_ref_scale}
        \end{figure}

        Scaling of the refinement operation is also studied by modifying the number of grid levels and the near-wall refinement distance to achieve a number of inserted child blocks ranging from $~7.7(10^2) \to ~3.3(10^5)$ (equivalent to $~61(10^3) \to \sim (26)10^5$ new cells). Execution times for the various steps (omitting the averaging, interpolation and reduction steps invoked prior to mesh adaptation) are stacked by area in Figure \ref{fig:tests_ref_scale} to reveal a linear scaling, from around 10 to 45 milliseconds at the largest insertion. Steps 7 and 8 play the most significant roles in scaling, exceeding the other steps by a full order of magnitude in the largest sample. The high cost of Step 7 is due to interpolation, while Step 8 requires subsequently more checks with neighbor block IDs as the number of blocks inserted in the vicinity of the physical boundary increases.

        \subsubsection{Efficiency versus $M_b, M_L$}

\begin{figure}
            \centering
            \begin{subfigure}[b]{0.49\textwidth}
                 \centering
                 \includegraphics[width=\textwidth]{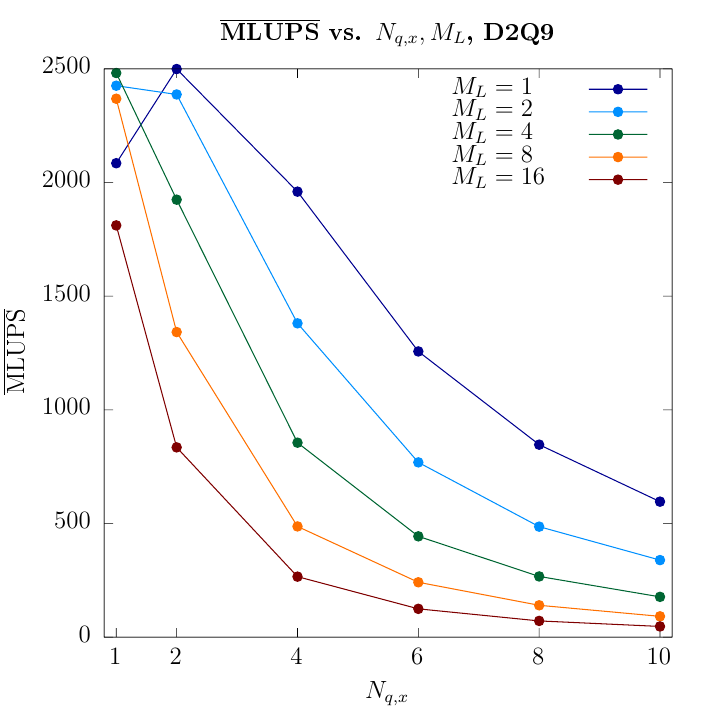}
                 \caption{D2Q9.}
                 \label{fig:tests_cellblock_size_A}
             \end{subfigure}
             \begin{subfigure}[b]{0.49\textwidth}
                 \centering
                 \includegraphics[width=\textwidth]{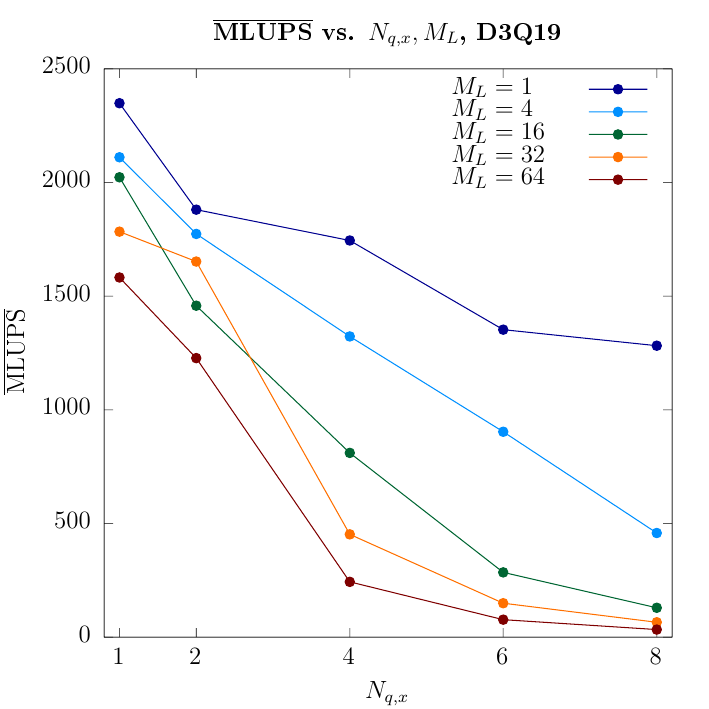}
                 \caption{D3Q19.}
                 \label{fig:tests_cellblock_size_B}
             \end{subfigure}
            \caption{Plots of $\overline{\text{MLUPS}}$ against cell-block size (via $N_{q,x}$) for different choices of $M_L$.}
            \label{fig:tests_cellblock_size}
        \end{figure}

        More work is generated per thread-block when cell-block sizes $M_b > M_t$ are used as more sub-blocks are processed, affecting performance. Additional overhead is also incurred in order to navigate the various sub-blocks. $M_L$, the number of cell-blocks processed per thread-block in the solver routine, is another factor that can add additional work. The efficiency of the solver, measured with $\overline{\text{MLUPS}}$, is studied with respect to $M_b$ (via $N_{q,x}$) and $M_L$. We set $L_{\text{max.}} = 1$ to specifically gauge the effects of work done per thread-block on performance without additional overhead that would be introduced, for example, by checking cell/block masks for grid communication. Values $N_{q,x} \in \{1,2,4,6,8,10\}$ and $M_L \in \{1,2,4,8,16\}$ are considered in 2D with $N_Q=9$, $N_x = 480$, while $N_{q,x} \in \{1,2,4,6,8\}$ and $M_L \in \{1,4,16,32,64\}$ are considered in 3D with $N_Q=19$ and $N_x=192$, respectively. Single-precision is used in both cases, and MLUPS are averaged over a period of $100$ s (in contrast with eq. \ref{eq:MLUPS_ave}). $\overline{\text{MLUPS}}$ vs. $N_{q,x}$ for the various choices of $M_L$ are displayed for the 2D and 3D simulations in Figures \ref{fig:tests_cellblock_size_A} and \ref{fig:tests_cellblock_size_B}, respectively.

        The solver is most efficient for smaller choices of $M_b$ and $M_L$. In 2D, a maximum of approximately 2500 $\overline{\text{MLUPS}}$ is obtained for $N_{q,x}=1$, $M_L \in \{2,4,8\}$ and $N_{q,x}=2, M_L \in \{1,2\}$. As $N_{q,x}$ and $M_L$ increase, $\overline{\text{MLUPS}}$ decreases sharply down to a minimum of approximately 45 at $N_{q,x}=10, M_L=16$. The only case where increasing $N_{q,x}$ leads to an increase in $\overline{\text{MLUPS}}$ is $M_L=1$. A similar trend is observed in 3D, where a maximum of approximately 2350 $\overline{\text{MLUPS}}$ is obtained with $N_{q,x}=1, M_L=1$. Unlike the 2D cases, increasing $N_{q,x}$ or $M_L$ always leads to a decrease in $\overline{\text{MLUPS}}$. The decrease is less substantial for $M_L=1$, down to 1300 $\overline{\text{MLUPS}}$ for $N_{q,x}=8$. When $M_L=4$ is chosen, the decrease is nearly linear. For all other choices of $M_L>1$, $\overline{\text{MLUPS}}$ decreases down to below 500. One outlier is observed at $N_{q,x}=2$, where the recorded $\overline{\text{MLUPS}}$ for $M_L=32$ is greater than that for $M_L=16$.

        \subsubsection{Cell vs. Block Arrangement}

        Two standalone scripts (\texttt{cell\_uniform.cu}, \texttt{block\_uniform.cu}) are provided, implementing cell- and block-based versions of the in-place LBM on a single grid level for the lid-driven cavity in 2D and 3D. Connectivity between cells/blocks is adjusted by three parameters: \texttt{N\_CONN\_TYPE} for calculation of neighbor indices vs. retrieval from global memory, \texttt{N\_SHUFFLE} to shuffle the indices and \texttt{N\_OCTREE} to specify quad-/octree index arrangement rather than structured. With the current mesh adaptation algorithm, block indices tend towards a shuffled ordering as blocks are inserted and removed in different parts of the domain, with indices of larger value eventually replacing earlier ones. The objective is to show that the block-based advancement scheme described in Section \ref{sec:solver_adv} is nearly independent of the index arrangement (i.e., near-constant execution time for a fixed grid size regardless of how far away neighboring cell data is in linear memory), remaining as efficient as in the case of a grid with trivial index arrangement.

\begin{figure}[h!]
            \centering
            \begin{subfigure}[b]{0.49\textwidth}
                \centering
                \begin{subfigure}[b]{0.3\textwidth}
                     \centering
                     \includegraphics[width=\textwidth]{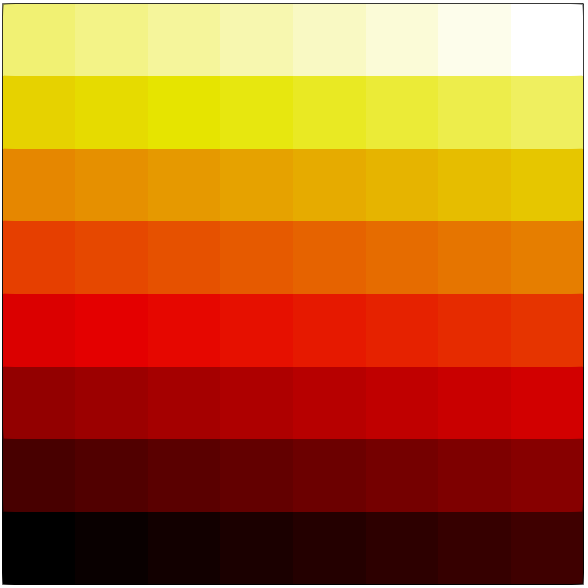}
                     \caption{Structured grid.}
                     \label{fig:tests_cellblock_param_A}
                 \end{subfigure}
                 \hspace{1.5cm}
                 \begin{subfigure}[b]{0.3\textwidth}
                     \centering
                     \includegraphics[width=\textwidth]{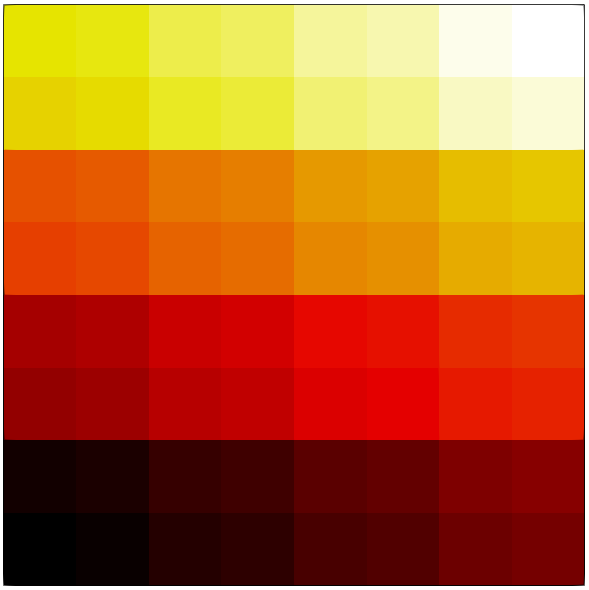}
                     \caption{Octree grid.}
                     \label{fig:tests_cellblock_param_B}
                 \end{subfigure}
                 \hspace{1.5cm}
                 \begin{subfigure}[b]{0.3\textwidth}
                     \centering
                     \includegraphics[width=\textwidth]{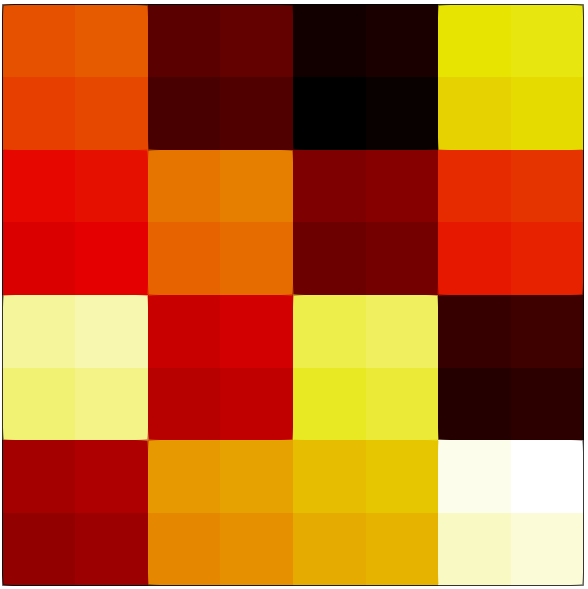}
                     \caption{Octree grid, shuffled.}
                     \label{fig:tests_cellblock_param_C}
                 \end{subfigure}
                \caption{Cell vs. block arrangements.}
                \label{fig:tests_cellblock_param_D}
            \end{subfigure}
            \begin{subfigure}[b]{0.49\textwidth}
                \centering
                \includegraphics[width=0.9\textwidth]{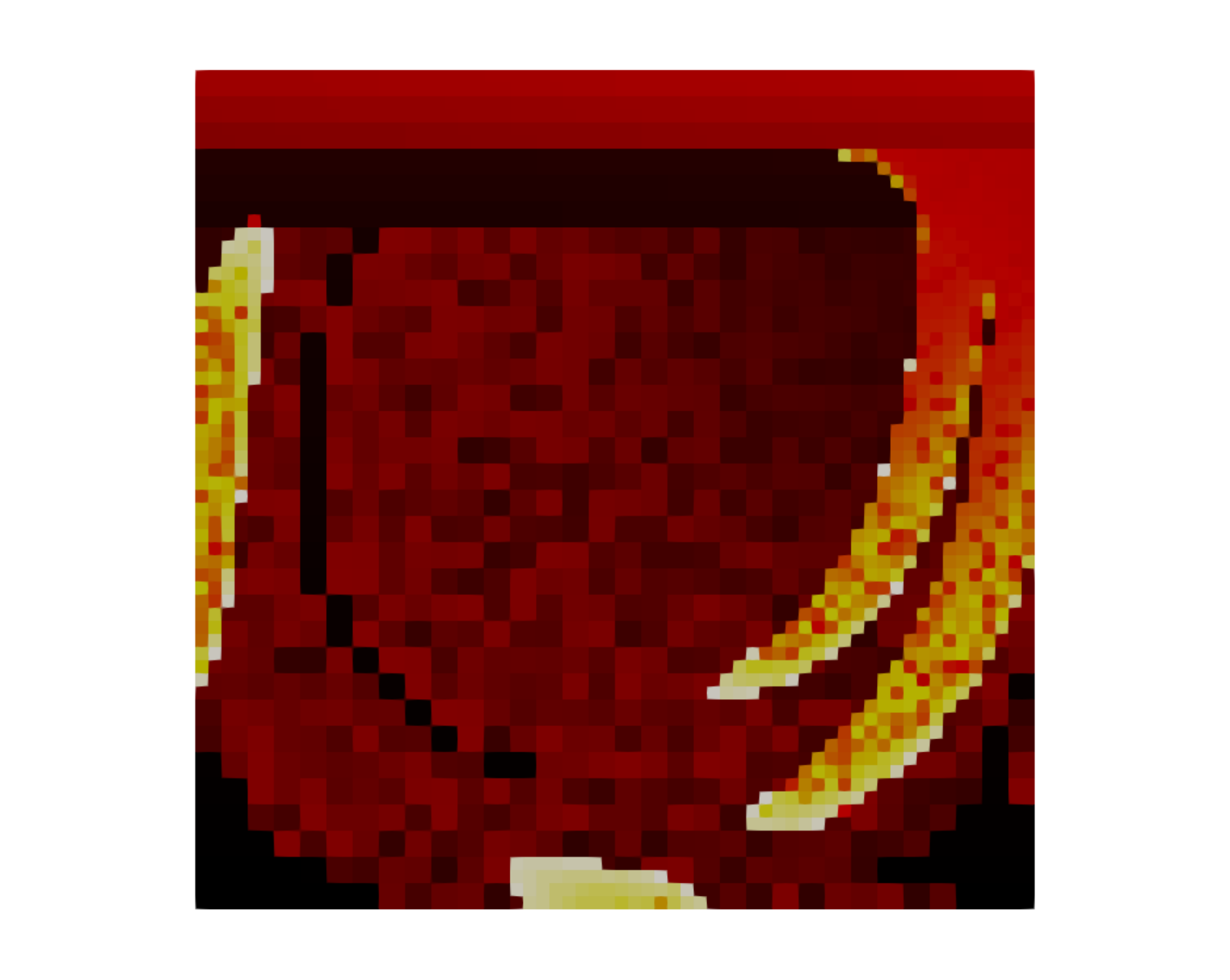}
                 \caption{Arrangement after lid-driven cavity simulation.}
                 \label{fig:tests_cellblock_param_E}
            \end{subfigure}
            \caption{Visualizations of the block ID distributions considered in the cell- and block-based solver comparison (left), and a sample of the block ID distribution in a lid-driven cavity test after completion (right).}
            \label{fig:tests_cellblock_param_o}
        \end{figure}

        Simulations with choices ($N_Q=9$, $N_x=2048^2$), ($N_Q=19$, $N_x=192^3$), and ($N_Q=27$, $N_x=192^3$) are performed for 10,000 iterations and averaged total execution times are recorded with 95\% confidence intervals. Four cases are considered for the cell- and block-based arrangements: 1) structured grid with calculated indices, 2) structured grid with retrieved indices, 3) octree grid, and 4) octree grid with 
        shuffled indices. These cases are presented in Figure \ref{fig:tests_cellblock_param_A}-\ref{fig:tests_cellblock_param_C} for sample 32x32 grids, and the corresponding results are displayed in Table \ref{tab:tests_cellblock_res}. A visualization of block index scattering observed with one of the lid-driven cavity tests is shown in Figure \ref{fig:tests_cellblock_param_E}. The data used to obtain the averages, and their decomposition among collision and streaming steps are included in the attached code repository.

        When the cell/block arrangement is well-ordered, the cell-based solver performs better than the block-based scheme. Collision and streaming were slightly slower in these cases, likely due to the additional processing of block IDs in the former and the various DDF arrangements that must be made in shared memory before the global write in the latter. The block-based scheme is consistently around 60-70\% slower in 2D and 10-50\% slower in 3D. When the indices are shuffled, the cell-based solver experiences a substantial increase in average execution time, ranging from 2.6-3.3x / 3.5-4.6x increases in 3D ($N_Q=19 / 27$, respectively) to a 4.5-5x increase in 2D. In contrast, the performance of the block-based solver remains nearly identical, increasing by about 12\% in 2D with respect to the ideal structured grid case with calculated indices and about 5\% in 3D. The block-based solver performs better than its cell-based counterpart for all $N_Q$, obtaining a 33-45\% relative reduction in average execution time.
        \begin{table}[h!]
            \centering
            \begin{tabular}{cc|rrr}
            \hline
            \hline
            \multirow{3}{*}{\begin{tabular}[c]{@{}c@{}}Grid\\ Arrangement\end{tabular}} & \multirow{3}{*}{\begin{tabular}[c]{@{}c@{}}Element\\ Arrangement\end{tabular}} & \multicolumn{3}{c}{Execution Times, $\mu$s} \\ \cline{3-5} 
             &  & \begin{tabular}[c]{@{}c@{}}$N_Q=9$\\ $N_x=2048$\end{tabular} & \begin{tabular}[c]{@{}c@{}}$N_Q=19$\\ $N_x=192$\end{tabular} & \begin{tabular}[c]{@{}c@{}}$N_Q=27$\\ $N_x=192$\end{tabular} \\ \hline
             \hline
            \multirow{3}{*}{\begin{tabular}[c]{@{}c@{}}Structured,\\ Calculated\end{tabular}} & Cell & $6,137 \pm 12$ & $23,236 \pm 7$ & $33,895 \pm 5$ \\
             & Block & $10,590 \pm 7$ & $32,319 \pm 9$ & $49,800 \pm 5$ \\
             & Ratio (B/C) & $1.73$ & $1.39$ & $1.47$ \\ \hline
            \multirow{3}{*}{\begin{tabular}[c]{@{}c@{}}Structured,\\ Retrieved\end{tabular}} & Cell & $6,868 \pm 5$ & $26,030 \pm 9$ & $37,803 \pm 5$ \\
             & Block & $10,852 \pm 7$ & $32,544 \pm 9$ & $50,526 \pm 5$ \\
             & Ratio (B/C) & $1.58$ & $1.25$ & $1.34$ \\ \hline
            \multirow{3}{*}{\begin{tabular}[c]{@{}c@{}}Octree,\\ Unshuffled\end{tabular}} & Cell & $6,881 \pm 5$ & $29,369 \pm 9$ & $44,306 \pm 6$ \\
             & Block & $10,884 \pm 7$ & $32,865 \pm 6$ & $50,960 \pm 5$ \\
             & Ratio (B/C) & $1.58$ & $1.12$ & $1.15$ \\ \hline
            \multirow{3}{*}{\begin{tabular}[c]{@{}c@{}}Octree,\\ Shuffled\end{tabular}} & Cell & $30,401 \pm 5$ & $76,350 \pm 12$ & $156,728 \pm 17$ \\
             & Block & $11,911 \pm 9$ & $33,914 \pm 5$ & $52,181 \pm 5$ \\
             & Ratio (B/C) & $0.39$ & $0.44$ & $0.33$ \\ \hline
             \hline
            \end{tabular}
            \caption{Results for the standalone cell- and block-based solvers.}
            \label{tab:tests_cellblock_res}
        \end{table}

    \begin{table}[h!]
        \centering
        \begin{tabular}{c|cc|cccc|c}
            \hline
            \hline
            \multicolumn{8}{c}{\textbf{Flow-Past-Square-Cylinder Simulations, Re=100}} \\
            \hline
            \hline
            \multirow{2}{*}{Label} & \multirow{2}{*}{$N_{\text{coarse}}^{N_d}$} & \multirow{2}{*}{$L_{\text{max.}}$} & \multicolumn{4}{c|}{Simulation Times (s)} & \multirow{2}{*}{GPU} \\ \cline{4-7}
            & & & Sol. & Ref. & Total & Eff. (\%) & \\
            \hline
            F-A$_{1}$ & \multirow{4}{*}{256} & 2 & 77 & 7 & 84 & 8.3 & \multirow{12}{*}{970M} \\
            F-A$_{2}$ &  & 3 & 259 & 11 & 270 & 4.1 &  \\
            F-A$_{3}$ &  & 4 & 960 & 16 & 976 & 1.6 &  \\
            F-A$_{4}$ &  & 5 & 4087 & 31 & 4118 & 0.8 &  \\
            F-B$_{1}$ & \multirow{4}{*}{512} & 1 & 157 & - & 157 & - &  \\
            F-B$_{2}$ &  & 2 & 413 & 21 & 434 & 4.8 &  \\
            F-B$_{3}$ &  & 3 & 1370 & 35 & 1405 & 2.5 &  \\
            F-B$_{4}$ &  & 4 & 5895 & 78 & 5973 & 1.3 &  \\
            F-C$_{1}$ & \multirow{3}{*}{1024} & 1 & 1157 & - & 1157 & - &  \\
            F-C$_{2}$ &  & 2 & 2133 & 74 & 2207 & 3.4 &  \\
            F-C$_{3}$ &  & 3 & 6592 & 147 & 6739 & 2.2 &  \\
            F-D & 2048 &  & 9327 & - & 9327 & - &  \\
            F-E & 4096 &  &  & - &  & - & A100 \\
            \hline
        \end{tabular}
        \caption{Summary of flow-past-square-cylinder simulation descriptions.}
        \label{tab:tests_sim_params_fpsc}
    \end{table}

    \subsection{Flow Past a Square Cylinder} \label{sec:tests_fpsc}

    A flow past a square cylinder (FPSC) is considered to compare total computation time against refinement level and the corresponding speedup relative to an equivalent uniform grid with resolution according to the finest level. A rectangular domain of length $32D$ and height $32D$ is used with a fixed square cylinder placed a length $10 D$ downstream. Although it is customary to employ spatial and temporal steps of unity with the LBM, we have retained the physical scales such that for a choice of $D=1/32 \ \text{m}$ (with channel length normalized to 1.0 m), the spatial and temporal steps are given by $1.0/N_x$. For boundary conditions, a specification similar to that of Fakhari and Lee \cite{Fakhari2014} is employed. The domain features an inlet on the left-hand-side supplying a uniform flow defined by a constant streamwise velocity $U_0$, and outlet condition downstream defined by a constant pressure and far-field conditions where the inlet velocity is also enforced (with zero normal velocity). The inlet/far-field and outflow conditions are enforced via bounce-back and anti-bounce-back, respectively. A Mach number of $\sim 0.1$ is enforced by setting the characteristic inlet velocity $U_0$ at $0.05 \ \text{m/s}$. The Reynolds number is enforced by setting viscosity according to $U_0, L$, and $D$. A flow regime with Re $=100$ has been considered in the current work, requiring $\nu=1.5625(10^{-5}) \ \text{m$^2$/s}$. Since a cell-centered grid is used in the current work, domain walls lie directly on cell edges. Hence, no adjustments need to be made to the expression for the dimensionless parameters as with Fakhari and Lee \cite{Fakhari2014}. This also means that the bounce-back naturally attains second-order accuracy with single-relaxation-time collision. The anti-bounce-back can attain the same order of accuracy, though this requires some form of extrapolation to estimate outflow wall velocity. This would incur additional costs during the collision step. Instead, the velocity at the cell center is used as the estimate, and a permanent layer of blocks is refined multiple times according to $L_{\text{max}}$ so that the first-order error is made negligible.

\begin{figure}[h!]
        \centering
        \begin{subfigure}[b]{0.475\textwidth}
             \centering
             \includegraphics[width=\textwidth]{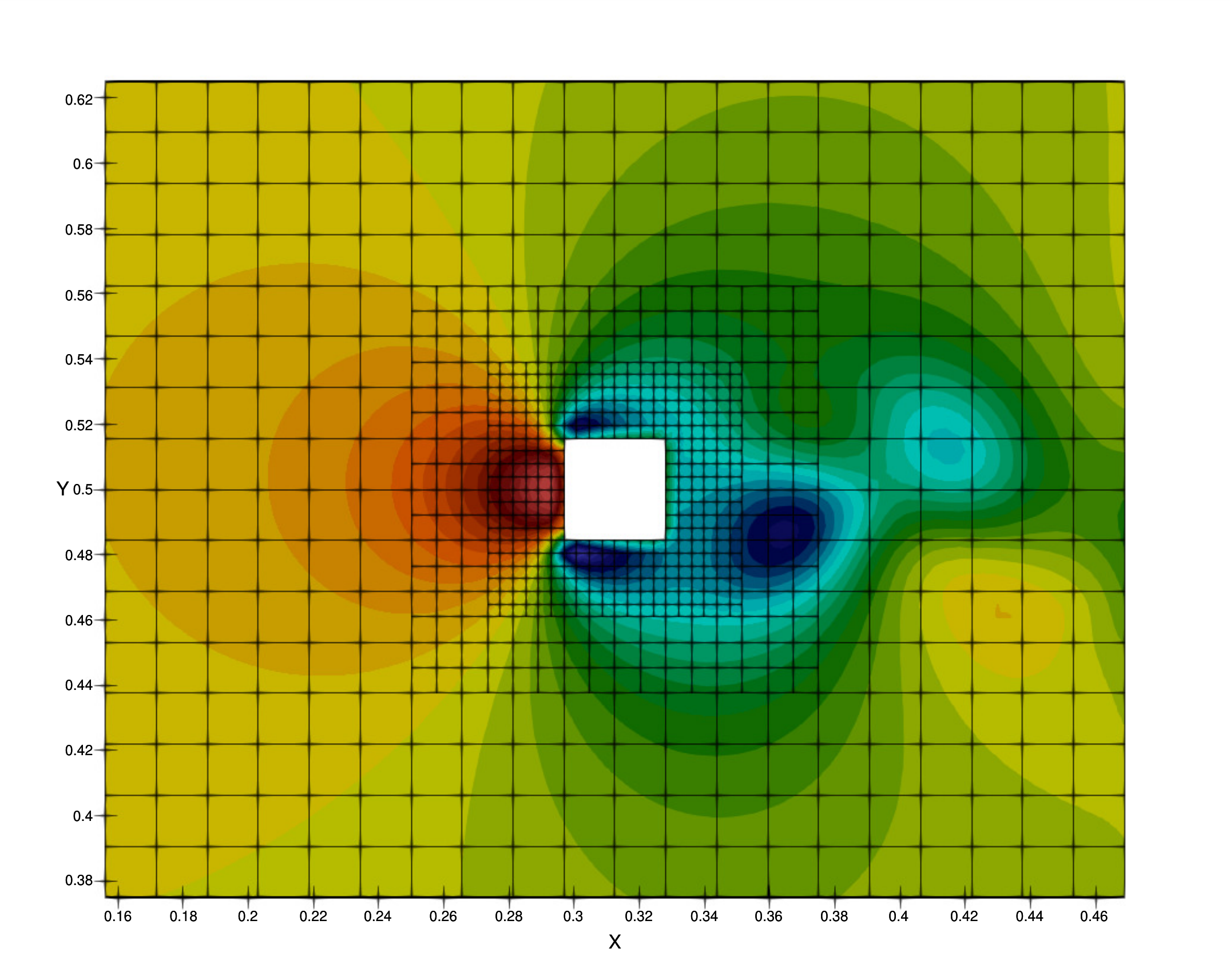}
             \caption{Simulation F-A$_2$ with linear interpolation.}
             \label{fig:tests_sim_contour_fpscA}
        \end{subfigure}
        \begin{subfigure}[b]{0.475\textwidth}
             \centering
             \includegraphics[width=\textwidth]{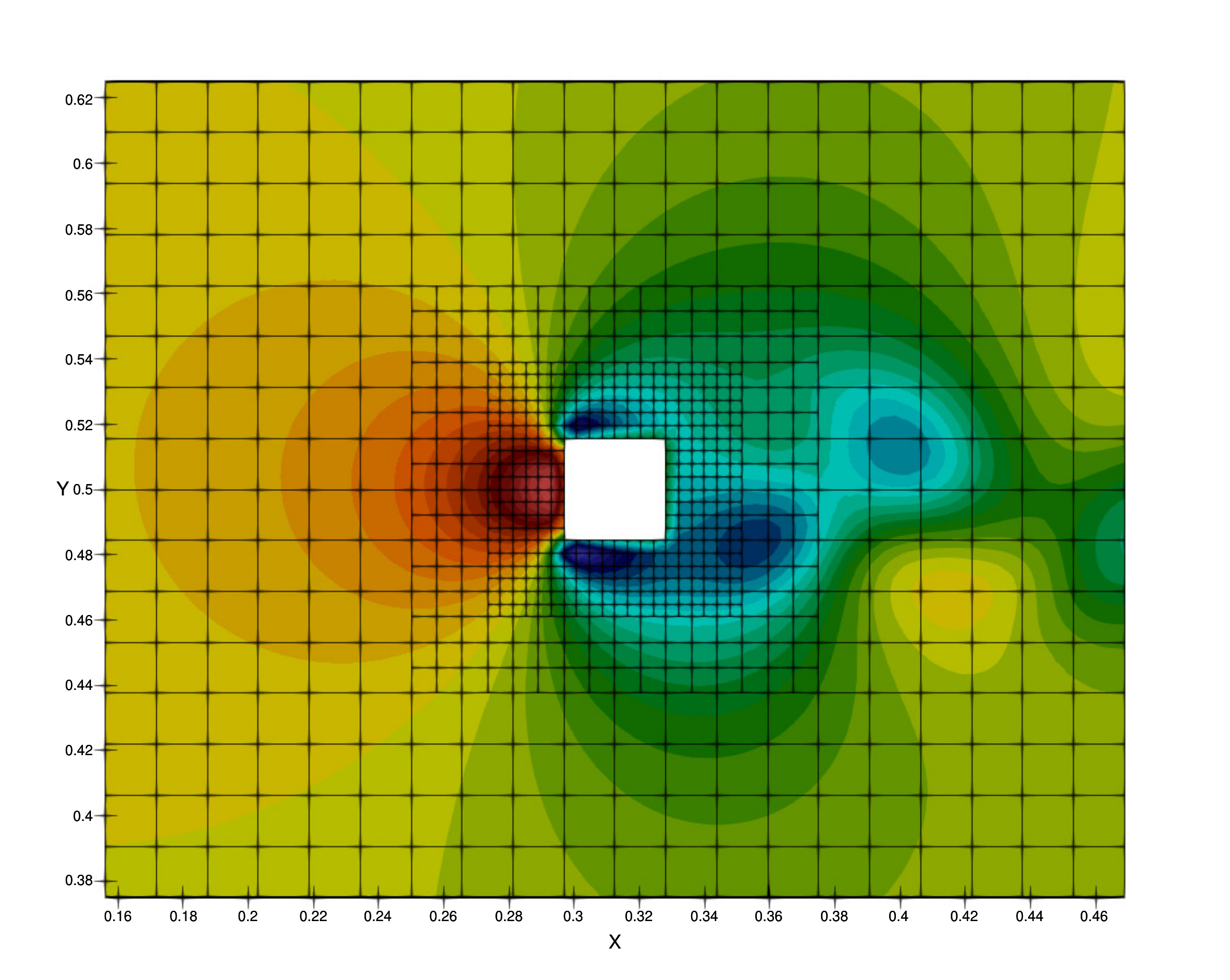}
             \caption{Simulation F-A$_2$ with cubic interpolation.}
             \label{fig:tests_sim_contour_fpscB}
        \end{subfigure}
           \caption{Pressure field obtained from simulation F-A$_2$ with linear (left) and cubic (right) interpolation.}
           \label{fig:tests_sim_contour_fpsc}
    \end{figure}

    Three sets of cases labeled F-X$_{L_{\text{max.}}}$ are considered in which the root grid size is set to 256, 512, and 1024 (for $X \in \{A, B,C\}$, respectively) and $L_{\text{max.}}$ is varied between 1 and 5, reaching a final effective resolution of 4096. Additional simulations are performed with uniform grids of $2048^2$ and $4096^2$. All simulations are run to a final time $t = 150 \ \text{s}$. Simulation parameters and corresponding advancement and refinement execution times with a choice of linear interpolation are displayed in Table \ref{tab:tests_sim_params_fpsc}. Cases F-A$_k|_{1\leq k \leq 4}$ are repeated using both linear and cubic interpolation to determine the potential effects on the pressure field and the time-averaged dimensionless quantities. Figures \ref{fig:tests_sim_contour_fpscA} and \ref{fig:tests_sim_contour_fpscB} display the contours of the pressure fields for case F-A$_2$ with linear and cubic interpolation, respectively. Fakhari and Lee \cite{Fakhari2014} reported that discontinuities in the pressure field resulting from linear interpolation were suppressed when the grid was adaptively refined, so we've chosen to employ a static grid that is refined twice only at the beginning to verify that the continuity of the current pressure field is inherent to the numerical scheme. Cell-blocks on level $L$ are refined if they are a distance of $0.15/2^L$ or less away from the cylinder or the outlet boundary.

    All simulations are performed on the 970M except for the $4096^2$ case, which is performed on the A100 instead due to memory limitations. Double-precision is used for all simulations to ensure high accuracy when comparing with results of the literature. Refinement parameters $N_{\text{start}}, N_{\text{inc.}}$ are set to -3 and -1, respectively, and the mesh is adapted every 32 iterations on the coarsest level as with the LDC test cases.
    
    The momentum exchange algorithm (MEA) is used to estimate the total force applied to the cylinder over time, proscribed for bounce-back conditions by \cite{Ladd1994, Fakhari2014, LBM_Book}:
    \begin{align}
        \textbf{F} &= 2\Delta x_{L_s}^{N_d} \sum_{\textbf{x}_b} \sum_p f_p^* \textbf{c}_p
    \end{align}

    \noindent where $L_s$ is the specified grid level and the post-collision $f_p^*$ considered in the calculation are such that $\textbf{c}_p$ forms a link between a boundary node and a solid node. The time-averaged drag coefficient $\overline{C_D}$, lift coefficient $C_L$, and Strouhal number \text{St},
    \begin{align}
        C_D &= \frac{2F_x}{\rho_0 U_0^2 D}, \quad C_L = \frac{2F_y}{\rho_0 U_0^2 D}, \quad \text{St} = \frac{f_s D}{U_0},
    \end{align}
    \begin{figure}[h!]
        \centering
        \begin{subfigure}[b]{\textwidth}
             \centering
             \includegraphics[width=0.475\textwidth]{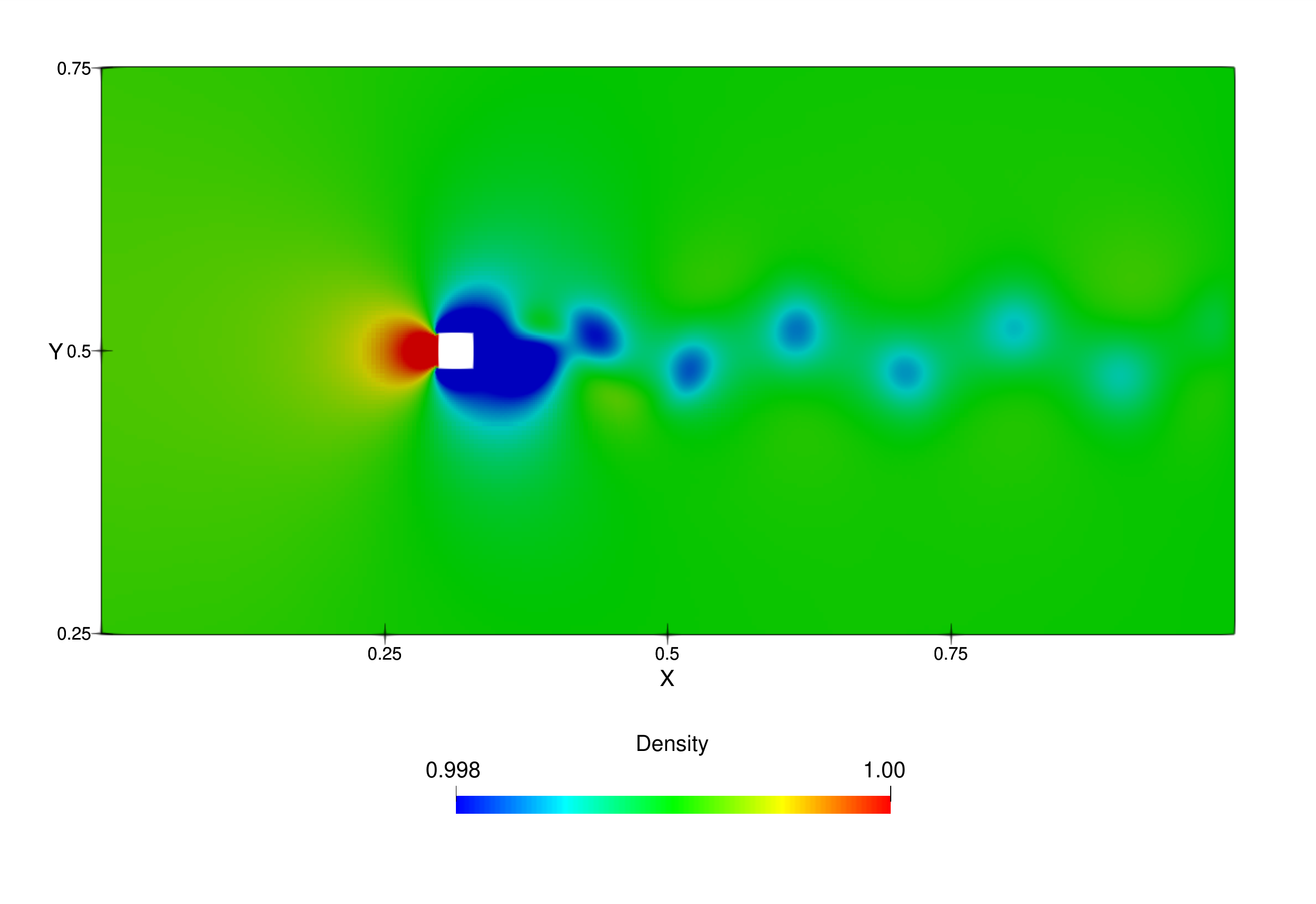}
             \includegraphics[width=0.475\textwidth]{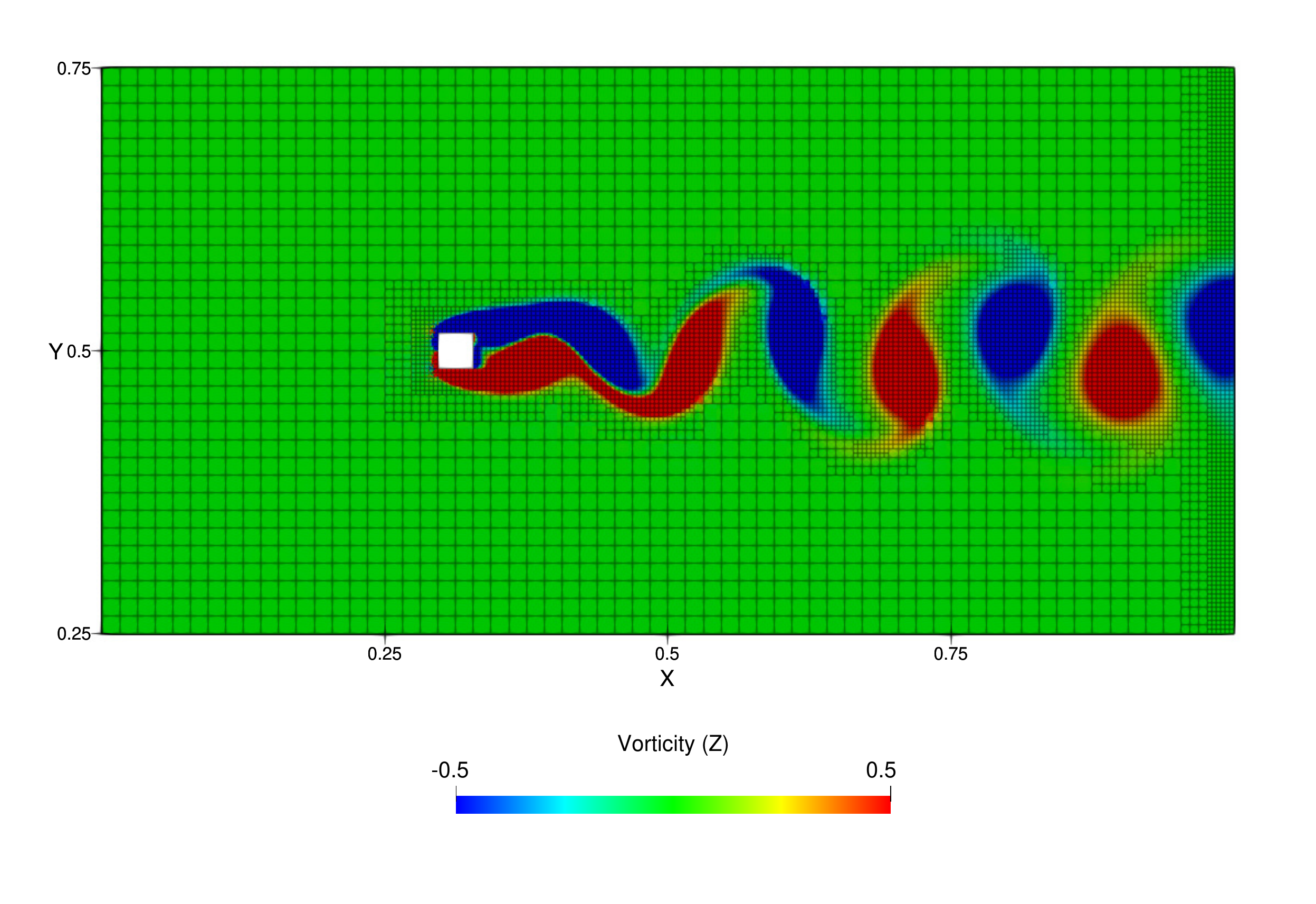}
             \caption{Simulation F-A$_2$ with $L_{\text{max.}}=3$.}
             \label{fig:y equals x}
        \end{subfigure}
        \begin{subfigure}[b]{\textwidth}
             \centering
             \includegraphics[width=0.475\textwidth]{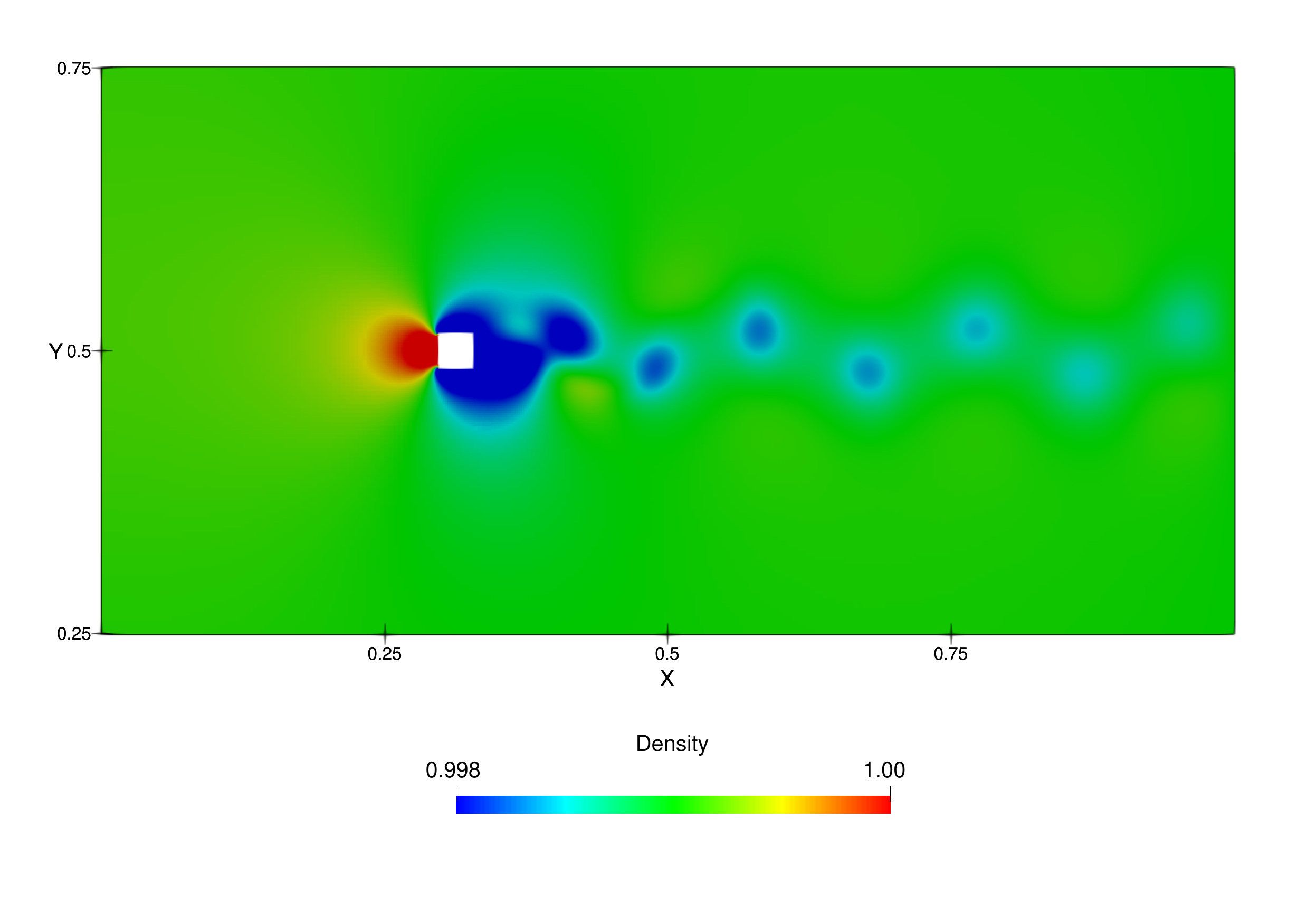}
             \includegraphics[width=0.475\textwidth]{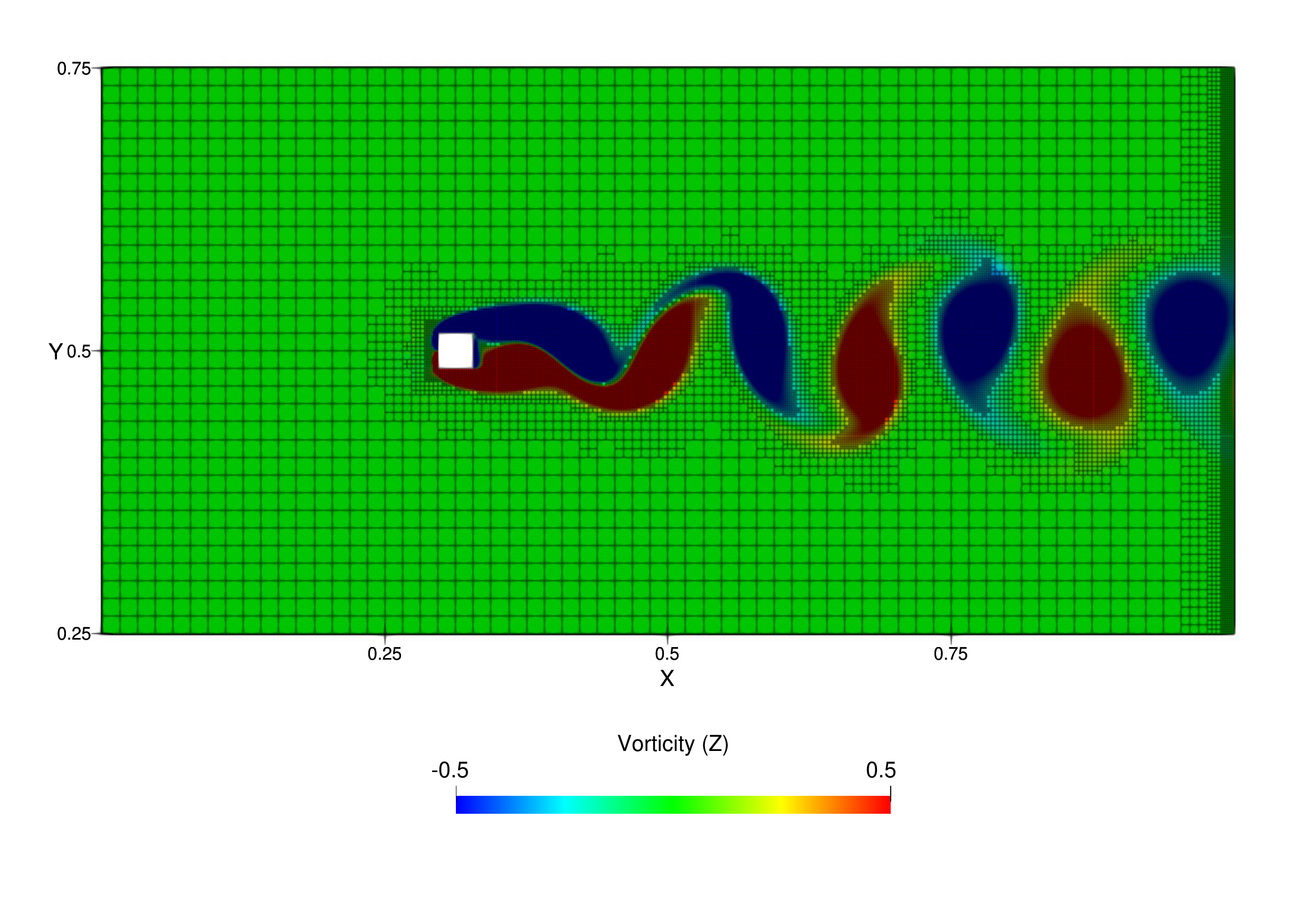}
             \caption{Simulation F-A$_3$ with $L_{\text{max.}}=4$.}
             \label{fig:y equals x}
        \end{subfigure}
        \begin{subfigure}[b]{\textwidth}
             \centering
             \includegraphics[width=0.475\textwidth]{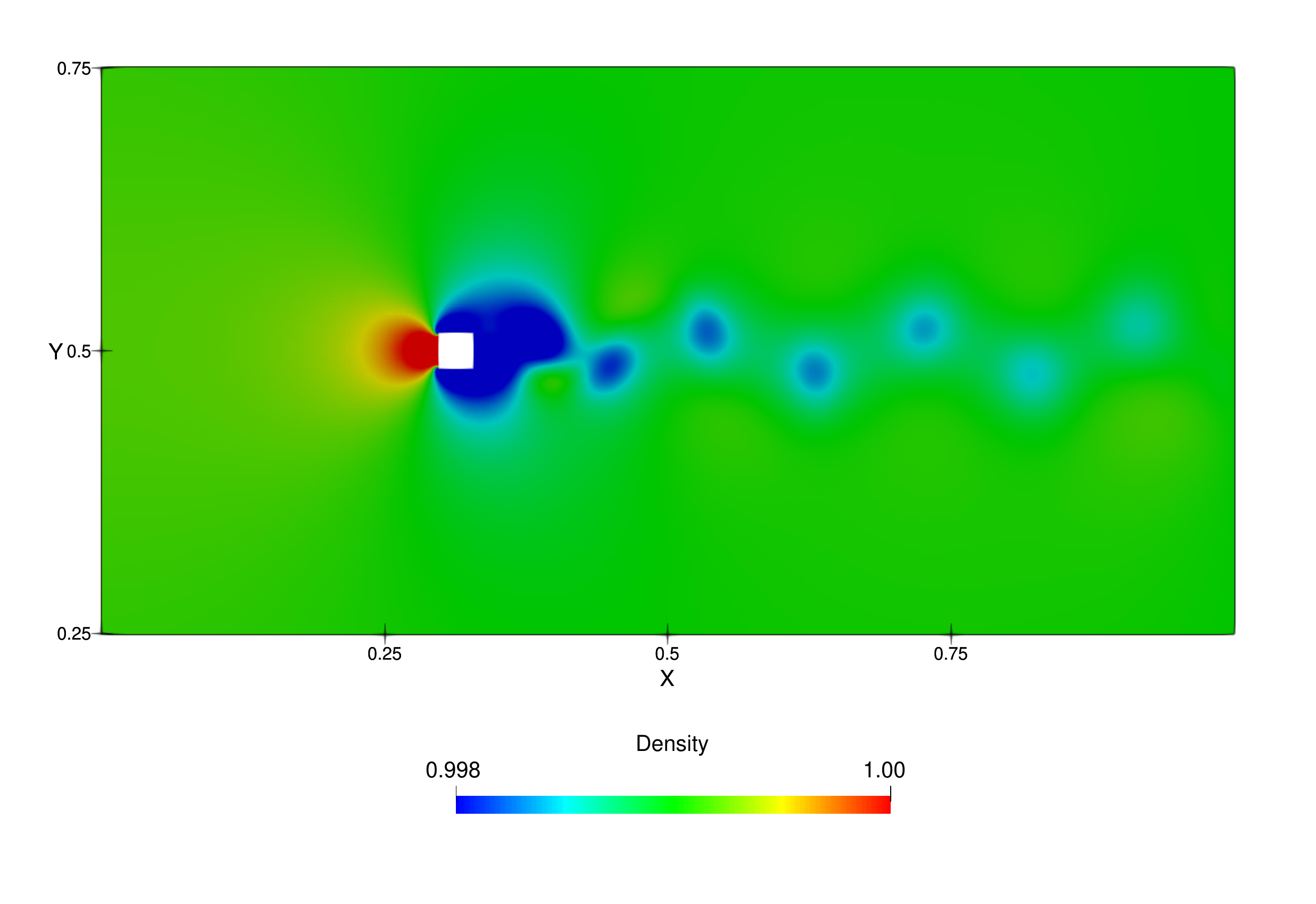}
             \includegraphics[width=0.475\textwidth]{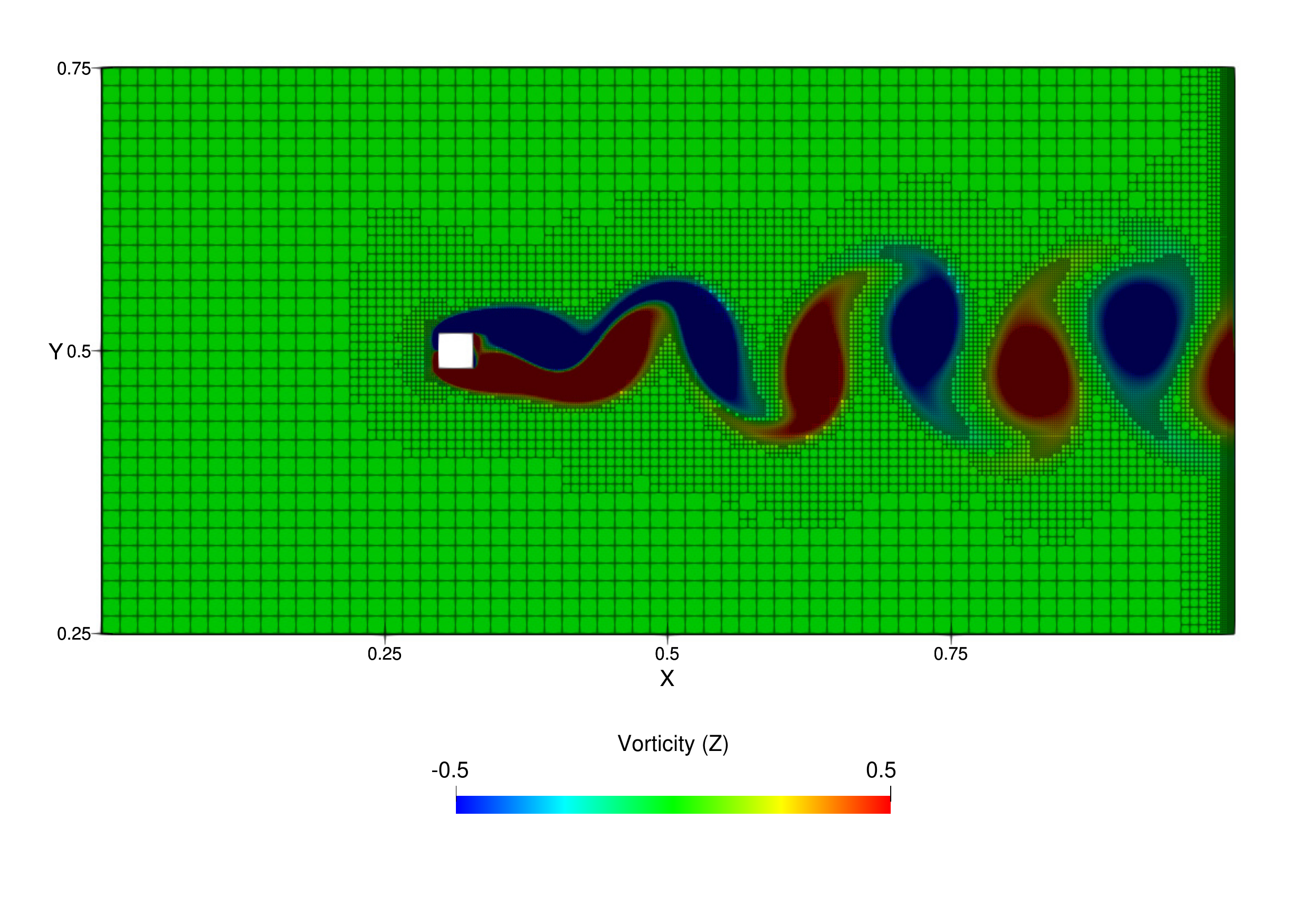}
             \caption{Simulation F-A$_4$ with $L_{\text{max.}}=5$.}
             \label{fig:y equals x}
        \end{subfigure}
           \caption{Density (left) and vorticity fields with overlaid grid outline (right) for simulations F-A$_3$, F-A$_4$ and F-A$_5$.}
           \label{fig:tests_sim_surface_fpsc}
    \end{figure}
    where $\rho_0$ is the characteristic density (set to unity for all simulations), and $f_s$ is the vortex-shedding frequency, are computed and used to validate the implementation of open boundaries. These are compared across root grid size and fine grid effective resolution in Tables \ref{tab:tests_sim_values_fpsc_A} and \ref{tab:tests_sim_values_fpsc_B}, respectively. $\overline{C_D}$ and St are also compared in value with literature data for the FPSC at Re$=100$ in Table \ref{tab:tests_sim_lit_fpsc}. Values are sampled at $140 \leq t \leq 150$ every 16 iterations on the root grid. Density and vorticity magnitude are plotted for simulations F-A$_k|_{2 \leq k \leq 4}$ in Figure \ref{fig:tests_sim_surface_fpsc} with the computational grid outline overlaid on the latter to reveal the capture of coherent structures.

    \begin{table}[b]
        \small
        \centering
            \begin{subtable}[h]{0.495\textwidth}
                \centering
                \begin{tabular}{c||cccc}
                    \hline
                    \multirow{2}{*}{$N_{\text{eff.}}$} & \multicolumn{4}{c}{Strouhal Number, St} \\ \cline{2-5}
                     & 256 & 512 & 1024 & Unif. \\
                    \hline
                    512 & 0.1469 & - & - & 0.1470 \\
                    1024 & 0.1472 & 0.1473 & - & 0.1473  \\
                    2048 & 0.1473 & 0.1472 & 0.1472 & 0.1473 \\
                    4096 & 0.1472 & 0.1472 & 0.1472 & 0.1472 \\
                    \hline
                \end{tabular}
                \caption{Strouhal numbers according to sinusoidal fitting.}
                \label{tab:tests_sim_values_fpsc_A}
            \end{subtable}
            \begin{subtable}[h]{0.495\textwidth}
                \centering
                \begin{tabular}{c||cccc}
                    \hline
                    \multirow{2}{*}{$N_{\text{eff.}}$} & \multicolumn{4}{c}{Drag Coefficient (Linear/Cubic Interp.) $\overline{C_D}$} \\ \cline{2-5}
                     & 256 & 512 & 1024 & Unif. \\
                    \hline
                    512 & 1.581/1.5010 & - & - & 1.513 \\
                    1024 & 1.584/1.503 & 1.513/1.500 & - & 1.501 \\
                    2048 & 1.553/1.500 & 1.518/1.500 & 1.518/1.499 & 1.501 \\
                    4096 & 1.508/1.498 & 1.511/1.499 & 1.511/1.499 & 1.499 \\
                    \hline
                \end{tabular}
                \caption{Time-averaged drag coefficients.}
                \label{tab:tests_sim_values_fpsc_B}
            \end{subtable}
        \caption{Summary of results for the FPSC simulations.}
        \label{tab:tests_sim_values_fpsc}
    \end{table}
    With linear interpolation, the value of $\overline{C_D}$ varies significantly across coarse grid resolutions but eventually converges to an approximate value of 1.511 upon refinement with AMR. In contrast, a value of about 1.500 is obtained when the grid is refined uniformly. A similar AMR result was obtained by Fakhari and Lee \cite{Fakhari2014} as they refined to an effective resolution of 2048. When cubic interpolation is selected, there is an appreciable difference in the values obtained for $\overline{C}_D$, bringing them more in line with the values obtained with uniform grids of equivalent effective resolution. Although a staggering of the fine and coarse cell centers ensures continuity of the pressure field, linear interpolation is seen to degrade accuracy relative to an equivalent uniform grid. However, cubic interpolation results in a value of $\overline{C}_D$ on the coarsest root grid and lowest effective resolution nearly equal to that obtained on the finest uniform grid. A sinusoidal fit is applied to $C_L$ to estimate $f_s$. The resulting Strouhal number is found to be approximately 0.147 across all simulations, converging to 0.1472 when the effective resolution of the finest level is 4096 and varies by $\sim 0.001$ otherwise. \text{St} is nearly identical when linear and cubic interpolation are used, regardless of the root grid size or the final effective resolution. The above values also agree with the tabulated literature values of Table \ref{tab:tests_sim_lit_fpsc}.

    \begin{table}[h!]
        \small
        \centering
        \begin{tabular}{lcl|cc}
            \hline
            \hline
            Reference & Year & Far-Field Type & $\overline{C_D}$ & St \\
            \hline
            \hline
            Saha et al. \cite{Saha2000} & 2000 & Slip Wall & 1.51 & 0.159 \\
            Singh et al. \cite{Singh2009} & 2009 & Slip Wall & 1.52 & 0.15 \\
            Fakhari and Lee \cite{Fakhari2014} & 2014 & Imposed Velocity & 1.51 & 0.149 \\
            González et al. \cite{Gonzalez2019} & 2019 & Slip Wall & 1.50 & 0.145 \\
            \hline
            Present & 2024 & Imposed Velocity & 1.51 & 0.147 \\
            \hline
        \end{tabular}
        \caption{Values of $\overline{C_D}$ and St reported in the literature for the FPSC at Re$=100$.}
        \label{tab:tests_sim_lit_fpsc}
    \end{table}
    The speedups obtained against simulations with uniform grid resolutions equivalent to resolutions on the finest grid are displayed in Figure \ref{tab:tests_sim_values_fpsc_C}. Linear interpolation is used for these simulations. Included are the CPU times reported by Fakhari and Lee \cite{Fakhari2014} for their simulations with the same fine-grid resolutions. The speedup with respect to uniform grids with the current solver are much more pronounced when the root grid has a coarser resolution. At an effective resolution of 2048, speedups of 9.6, 6.6 and 4.2 are found for starting grids 256$^{N_d}$, 512$^{N_d}$, and 1024$^{N_d}$, respectively. When compared with the results of the CPU-based\footnote{The GeForce 970M GPU was released in 2014, the same year in which the work of Fakhari and Lee \cite{Fakhari2014} was published.} AMR and uniform grid simulations, the speedups become 12.4, 8.6, 5.5, and 48.0, 33.2, 21.1 for the three choices of starting grid.

    \begin{table}[h]
        \footnotesize
        \centering
        \begin{tabular}{c||ccc|c|cc}
            \hline
            \hline
            \multirow{2}{*}{$N_{\text{eff.}}$} & \multicolumn{4}{c|}{Present Total Times (s)} & \multicolumn{2}{c}{Fakhari and Lee \cite{Fakhari2014}} \\ \cline{2-7}
             & 256 & 512 & 1024 & Uniform & CPU (AMR) & CPU (unif.) \\
            \hline
            \hline
            512 & \phantom{0}84 (x1.9) & - & - & 157 & 331 & 701 \\
            1024 & 270 (x4.3) & 434 (x2.7) & - & 1157 & 1827 & 6074 \\
            2048 & 976 (x9.6) & 1405 (x6.6) & 2207 (x4.2) & 9327 & 12081 & 46652 \\
            \hline
        \end{tabular}
        \caption{Total simulation times and associated speedup with respect to the uniform grid equivalent. CPU times reported by Fakhari and Lee \cite{Fakhari2014} are included.}
        \label{tab:tests_sim_values_fpsc_C}
    \end{table}

%% file: p4_conclusion.tex
\section{Conclusion} \label{sec:conc}

This paper presents an algorithm that enables efficient GPU-native adaptive mesh refinement on block-based, cell-centered grids managed as a forest of octrees. An open-source C++/CUDA code implements the algorithm together with a solver based on the Lattice Boltzmann Method. Integer index lists represent the forest of octrees, with indices referring to locations in the solution and mesh metadata arrays. Fixed memory allocation, data structure selection, and ID tracking tackle the challenges posed in managing and adapting a mesh with general-purpose GPU (GPGPU) programming, such as variability of the mesh size, lack of a recursive data structure, and fragmentation due to coarsening. The paper details the implementation of the individual steps in the AMR algorithm, along with routines for coarse-fine grid communication and grid advancement for the LBM with a block-based data arrangement and in-place streaming.

The 2D/3D lid-driven cavity (LDC) problem benchmarks the implementation with a variation of the Reynolds number $1000 \leq \text{Re} \leq 10000$, number of dimensions $N_d \in \{2,3\}$, coarse mesh resolution $N_x \in \{64,128,512\}$, velocity set $N_Q \in \{9,19,27\}$, grid hierarchy size $1 \leq L_{\text{max.}} \leq 4$, and floating-point precision word length $N_p \in \{4,8\}$ (in bytes). This benchmark utilizes three GPUs (GTX 970M, V100, and A100) to demonstrate feasibility on older hardware and high performance on newer hardware. The flow past a square cylinder (FPSC) problem enables a comparison of the speedup provided by the current implementation against uniform grids with an equivalent effective resolution with CPU-based literature data. A comparison of the LDC velocity profiles with literature data validates the implementation. Streamline plots alongside the AMR grids colored by level illustrate the detection of the primary vortex and the successful capture of corner vortices. The current solver produces profiles that agree with the literature, with slight deviations at inflection points corresponding to regions of less refinement. A comparison of the time-averaged drag coefficient and Strouhal number from the current FPSC simulations with literature values demonstrates appropriate convergence of the refined grids, with respective values of $1.51$ and $0.1472$. Grid outlines overlaid on vorticity magnitude plots show that our simulations successfully capture the coherent structures in the vortex street formed past the cylinder. Cubic interpolation produces the most accurate values for the drag coefficient relative to the values from a uniform grid with the highest resolution. Linear interpolation produces pressure fields that are continuous across grid refinement interfaces.

Total execution times and estimated updates per second via the Million/Mega Lattice node Updated per Second (MLUPS) metric over time both assess solver performance. MLUPS increases as the total number of blocks in the grid hierarchy increases either by refinement over time or by the selection of a coarse grid with higher resolution, achieving maximums of 161, 1138, and 1753 on the 970M, V100, and A100, respectively. The speedup provided by choosing single- over double-precision in 3D ranges from 2.4 times on the 970M to 1.4/1.5 on the V100/A100. Simulations in 2D terminate too quickly with the given grid hierarchy size, resulting in similar performance for both choices.

A comparison of execution times with a standalone script implementing various grid arrangements highlights the impact of the index order, which tends toward a shuffled arrangement as the mesh adapts over time. We simulated the lid-driven cavity using cell- and block-based solvers with four grid arrangements: 1) a structured grid with computed neighbor indices, 2) a structured grid with neighbor indices retrieved from global memory, 3) an octree grid with un-shuffled indices, and 4) an octree grid with shuffled indices. These tests include all three velocity sets. The block-based code produces a total simulation time independent of the grid arrangement. In contrast, the cell-based code produces a total time lower than the block-based counterpart when an unshuffled arrangement is used (both structured and octree), with a speedup ranging from 1.12-1.73. However, the block-based code performs better with shuffled indices, with a speedup ranging from 2.25 to 3.00. 

We assess the performance of the GPU-native AMR implementation by analyzing the distributions of execution times by step for several simulations and the computed total time taken to adapt the mesh via refinement and coarsening as a fraction of the total solver time. Most LDC simulations maintain this fraction under 2.00\%. Efficiency improves for the FPSC simulations as the grid hierarchy size increases but remains below 10\% in all cases. The 2D cases across all GPUs present an exception, where smaller grids advance too quickly relative to the fixed cost of resetting intermediate arrays in preparation for mesh adaptation. In 3D, the pre-computation step, which prepares the mesh for inserting valid data in new children and scales with interpolation and averaging, requires the most time to resolve. Other steps remain similar by order of magnitude otherwise. Scaling tests, performed by increasing the grid hierarchy size and expanding the near-wall refinement zone to reach approximately 2.5 million simultaneously inserted blocks, demonstrate linear scaling, requiring up to 45 ms on the V100.

Future work will explore the feasibility of a multi-GPU and/or multi-node extension to the current refinement and coarsening algorithm, including a suitable load-balancing strategy (e.g., based on a space-filling curve) that efficiently partitions the grid dynamically for distribution to the various devices. The fixed array-reset and interpolation/averaging costs create bottlenecks in the mesh adaptation procedure, necessitating optimization. Although cubic interpolation greatly benefits accuracy in the FPSC test cases, its computational cost increases in 3D and for larger cell-block sizes due to the numerous accesses to global and shared memory during run-time computation of the interpolation weights. Averaging is less expensive than cubic interpolation but appreciably more expensive than linear interpolation owing to the separated global writes to parent cells and the memory access concerns similar to interpolation. Simultaneous loads of child cell DDFs are required when computing equilibrium distributions for re-scaling to avoid repeated global memory loads, which limits the number of parent DDFs that can be stored. A new class of Lattice Boltzmann Methods referred to as the simplified LBM \cite{Shu2014, Chen2017, Chen2018, Qin2022} has emerged in recent years, which recasts the Chapman-Enskog analysis to derive governing equations that update the macroscopic properties directly (as opposed to storing numerous DDFs from which these properties are recovered). Switching to the simplified LBM could benefit the grid communication routines since macroscopic properties no longer require re-scaling. The total number of macroscopic properties is fewer in count than DDFs, so there would be fewer global and shared memory accesses overall.

%% file: pA_appendix.tex
\section{Case Studies - Configuration Files}

The configuration variables used to produce the test cases considered in Section \ref{sec:tests} have been made available in Table \ref{tab:app_conf}. A single configuration file \texttt{confmake.sh} defines the solver parameters and invokes the Makefile for compilation. Modification of any of these parameters requires recompilation in the current implementation. Other unused parameters, fixed in value for all simulations or require customized adjustment, are listed in Table \ref{tab:app_conf_fixed}.

\tiny
\newpage
\begin{sidewaystable}
    \centering
    \begin{tabular}{l|ccccccccc}
        \hline
        Parameter&G/V/A$_1$&G/V/A$_1$&G/V/A$_1$&G/V/A$_1$&G/V/A$_1$&G/V/A$_1$&V-S$_1$&V-S$_2$&V-S$_3$\\
        \hline
        \texttt{N\_PRECISION}&0&1&0&1&0&1&0&0&0\\
        \texttt{N\_Q}&9&9&19&19&27&27&9&9&9\\
        \texttt{MAX\_LEVELS}&4&4&38050&38050&38050&38050&4&4&3\\
        \texttt{L\_c}&1&1&1&1&1&1&1&1&1\\
        \texttt{L\_fy}&1&1&1&1&1&1&1&1&1\\
        \texttt{v0}&0.00005&0.00005&0.00005&0.00005&0.00005&0.00005&0.000015625&0.00001&0.000005\\
        \texttt{Nx}&128&128&64&64&64&64&64&64&512\\
        \texttt{N\_CASE}&0&0&0&0&0&0&0&0&0\\
        \texttt{N\_REFINE\_START}&-2&-2&-2&-2&-2&-2&-2&-2&-2\\
        \texttt{N\_PROBE\_FORCE}&0&0&0&0&0&0&0&0&0\\
        \texttt{N\_PROBE\_AVE}&0&0&0&0&0&0&0&0&0\\
        \texttt{P\_PRINT}&1000*$N_x$&1000*$N_x$&1000*$N_x$&1000*$N_x$&1000*$N_x$&1000*$N_x$&1000*$N_x$&1000*$N_x$&1000*$N_x$\\
        \hline
        Parameter&V-S$_4$&F-A$_{1,...,4}$&F-B$_{1,...,4}$&F-C$_{1,...,3}$&F-D&F-E&-&-&-\\
        \hline
        \texttt{N\_PRECISION}&0&1&1&1&1&1&-&-&-\\
        \texttt{N\_Q}&19&9&9&9&9&9&-&-&-\\
        \texttt{MAX\_LEVELS}&4&2,...,5&1,...,4&1,...,3&1&1&-&-&-\\
        \texttt{L\_c}&1&1&1&1&1&1&-&-&-\\
        \texttt{L\_fy}&1&0.5&0.5&0.5&0.5&0.5&-&-&-\\
        \texttt{v0}&0.000015625&0.000015625&0.000015625&0.000015625&0.000015625&0.000015625&-&-&-\\
        \texttt{Nx}&128&256&512&1024&2048&4096&-&-&-\\
        \texttt{N\_CASE}&0&1&1&1&1&1&-&-&-\\
        \texttt{N\_REFINE\_START}&-2&-3&-3&-3&-3&-3&-&-&-\\
        \texttt{N\_PROBE\_FORCE}&0&1&1&1&1&1&-&-&-\\
        \texttt{N\_PROBE\_AVE}&0&0&0&0&0&0&-&-&-\\
        \hline
    \end{tabular}
    \caption{Solver parameters for the case studies of Section \ref{sec:tests}.}
    \label{tab:app_conf}
\end{sidewaystable}
\newpage
\normalsize

\begin{table}[h!]
    \centering
    \begin{tabular}{lcc}
        \hline
        Parameter & Value & Desc. \\
        \hline
        \texttt{PERIODIC\_X} &  & \multirow{9}{*}{Unused} \\
        \texttt{PERIODIC\_Y} &  & \\
        \texttt{PERIODIC\_Z} &  & \\
        \texttt{S\_INTERP\_TYPE} & 1 & \\
        \texttt{S\_INIT\_TYPE} & 0 & \\
        \texttt{N\_PROBE} & 0 & \\
        \texttt{N\_PROBE\_DENSITY} & 4 & \\
        \texttt{N\_PROBE\_FREQUENCY} & 32 & \\
        \texttt{V\_PROBE\_TOL} & 0.0001 & \\
        \hline
        \texttt{MAX\_LEVELS\_INTERIOR} & \texttt{MAX\_LEVELS} & \multirow{12}{*}{Unused} \\
        \texttt{L\_fz} & 1 &  \\
        \texttt{B\_TYPE} & 1 & \\
        \texttt{S\_LES} & 0 & \\
        \texttt{P\_REFINE} & 32 & \\
        \texttt{N\_REFINE\_INC} & 1 & \\
        \texttt{N\_CONN\_TYPE} & 1 & \\
        \texttt{P\_SHOW\_REFINE} & 1 & \\
        \texttt{N\_PROBE\_F\_FREQUENCY} & 16 & \\
        \texttt{N\_PROBE\_AVE\_FREQUENCY} & 10 & \\
        \texttt{N\_PROBE\_AVE\_START} & 100*\texttt{Nx} & \\
        \texttt{N\_PRINT} & 1 & \\
        \hline
        \texttt{N\_LEVEL\_START} & 0 & \multirow{6}{*}{Custom} \\
        \texttt{P\_DIR\_NAME} & ../out/ & \\
        \texttt{N\_PRINT\_LEVELS} & 1 & \\
        \texttt{P\_SHOW\_ADVANCE} & 0 & \\
        \texttt{P\_PRINT\_ADVANCE} & 0 & \\
        \texttt{N\_REGEN} & 1 & \\
        \hline
    \end{tabular}
    \caption{Solver parameters which were unused, fixed in value or customized depending on desired output.}
    \label{tab:app_conf_fixed}
\end{table}

%% file: nomenclature.tex
\nomenclature{AMR}{Adaptive Mesh Refinement}
\nomenclature{CFD}{Computational Fluid Dynamics}
\nomenclature{CPU}{Central Processing Unit}
\nomenclature{DDF}{Density Distribution Function}
\nomenclature{GPU}{Graphics Processing Unit}
\nomenclature{GPGPU}{General-Purpose Graphics Processing Unit}
\nomenclature{LBM}{Lattice Boltzmann Method}
\nomenclature{MPI}{Message Passing Interface}
\nomenclature{SAMR}{Structured Adaptive Mesh Refinement}
\nomenclature{SPC}{Space-Filling Curve}
\nomenclature{$\textbf{c}_p$}{Particle velocity vector}
\nomenclature{$c_s$}{Lattice speed of sound}
\nomenclature{$\textbf{e}_p$}{Unit vector in the direction of particle velocity}
\nomenclature{$f_p$}{Density distribution function}
\nomenclature{$f_p^{\text{eq}}$}{Equilibrium distribution function}
\nomenclature{$p$}{Pressure}
\nomenclature{$\textbf{u}$}{Macroscopic velocity vector}
\nomenclature{$w_p$}{Quadrature weight}
\nomenclature{$\Delta t$}{Time step}
\nomenclature{$\Delta x$}{Lattice spacing}
\nomenclature{$\Omega$}{Collision operator}
\nomenclature{$\rho$}{Density}
\nomenclature{$\tau$}{Relaxation time}
\nomenclature{$St$}{Strouhal number}
\nomenclature{$\overline{C}_D$}{Time-averaged drag coefficient}
\nomenclature{$t$}{Time in seconds}
\nomenclature{$U$}{Streamwise velocity}
\nomenclature{$U_\infty$}{Freestream velocity}
\nomenclature{$x$}{Cartesian coordinate axis}
\nomenclature{$y$}{Cartesian coordinate spanwise axis}
\nomenclature{$z$}{Cartesian coordinate vertical axis}
\nomenclature{$\mathcal{O}$}{Order of magnitude}
\nomenclature{AMR}{Adaptive Mesh Refinement}
\nomenclature{BGK}{Bhatnagar-Gross-Krook}
\nomenclature{CFD}{Computational Fluid Dynamics}
\nomenclature{CPU}{Central Processing Unit}
\nomenclature{CUDA}{Compute Unified Device Architecture}
\nomenclature{DDF}{Density Distribution Function}
\nomenclature{FPSC}{Flow Past a Square Cylinder}
\nomenclature{GPU}{Graphics Processing Unit}
\nomenclature{GPGPU}{General-Purpose Graphics Processing Unit}
\nomenclature{LBM}{Lattice Boltzmann Method}
\nomenclature{LDC}{Lid-Driven Cavity}
\nomenclature{MLUPS}{Million Lattice node Updates Per Second}
\nomenclature{$\textbf{c}_p$}{Particle velocity vector}
\nomenclature{$c_s$}{Lattice speed of sound}
\nomenclature{$d$}{Distance from the nearest wall}
\nomenclature{$f_p$}{Density distribution function}
\nomenclature{$f_p^*$}{Post-collision density distribution function}
\nomenclature{$f_p^{\text{eq}}$}{Equilibrium distribution function}
\nomenclature{$f_p^{\text{neq}}$}{Non-equilibrium distribution function}
\nomenclature{$L$}{Grid level}
\nomenclature{$L_{\text{max.}}$}{Maximum grid level}
\nomenclature{$L_{\text{des.}}$}{Desired grid level}
\nomenclature{$N_d$}{Number of dimensions}
\nomenclature{$N_p$}{Floating-point precision word length (in bytes)}
\nomenclature{$N_Q$}{Number of discrete velocities}
\nomenclature{$N_x$}{Root grid resolution}
\nomenclature{$N_{\text{start}}$}{Start parameter for refinement criterion}
\nomenclature{$N_{\text{inc.}}$}{Incremental parameter for refinement criterion}
\nomenclature{$\text{Re}$}{Reynolds number}
\nomenclature{$\textbf{u}$}{Macroscopic velocity vector}
\nomenclature{$\textbf{x}$}{Spatial location}
\nomenclature{$\epsilon$}{Logarithmic refinement criterion}
\nomenclature{$\nu$}{Kinematic viscosity}
\nomenclature{$\Pi_{\alpha\beta}$}{Second moment tensor}
\nomenclature{$\rho$}{Density}
\nomenclature{$\tau$}{Relaxation time}
\nomenclature{$\omega$}{Relaxation parameter}
\nomenclature{$\omega'$}{Complementary relaxation parameter}